\documentclass[aps,prb,twocolumn,superscriptaddress]{revtex4}
\usepackage{graphicx}
\usepackage{siunitx}
\usepackage{bm}
\usepackage[caption=false]{subfig}
\usepackage{mathtools}
\usepackage{multirow}
\usepackage{amssymb}
\usepackage{braket}
\usepackage{amsmath}
\usepackage{rotating}
\usepackage{placeins}
\usepackage{cancel}
\usepackage{siunitx}
\usepackage{xcolor}
\usepackage{adjustbox}
\usepackage{afterpage}
\usepackage{spreadtab}
\usepackage{tabularx}
\newcommand\setrow[1]{\gdef\rowmac{#1}#1\ignorespaces}
\newcommand\clearrow{\global\let\rowmac\relax}

\clearrow

\definecolor{Gray}{gray}{0.9}
\newcommand{\bb}{\setrow{\bfseries}}

\begin{document}

\title{First-principles predictions of Hall and drift mobilities in semiconductors}

\author{Samuel Ponc\'e}
\email{samuel.ponce@epfl.ch}
\affiliation{%
Theory and Simulation of Materials (THEOS), \'Ecole Polytechnique F\'ed\'erale de Lausanne,
CH-1015 Lausanne, Switzerland
}%
\author{Francesco Macheda}
\affiliation{%
Department of Physics, King's College London, Strand, London WC2R 2LS, United Kingdom
}
\author{Elena Roxana Margine}
\affiliation{%
Department of Physics, Applied Physics and Astronomy, Binghamton University-SUNY, Binghamton, NY 13902, USA
}
\author{Nicola Marzari}
\affiliation{%
Theory and Simulation of Materials (THEOS), \'Ecole Polytechnique F\'ed\'erale de Lausanne,
CH-1015 Lausanne, Switzerland
}%
\author{Nicola Bonini}
\affiliation{%
Department of Physics, King's College London, Strand, London WC2R 2LS, United Kingdom
}
\author{Feliciano Giustino}
\affiliation{%
Oden Institute for Computational Engineering and Sciences, The University of Texas at Austin, Austin, Texas 78712, USA
}%
\affiliation{%
Department of Physics, The University of Texas at Austin, Austin, Texas 78712, USA
}%

\date{\today}

\begin{abstract}
Carrier mobility is one of the defining properties of semiconductors.
Significant progress on parameter-free calculations of carrier mobilities in real materials has been made during the past decade; however,
the role of various approximations remains unclear and a unified methodology is lacking.
Here, we present and analyse a comprehensive and efficient approach to compute the intrinsic, phonon-limited drift and Hall carrier mobilities of semiconductors, within the framework of the first-principles Boltzmann transport equation.
The methodology exploits a novel approach for estimating quadrupole tensors and including them in the electron-phonon interactions, and capitalises on a rigorous and efficient procedure for numerical convergence.
The accuracy reached in this work allows to assess common approximations, including the role of exchange and correlation functionals, spin-orbit coupling, pseudopotentials, Wannier interpolation, Brillouin-zone sampling, dipole and quadrupole corrections, and the relaxation-time approximation.
A detailed analysis is showcased on ten prototypical semiconductors, namely diamond, silicon, GaAs, 3C-SiC, AlP, GaP, c-BN, AlAs, AlSb, and SrO.
By comparing this extensive dataset with available experimental data, we explore the intrinsic nature of phonon-limited carrier transport and magnetotransport phenomena in these compounds.
We find that the most accurate calculations predict Hall mobilities up to a factor of two larger than experimental data;
this could point to promising experimental improvements in the samples quality, or to the limitations of density-function theory in predicting the
carrier effective masses and overscreening the electron-phonon matrix elements.
By setting tight standards for reliability and reproducibility, the present work aims to facilitate validation and verification of data and software towards predictive calculations of transport phenomena in semiconductors.
\end{abstract}

\maketitle

\section{Introduction}

The ability of metals and semiconductors to transport electrical charges is a fundamental property in manifold applications, ranging from solar cells, light-emitting devices, thermoelectric, transparent conductors, photodetectors, photo-catalysts, and
transistors~\cite{Sirringhaus1999,Ellmer2012,Green2014,Loevvik2018,Li2019b}.
Predicting transport properties from first-principles~\cite{Ponce2020} thus offers a powerful tool for the design of new materials and devices.

The aim of this work is to to assess the reliability and predictive power of novel first-principles methods, using III-V and group IV semiconductors as a representative benchmark.
Already in 1966, in a seminal work Cohen and Bergstresser~\cite{Cohen1966} computed the band structure of 14 semiconductors with the diamond and zincblende structure.
In 2013, Malone and Cohen~\cite{Malone2013} repeated this task computing the quasiparticle bandstructure in the many-body $GW$ approximation, including spin-orbit coupling (SOC) effects.
Last year, Miglio \textit{et al.}~\cite{Miglio2020} studied the zero-point renormalization of the electronic band gap of 30 semiconductors, mostly of the zincblende, wurtzite and rocksalt crystal structure.
Here, we deliver an accurate and efficient calculation method for carrier mobilities, including spin-orbit coupling, dynamical quadrupoles and magnetic Hall effects, focusing on prototypical semiconductors.
Our benchmark includes the following ten semiconductors: diamond, Si, GaAs, 3C-SiC, AlP, GaP, cubic BN, AlAs, AlSb and SrO.
For five of these no first-principles calculations of carrier mobility have been reported to date.

Due to the complexity of computing carrier mobilities fully from first principles, the first calculation appeared only in 2009~\cite{Restrepo2009}.
Since then, only 19 bulk semiconductors have been investigated:
Si~\cite{Restrepo2009,Li2015,Fiorentini2016,Ma2018,Ponce2018,Park2020a,Brunin2020a},
Diamond~\cite{Macheda2018},
GaAs~\cite{Zhou2016,Liu2017,Ma2018,Lee2020,Protik2020,Brunin2020},
GaN~\cite{Ponce2019b,Ponce2019c,Jhalani2020},
3C-SiC~\cite{Meng2019,Protik2020a},
GaP~\cite{Brunin2020},
GeO$_2$~\cite{Bushick2020a},
SnSe~\cite{Ma2018a},
SnSe$_2$~\cite{Jalil2020},
BAs~\cite{Liu2018,Bushick2020},
PbTe~\cite{Cao2018a,DSouza2020},
naphtalene~\cite{Lee2018a,BrownAltvater2020},
Bi$_2$Se$_3$~\cite{Cepellotti2020},
Ga$_2$O$_3$~\cite{Ghosh2016,Mengle2019,Ma2020,Ponce2020a},
TiO$_2$~\cite{Kang2019},
SrTiO$_3$~\cite{Zhou2018a,Zhou2019a},
PbTiO$_3$~\cite{Park2020a},
Rb$_3$AuO~\cite{Zhao2020}, and
CH$_3$NH$_3$PbI$_3$~\cite{Ponce2019}.

In this work, we perform an in-depth analysis of the predictive accuracy of the first-principles Boltzmann transport equation (BTE) for computing the carrier drift mobility and the Hall mobility, and we compare directly to experimental data.
In brief, we find that spin-orbit coupling plays an important role for hole mobilities, that a local approximation to the velocity matrix elements (also used by us in earlier work) is not accurate enough, and that optimizing the construction of the Wigner-Seitz cell used in Wannier interpolations can accelerate the convergence of the Brillouin zone integrals.
We confirm the recent findings~\cite{Brunin2020a,Brunin2020,Jhalani2020} that dynamical quadrupoles, the next order of correction to dynamical dipoles, can
significantly affect carrier mobilities, and we show that this effect arises predominantly from the change of the vibrational eigenmodes induced by the quadrupole term of the dynamical matrix.

In this work we also find that Wannier-function interpolations (especially those associated with the conduction bands) converge more slowly than previously estimated with respect to the sampling of the coarse grid employed in the Wannier representation.
As a result, in order to obtain carrier mobilities with an accuracy of 1\% for fixed fine grids, it is often necessary to use coarse grids including up to 18$^3$ points, still providing a speed-up of two to three orders of magnitude with respect to brute force sampling.
We also find that a good indicator of convergence is provided by the mobility effective mass, as defined below.
The sampling of Wannier-interpolated quantities require between 80$^3$ and 250$^3$ \textbf{k} and \textbf{q} grids, with the exception of the electron mobility of GaAs which requires a 500$^3$ grid due to its very small electron effective mass.
We also show that the mobility and Hall factor converge linearly with grid spacing, allowing for a simple extrapolation~\cite{Ponce2015} which achieves an accuracy of 1\%.

This work answers crucial methodological and physical questions.
On the technical side, it establishes well-defined criteria for high-accuracy calculations of transport coefficients, clarifying the role and the level of (i) the approximations in the underlying first-principles calculation to extract materials parameters (exchange-and-correlation functionals, pseudopotentials, lattice parameters and SOC),
(ii) the interpolation procedures for quasiparticle dispersions and interactions, and
(iii) the numerical techniques that ensure the efficient convergence of transport coefficients.
On the physical side, the ability to describe with high accuracy transport phenomena provides detailed access to the carrier dynamics and to the inherent limits of the electrical transport properties of key prototypical semiconductors.
Crucially, since usually it is the Hall mobility that is measured rather than the drift mobility, this work allows a detailed and unambiguous comparison with experimental data.

The manuscript is organized as follows.
In Section~\ref{sec:theory} we briefly present the theory for computing phonon-limited drift mobilities.
We then extend the theory to include a small finite external magnetic field, to access the Hall mobility.
We also show how to efficiently compute mobility using interpolations based on maximally-localized Wannier functions (MLWF)~\cite{Marzari2012}.
In particular, we discuss how to compute exact velocities and estimate quadrupole tensors, and include dipole-dipole, dipole-quadrupole and quadrupole-quadrupole corrections during the Fourier interpolation of the interatomic force constant, as well as how to include dynamical quadrupoles in the interpolation of the electron-phonon matrix element.
Finally, we discuss the use of adaptive broadening for improved convergence.

In Section~\ref{computsection}, we present the computational methodology.
We first report the methodology and parameters used, then outline the Wannier interpolation procedure for electrons, phonons, and electron-phonon matrix elements.
Next, we report our interpolation of the electron-phonon matrix elements and deformation potential including dynamical quadrupole, and show that the interpolated values reproduce density functional perturbation theory calculations.
We then discuss typical coarse and fine grid convergence rates.
We conclude the section by showing that the mobility and the Hall factor converge linearly with the spacing of the fine grid, allowing for a linear extrapolation to the limit of exact momentum integration.

In Section~\ref{resultsection} we present and discuss our results, starting with analysis of the band structures, phonon dispersions, and effective masses.
We then assess the quality of various popular approximations, such as the neglect of the non-local pseudopotential contribution to the electron velocity, the neglect of SOC, the neglect of quadrupole correction, the use of the relaxation time approximation, the effect of the isotropic Hall factor approximation, the effect of the exchange-correlation functional, the effect of lattice parameters and pseudopotential choice.
In addition, we identify the dominant scattering mechanisms responsible for limiting the carrier mobility.
Finally, we conduct an in-depth comparison with available experimental mobility data, and assess the overall predictive power of the BTE.
We offer our conclusions in Section~\ref{sec:conclusion}.

\section{Theory}\label{sec:theory}

In this Section we first present the main equations for calculating the drift mobility, and their generalization to the case of vanishing magnetic field, yielding the Hall mobility.
We then present the Wannier interpolation scheme employed here, and discuss how to obtain accurate electron velocities.
We also discuss the multipole expansion of the long-range part of the dynamical matrix and electron-phonon matrix elements.
Finally, we describe how we use adaptive broadening in practical calculations.

\subsection{Drift mobility}\label{sec:drift_mob}

The low-field phonon-limited charge carrier mobility is calculated as~\cite{Restrepo2009, Li2015, Fiorentini2016, Zhou2016, Ghosh2016, Liu2017, Ma2018, Ponce2018, Macheda2018, Ma2018a, Cao2018a, Lee2018a, Zhou2018a, Liu2018, Kang2019, Zhou2019a, Mengle2019, Ponce2019b, Ponce2019c, Ponce2019, Park2020a,Brunin2020a,Brunin2020, Lee2020, Protik2020, Jhalani2020, Protik2020a, Bushick2020a, Jalil2020, Bushick2020, DSouza2020, BrownAltvater2020, Cepellotti2020, Ma2020, Ponce2020a, Park2020a, Zhao2020}:
\begin{equation}\label{eq:btemob}
\mu_{\alpha\beta} = \frac{-1}{V_{\rm uc}n_{\rm c}}\sum_n \int \frac{\mathrm{d}^3k}{\Omega_{\rm BZ}} v_{n\mathbf{k}\alpha} \partial_{E_\beta} f_{n\mathbf{k}}.
\end{equation}
Here $\alpha$, $\beta$ run over the three Cartesian directions and  $\partial_{E_{\beta}} f_{n\mathbf{k}} \equiv (\partial f_{n\mathbf{k}}/\partial E_\beta)|_{\mathbf{E}=\mathbf{0}}$ is the linear variation of the electronic occupation function $f_{n\mathbf{k}}$ in response to the electric field $\mathbf{E}$, $V_{\rm uc}$ is the unit cell volume, $\Omega_{\rm BZ}$ the first Brillouin zone volume, $v_{n\mathbf{k}\alpha} = \hbar^{-1} \partial \varepsilon_{n\mathbf{k}}/\partial k_{\alpha}$ is the band velocity for the Kohn-Sham state $\varepsilon_{n\mathbf{k}}$, and $n_{\rm c} = (1/V_{\rm uc})\sum_n \int (\mathrm{d}^3 k/ \Omega_{\rm BZ}) f^0_{n\mathbf{k}}$ is the carrier concentration.
$f_{n\mathbf{k}}^0$ is the Fermi-Dirac occupation function at equilibrium (in the absence of fields).
We follow the notation of Refs.~\onlinecite{Ponce2018, Ponce2020}.
The BTE describes the detailed balance between the carriers' populations moving through phase space under the action of the driving electric field \textbf{E} and the carriers scattered by phonons.
It can be derived from the Kadanoff-Baym equation of motion by approximating the Hartree and exchange-correlation potential, taking the diagonal Bloch state
projection and assuming a spatially homogenous electric field~\cite{Ponce2020}.
One then obtains the quantum time-dependent BTE that can be further simplified by considering a time-independent external field (DC), static electron-one-phonon interactions and adiabatic phonons to give:
\begin{multline}\label{eq:fullBTE}
e \mathbf{E}\cdot \frac{1}{\hbar} \frac{\partial f_{n\mathbf{k}}}{\partial \mathbf{k}} = \frac{2\pi}{\hbar} \sum_{m,\nu} \int \frac{d^3q}{\Omega_{\rm BZ}} |g_{mn\nu}(\mathbf{k,q})|^2 \\
\times \big[ f_{n\mathbf{k}} (1-f_{m\mathbf{k+q}})\delta(\varepsilon_{n\mathbf{k}}-\varepsilon_{m\mathbf{k+q}}+\hbar\omega_{\mathbf{q}\nu})n_{\mathbf{q}\nu} \\
+ f_{n\mathbf{k}} (1-f_{m\mathbf{k+q}})\delta(\varepsilon_{n\mathbf{k}}-\varepsilon_{m\mathbf{k+q}} - \hbar\omega_{\mathbf{q}\nu})(n_{\mathbf{q}\nu} + 1) \\
- (1- f_{n\mathbf{k}})f_{m\mathbf{k+q}}\delta(\varepsilon_{m\mathbf{k+q}} - \varepsilon_{n\mathbf{k}} - \hbar\omega_{\mathbf{q}\nu})(n_{\mathbf{q}\nu} + 1) \\
- (1- f_{n\mathbf{k}})f_{m\mathbf{k+q}}\delta(\varepsilon_{m\mathbf{k+q}} - \varepsilon_{n\mathbf{k}} + \hbar\omega_{\mathbf{q}\nu})n_{\mathbf{q}\nu} \big],
\end{multline}
where $n_{\mathbf{q}\nu}$ is the Bose-Einstein distribution.
The electron-phonon matrix elements $g_{mn\nu}(\mathbf{k,q})$ are the amplitude for scattering from an initial state $n\mathbf{k}$ to a final state $m\mathbf{k+q}$ via the emission or absorption of a phonon of frequency $\omega_{\mathbf{q}\nu}$.
We obtain $\partial_{E_\beta} f_{n\mathbf{k}}$ required for Eq.~\eqref{eq:btemob} by taking the field derivative of Eq.~\eqref{eq:fullBTE} at vanishing field.
This yields the linearized Boltzmann transport equation (BTE)~\cite{Ponce2018,Ponce2020}:
\begin{align}\label{eq:iter}
\partial_{E_{\beta}} f_{n\mathbf{k}} =& e v_{n\mathbf{k}\beta} \frac{\partial f_{n\mathbf{k}}^0}{\partial \varepsilon_{n\mathbf{k}}} \tau_{n\mathbf{k}} + \frac{2\pi\tau_{n\mathbf{k}}}{\hbar}
  \sum_{m\nu} \!\int\! \frac{\mathrm{d}^3 q}{\Omega_{\mathrm{BZ}}} | g_{mn\nu}(\mathbf{k},\mathbf{q})|^2 \nonumber \\
 \times & \Big[(n_{\mathbf{q}\nu}+1-f_{n\mathbf{k}}^0)\delta(\varepsilon_{n\mathbf{k}} - \varepsilon_{m\mathbf{k+q}}  + \hbar \omega_{\mathbf{q}\nu} ) \nonumber  \\
  +  (n_{\mathbf{q} \nu} +& f_{n\mathbf{k}}^0)\delta(\varepsilon_{n\mathbf{k}} - \varepsilon_{m\mathbf{k+q}}  - \hbar \omega_{\mathbf{q}\nu}) \Big] \partial_{E_{\beta}} f_{m\mathbf{k}+\mathbf{q}} ,
\end{align}
with  $\tau_{n\mathbf{k}}$ being the total scattering lifetime and the inverse $\tau_{n\mathbf{k}}^{-1}$  is the scattering rate given by:
\begin{align}\label{eq:scattering_rate}
  \tau_{n\mathbf{k}}^{-1} =& \frac{2\pi}{\hbar} \sum_{m\nu} \!\int\! \frac{d^3 q}{\Omega_{\text{BZ}}} | g_{mn\nu}(\mathbf{k,q})|^2 \nonumber \\
  \times & \big[ (n_{\mathbf{q}\nu} +1 - f_{m\mathbf{k+q}}^0) \delta( \varepsilon_{n\mathbf{k}} - \varepsilon_{m\mathbf{k+q}}   -  \hbar \omega_{\mathbf{q}\nu}) \nonumber \\
   + &  (n_{\mathbf{q}\nu}  +   f_{m\mathbf{k+q}}^0 )\delta(\varepsilon_{n\mathbf{k}} - \varepsilon_{m\mathbf{k+q}}  +  \hbar \omega_{\mathbf{q}\nu}) \big].
\end{align}
A common approximation that we refer to as the self-energy relaxation time approximation (SERTA) consists in neglecting the second term on the right-hand side of Eq.~\eqref{eq:iter}.
The mobility then takes the simpler form~\cite{Ponce2018}:
\begin{equation}\label{eq:serta}
\mu_{\alpha\beta}^{\rm SERTA} = \frac{-e}{V_{\rm uc} n_{\rm c}}\sum_n \int \frac{\mathrm{d}^3 k}{\Omega_{\rm BZ}} \frac{\partial f_{n\mathbf{k}}^0}{\partial \varepsilon_{n\mathbf{k}}} v_{n\mathbf{k}\alpha} v_{n\mathbf{k}\beta} \tau_{n\mathbf{k}}.
\end{equation}

In order to analyze the role of band curvature in mobility calculations, we introduce a ``mobility effective mass''
as follows:
\begin{equation}\label{eq:effectivemassmob}
\frac{1}{m_{\rm mob}^*} = \frac{-1}{3 V_{\rm uc}n_{\rm c}} \! \sum_{\alpha n} \!\! \int \frac{\mathrm{d}^3k}{\Omega_{\rm BZ}} \frac{\partial f_{n\mathbf{k}}^0}{\partial \varepsilon_{n\mathbf{k}}} v^2_{n\mathbf{k}\alpha}.
\end{equation}
This definition is obtained by setting $\tau_{n\mathbf{k}}$ to a constant value $\tau$ in Eq.~\eqref{eq:serta},
and by requiring that the resulting mobility can be expressed in the standard Drude form with the
same relaxation time $\tau$:
\begin{equation}\label{eq:Drudeformula}
\mu = \frac{e \tau}{m_{\rm mob}^*}.
\end{equation}
Lastly, we introduce an approximate version of the SERTA mobility, which is useful to disentangle the convergence of electron band structures from that of the phonon energies and electron-phonon matrix elements.
To this aim we define the ``constant adiabatic relaxation time approximation'' (CARTA) as the mobility obtained by taking the SERTA of Eq.~\eqref{eq:serta} with a constant electron-phonon matrix elements and vanishing phonon frequencies: $|g_{mn\nu}(\mathbf{k},\mathbf{q})|=g^2$, with $g$ an arbitrary constant with units of energy, and $\omega_{\mathbf{q}\nu} = 0$:
\begin{equation}\label{eq:carta}
\! \mu_{\alpha\beta}^{\rm CARTA} \!\!=\! \frac{-e\hbar g^2}{2\pi V_{\rm uc} n_{\rm c}}\! \! \sum_n \!\! \int \frac{\mathrm{d}^3 k}{\Omega_{\rm BZ}} \frac{\frac{\partial f_{n\mathbf{k}}^0}{\partial \varepsilon_{n\mathbf{k}}} v_{n\mathbf{k}\alpha} v_{n\mathbf{k}\beta}}{
 \! \sum_{m}\! \! \int \frac{d^3 q}{\Omega_{\rm BZ}} \delta(\varepsilon_{n\mathbf{k}}\!-\!\varepsilon_{m\mathbf{k+q}})}.
\end{equation}

\subsection{Hall mobility}\label{sec:hall_mob}
Calculation of the direct current Hall coefficient has seen a resurgence of interest~\cite{Macheda2018,Macheda2020,Wang2020a,Desai2021}.
Experimentally, Hall mobility measurements are more common than time-of-flight measurements of drift mobilities due to their superior accuracy and simplicity.
This is reflected in the literature by a significantly higher number of available Hall mobility data compared to drift mobilities.
However Hall measurements are performed in the presence of an external finite magnetic field, which introduces an additional Lorentz force on the carriers, thereby altering the mobility.
To compare with experiment, we have to augment the BTE of Eq.~\eqref{eq:iter} to account for the additional magnetic field \textbf{B}~\cite{Macheda2018,Ponce2020}:
\begin{multline}\label{eq:iterwithbimpl}
 \Big[ 1 - \frac{e}{\hbar}\tau_{n\mathbf{k}} ({\bf v}_{n\mathbf{k}} \times {\bf B}) \cdot \nabla_{\bf k}
\Big]\partial_{E_{\beta}} f_{n\mathbf{k}} = e v_{n\mathbf{k}\beta} \frac{\partial f_{n\mathbf{k}}^0}{\partial \varepsilon_{n\mathbf{k}}} \tau_{n\mathbf{k}} \\
+ \frac{2\pi\tau_{n\mathbf{k}}}{\hbar}
  \sum_{m\nu} \!\int\! \frac{\mathrm{d}^3 q}{\Omega_{\mathrm{BZ}}} | g_{mn\nu}(\mathbf{k},\mathbf{q})|^2 \\
 \times  \Big[(n_{\mathbf{q}\nu}+1-f_{n\mathbf{k}}^0)\delta(\varepsilon_{n\mathbf{k}} - \varepsilon_{m\mathbf{k+q}}  + \hbar \omega_{\mathbf{q}\nu} )   \\
  +  (n_{\mathbf{q} \nu} + f_{n\mathbf{k}}^0)\delta(\varepsilon_{n\mathbf{k}} - \varepsilon_{m\mathbf{k+q}}  - \hbar \omega_{\mathbf{q}\nu}) \Big] \partial_{E_{\beta}} f_{m\mathbf{k}+\mathbf{q}},
\end{multline}
with $\tau_{n\mathbf{k}}$ being the total scattering lifetime defined in Eq.~\eqref{eq:scattering_rate}.
Taking derivatives on both sides of Eq.~\eqref{eq:iterwithbimpl} with respect to the Cartesian components of the magnetic field, $B_{\gamma}$, at zero field, yields an equation for the linear response coefficients
$\partial_{E_{\beta}}f_{n\mathbf{k}}$ and $\partial_{E_{\beta}}\partial_{B_{\gamma}}f_{n\mathbf{k}}$:
\begin{multline}\label{eq:secondderivofB}
\partial^2_{E_\beta,B_\gamma} f_{n\mathbf{k}} = \frac{-e}{\hbar}\tau_{n\mathbf{k}}({\bf v}_{n\mathbf{k}} \times \nabla_{\bf k})_\gamma
\partial_{E_\beta} f_{n\mathbf{k}} \\
  +  \frac{2\pi\tau_{n\mathbf{k}}}{\hbar} \sum_{m\nu} \int \frac{\mathrm{d}^3 q}{\Omega_{\mathrm{BZ}}} | g_{mn\nu}(\mathbf{k},\mathbf{q})|^2  \\
 \times  \Big[(n_{\mathbf{q}\nu}+1-f_{n\mathbf{k}}^0)\delta(\varepsilon_{n\mathbf{k}} - \varepsilon_{m\mathbf{k+q}}  + \hbar \omega_{\mathbf{q}\nu} )   \\
 +  (n_{\mathbf{q} \nu} + f_{n\mathbf{k}}^0)\delta(\varepsilon_{n\mathbf{k}} - \varepsilon_{m\mathbf{k+q}}  - \hbar \omega_{\mathbf{q}\nu}) \Big] \partial^2_{E_{\beta},B_\gamma} f_{m\mathbf{k}+\mathbf{q}}.
\end{multline}

The Hall conductivity tensor is obtained from the second derivatives of the current density with respect to electric and magnetic field for vanishing fields:
\begin{equation}\label{eq:conductivityhall}
  \sigma^{\mathrm{H}}_{\alpha\beta\gamma} = \frac{-e}{V_{\mathrm{uc}}} \sum_n \int \frac{\mathrm{d}^3 k}{\Omega_{\mathrm{BZ}}} \, v_{n\mathbf{k}\alpha} \, B_{\gamma}\, \partial^2_{E_{\beta},B_{\gamma}}f_{n\mathbf{k}},
\end{equation}
or equivalently directly from Eq.~\eqref{eq:iterwithbimpl} as:
\begin{equation}\label{eq:conductivityhall2}
  \sigma^{\mathrm{H}}_{\alpha\beta\gamma} = \frac{-e}{V_{\mathrm{uc}}} \sum_n \int \frac{\mathrm{d}^3 k}{\Omega_{\mathrm{BZ}}} \, v_{n\mathbf{k}\alpha}  \partial_{E_{\beta}}f_{n\mathbf{k}}(B_{\gamma}).
\end{equation}
Besides the drift and Hall conductivity and their mobility analogues, a commonly reported quantity is the dimensionless Hall tensor, which is
defined as the ratio between the Hall conductivity and the drift conductivity~\cite{Reggiani1983,Popovic1991}:
\begin{align}\label{eq:hallfactor}
  r^{\mathrm{H}}_{\alpha\beta\gamma} \equiv &  n_{\rm c} e \sum_{\delta\epsilon} \frac{\sigma_{\alpha\delta}^{-1} \, \sigma^{\rm H}_{\delta\epsilon\gamma} \, \sigma_{\epsilon\beta}^{-1}}{B_{\gamma}},  \\
   =&  \sum_{\delta\epsilon} \frac{\mu_{\alpha\delta}^{-1} \, \mu^{\rm H}_{\delta\epsilon\gamma} \, \mu_{\epsilon\beta}^{-1}}{B_{\gamma}},
\end{align}
where Eq.~\eqref{eq:hallfactor} is the tensorial generalization of Eq.~(1) from Ref.~\onlinecite{Reggiani1983}.
%
%
For the cubic materials that we studied here, Eq.~\eqref{eq:hallfactor} has only one non-equivalent component, $r^{\mathrm{H}}_{123}$.

A popular approximation to Eq.~\eqref{eq:hallfactor} consists in assuming a parabolic and non-degenerate band extremum, following Ref.~\onlinecite{Wiley1975}, p.~118 and Ref.~\onlinecite{Price1957}, Eq.~3.12.
Within this approximation, the isotropic and temperature-dependent Hall factor is given by~\cite{Beer1963}:
\begin{equation}\label{eq65}
  r^{\mathrm{iso}} = \frac{\langle \tau^2 \rangle}{\langle \tau \rangle^2},
\end{equation}
with
\begin{equation}\label{eq66}
  \langle \tau^n \rangle \equiv \frac{\int_0^\infty \tau^n(x k_{\mathrm{B}}T) x^{3/2}
e^{-x} \mathrm{d} x }{\int_0^\infty x^{3/2} e^{-x} \mathrm{d} x}.
\end{equation}
Here, $x = \varepsilon/(k_{\mathrm{\rm B}}T)$ and we introduced the distribution function of the total decay rate:
\begin{equation}
\tau(\varepsilon ) = \sum_n \int \frac{\mathrm{d}^3 k}{\Omega_{\mathrm{BZ}}}
\delta(\varepsilon -\varepsilon_{n\mathbf{k}}) \tau_{n\mathbf{k}}.
\end{equation}

The numerical solution of the BTE requires the evaluation of a large number of electron-phonon matrix elements in order to converge the double momentum integrals (\textbf{k} and \textbf{q}).
Various schemes have been developed to deal with this challenge including models with parameters computed from first principles~\cite{Ganose2020}, direct evaluation of the electron-phonon matrix elements using density-functional perturbation theory (DFPT)~\cite{Restrepo2009}, linear interpolation of the scattering rates~\cite{Li2015,Sohier2018}, local orbital implementations~\cite{Gunst2016}, smoothened Fourier interpolation~\cite{Madsen2006,Madsen2018}, Fourier interpolation of the perturbed potential~\cite{Gonze2019,Brunin2020,Brunin2020a} and interpolation based on MLWFs~\cite{Ponce2016a,Zhou2020}.
The use of MLWFs makes the calculations affordable and is the method of choice in this work, where we rely on the EPW software~\cite{Giustino2007,Ponce2016a}.
Such interpolation implies subtleties that need to be dealt with carefully, as discussed in Section~\ref{sec:interpolation}.

\subsection{Interpolation of the electron bands, phonon dispersions, and electron-phonon matrix elements}\label{sec:interpolation}

The interpolation of the various quantities required to compute the drift mobility on ultra-dense grids relies on a discrete Fourier transform followed by a Fourier interpolation at arbitrary crystal momenta.
Due to the fact that the discrete Fourier transform is in practice performed on a uniform finite grid, the quality of the interpolation depends on the localization in real space of each quantity.

In the case of the electronic Hamiltonian, we can leverage the phase freedom of the Bloch orbitals to create Wannier functions that are maximally localized in real space~\cite{Marzari2012}.
Since the interatomic force constants do not have such gauge freedom, we use direct Fourier interpolation.
This interpolation requires some care in the case of polar materials, in order to correctly describe long-range interactions and the splitting of longitudinal-optical (LO) and transvers-optical (TO) modes~\cite{Gonze1997a,Baroni2001}.
Electronic Wannier and Bloch states are related by~\cite{Marzari2012}:
\begin{align}
|\Psi_{m\mathbf{R}_p} \rangle &= \frac{1}{N_p} \sum_{n\mathbf{k}} e^{-i \mathbf{k} \cdot \mathbf{R}_p} U_{nm\mathbf{k}} | \Psi_{n\mathbf{k}} \rangle \\
|\Psi_{n\mathbf{k}} \rangle &=  \sum_{m\mathbf{R}_p} e^{i\mathbf{k}\cdot \mathbf{R}_p} U_{mn\mathbf{k}}^\dagger | \Psi_{m \mathbf{R}_p} \rangle,
\end{align}
where $\mathbf{R}_p$ is a lattice vector, $N_p$ is the number of unit cells in the Born-von K\'arman supercell, corresponding to the number of $\mathbf{k}$-points, and $U_{nm,\mathbf{k}}$  is a unitary rotation matrix that transforms the Bloch wavefunctions to a Wannier gauge:
\begin{equation}\label{eq:WannierGauge}
| \Psi_{n\mathbf{k}}^{\rm W} \rangle  = \sum_m  U_{nm\mathbf{k}}| \Psi_{m\mathbf{k}} \rangle.
\end{equation}
In particular, the unitary transformation $U$ can be chosen to obtain MLWFs that minimize the spatial spread functional~\cite{Marzari2012}.

In this context, the Hamiltonian $H$ and dynamical matrices $D$ can be transformed from the Bloch to Wannier representation as:
\begin{align}\label{eq:hamiltonian}
  H_{mn}(\mathbf{R_p}) =& \frac{1}{N_p} \sum_{m'n'\mathbf{k}} e^{-i\mathbf{k}\cdot \mathbf{R}_p} \nonumber \\
                                         & \times U_{mm'\mathbf{k}}^{\dagger} H_{m'n'\mathbf{k}}  U_{n'n\mathbf{k}} \\
  D_{\kappa\alpha,\kappa'\beta}(\mathbf{R_{p'}}) =& \frac{1}{N_{p'}} \sum_{\mathbf{q}\mu\nu} e^{-i\mathbf{q}\cdot \mathbf{R}_{p'}} \nonumber \\\label{eq:dynamical}
                                                           & \times e_{\kappa\alpha,\mathbf{q}\mu}^{\dagger} D_{\mu\nu}(\mathbf{q}) e_{\kappa'\beta,\mathbf{q}\nu},
\end{align}
where $N_p$, $N_{p'}$ are the number of unit cells in the Born-von K\'arman supercells for the electrons and phonons, respectively and $e_{\kappa\alpha,\mathbf{q}\nu}$ are the eigendisplacement vectors corresponding to the atom $\kappa$ in the Cartesian direction $\alpha$ for a collective phonon mode $\nu$ of momentum $\mathbf{q}$.

The extension of these two concepts to interpolate the electron-phonon matrix elements has been derived in Ref.~\onlinecite{Giustino2007} and yields:
\begin{multline} \label{eq:gkk}
      g_{mn\kappa\alpha}(\mathbf{R}_p, \mathbf{R}_{p'}) = \frac{1}{N_pN_{p'}} \sum_{\mathbf{k,q}}  e^{-i(\mathbf{k}\cdot \mathbf{R}_p+\mathbf{q}\cdot\mathbf{R}_{p'})} \sum_{m'n'\nu} \\
     \times  \sqrt{\frac{2M_\kappa \omega_{\mathbf{q}\nu}}{\hbar}} e_{\kappa\alpha,\mathbf{q}\nu}^{\dagger} U_{mm'\mathbf{k+q}}^{\dagger}  g_{m'n'\nu}(\mathbf{k,q})  U_{n'n\mathbf{k}}.
\end{multline}

In practice, we compute the electron-phonon matrix elements $g_{nm\nu}(\mathbf{k,q})$ on a coarse momentum grid and rely on crystal symmetries to reduce the number of calculations to be performed.
To carry out the Wannier transformation from coarse reciprocal space to real space, we first evaluate the matrix elements on the entire Brillouin zone using symmetries.
As pointed out in a recent preprint~\cite{Zhou2020}, in the case of SOC the spinor wavefunctions need to be rotated using the SU(2) unitary group.
There was a bug in version 5.2 of the EPW software~\cite{Ponce2016a} related to this rotation.
The bug has been eliminated in version 5.3, and for the systems tested so far we found a negligible impact.
For example, we recalculated the mobilities of silicon, GaAs, and CsPbI$_3$, and found differences smaller than 0.4\% in all cases.

There are additional subtleties related to the Wannier interpolation of the various quantities required to compute the mobility.
These aspects will be discussed in Section~\ref{resultsection}.

\subsection{Band velocities}

\begin{figure*}[ht]
  \centering
  \includegraphics[width=0.95\linewidth]{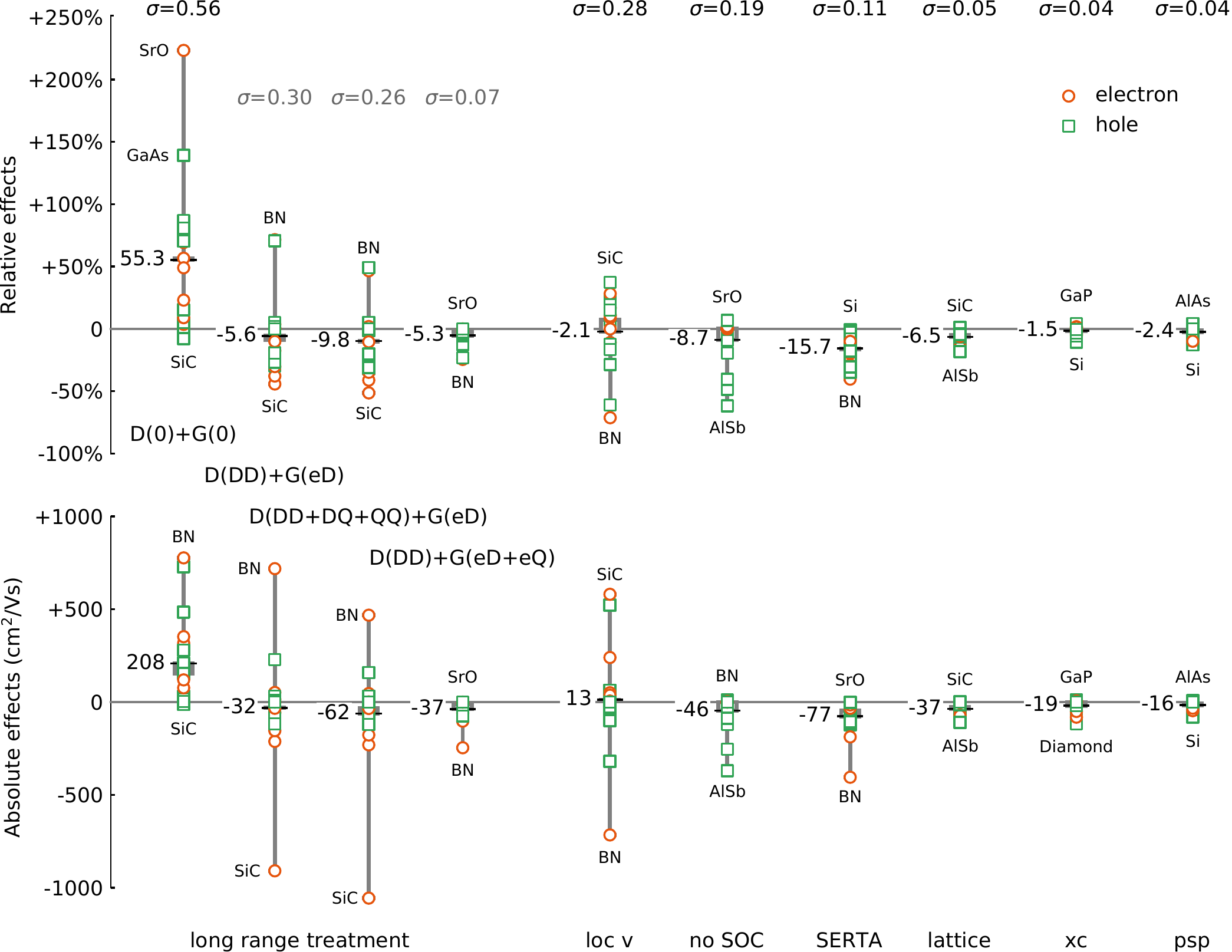}
  \caption{\label{fig:effects}
(a) Relative deviation of the calculated carrier mobility with respect to our most accurate calculations when using one of the approximations described in Sec.~\ref{sec:effects}.
Electron mobilities are in orange, and holes are in green.
In each case, only one effect is considered, while all other settings are the same as for the reference calculation.
We consider the following effects:
(i) Long range part of the dynamical matrix (D) and electron-phonon matrix elements (G).
The label ``0" means that no long range part has been considered, and ``DD", ``DQ", and ``QQ" mean that dipole-dipole, dipole-quadrupole and quadrupole-quadrupole contributions have been included.
The labels ``eD" and ``eQ" mean that monopole-dipole and monopole-quadrupole interactions are included, respectively.
(ii) Using the velocity computed in the local approximation via Eq.~\eqref{finalvelo} instead of Eq.~\eqref{eq:final_velo}.
(iii) Neglecting SOC.
(iv) Employing the SERTA in Eq.~\eqref{eq:serta} instead of Eq.~\eqref{eq:btemob}.
(v) Using the relaxed lattice parameter of the PBE exchange-correlation functional.
(vi) Using the experimental lattice parameter with the LDA exchange-correlation functional.
(vii) Using the PBE exchange-correlation functional and experimental lattice parameter, but with a different
parametrization of the pseudopotentials.
The gray boxes represent the first and third quartile, while the average values are shown with
a small horizontal black line.
The standard deviations are reported at the top of the figure.
The electron mobility of GaAs is not reported in this figure since the values are off the chart.
(b) Same as in (a), but the deviations are reported as absolute values in cm$^2$/Vs.
  }
\end{figure*}
A key ingredient required in the calculation of carrier mobility is the carrier velocity as can be seen in Eq.~\eqref{eq:btemob}.
The velocity operator can be expressed in terms of the commutator between the Hamiltonian of the system and the position operator, $\hat{\mathbf{v}} = (i/\hbar)[\hat{H},\hat{\mathbf{r}}]$.
The matrix elements of $\hat{\mathbf{v}}$ in the presence of non-local pseudopotentials are~\cite{Starace1971,Read1991,Ismail-Beigi2001,Pickard2003}:
\begin{align}
\mathbf{v}_{nm\mathbf{k}} &= \langle \psi_{m\mathbf{k}} | \hat{\mathbf{p}}/m_e + (i/\hbar)[\hat{V}_{\rm NL},\hat{\mathbf{r}}] | \psi_{n\mathbf{k}}  \rangle, \label{basicvelo}
\end{align}
where $\hat{\mathbf{p}} = -i\hbar \partial/\partial \mathbf{k}$ is the momentum operator, $m_e$ is the electron mass, and $\hat{V}_{\rm NL}$ is the non-local part of the pseudopotential.
Neglecting the term with $\hat{V}_{\rm NL}$ in the case of a non-local pseudopotential is what we call the ``local approximation''~\cite{Ponce2018}.
Within this approximation, Eq.~\eqref{basicvelo} reads:
\begin{equation}\label{eqvelocity}
\mathbf{v}^{\rm loc}_{nm\mathbf{k}}=
 \frac{\hbar \mathbf{k}}{m_e} \delta_{mn} -i\frac{\hbar}{m_e} \int d\mathbf{r} u_{m\mathbf{k}}^*(\mathbf{r}) \frac{\partial}{\partial \mathbf{k}} u_{n\mathbf{k}}(\mathbf{r}),
\end{equation}
where $u_{n\mathbf{k}}$ are the periodic part of the Bloch wavefunction.
If in the second part of Eq.~\eqref{eqvelocity} the terms corresponding to the periodic part of the Bloch function are expanded into planewaves, $u_{n\mathbf{k}}= \sum_{\mathbf{G}} c_{n\mathbf{k}}(\mathbf{G})e^{i\mathbf{G}\cdot \mathbf{r}}$, we obtain the simple velocity expression in the \textit{local approximation}:
\begin{equation}\label{finalvelo}
\mathbf{v}^{\rm loc}_{nm\mathbf{k}} =  \frac{\hbar{\bf k}}{m_e} \delta_{mn} +
   \sum_{\mathbf{G}} \frac{\hbar {\mathbf{G}}}{m_e}
    c_{m\mathbf{k}}^*(\mathbf{G})c_{n\mathbf{k}}(\mathbf{G}).
\end{equation}
We note that this expression is only valid in the case of local norm-conserving pseudopotentials.
In practice, we will show that this local approximation is too crude and one needs to compute the matrix elements of the commutator in Eq.~\eqref{basicvelo}.
We carry out this procedure in the Wannier representation, as described in Refs.~\onlinecite{Blount1962,Wang2006,Yates2007,Pizzi2020}.
In particular, we construct the velocities interpolated at an arbitrary momentum $\mathbf{k'}$ following Eq.~(31) of Ref.~\onlinecite{Wang2006}:
\begin{equation}\label{eq:final_velo}
\mathbf{v}_{nm\mathbf{k'}} = \frac{1}{\hbar} \frac{\partial}{\partial \mathbf{k'}} H_{nm\mathbf{k'}} - \frac{i}{\hbar}(\varepsilon_{n\mathbf{k'}}-\varepsilon_{m\mathbf{k'}}) \mathbf{A}_{nm\mathbf{k'}},
\end{equation}
where $\mathbf{A}_{nm\mathbf{k'}}$ is the Berry connection defined as:
\begin{equation}
\mathbf{A}_{nm\mathbf{k'}} = i \langle u_{n\mathbf{k'}} | \partial u_{m\mathbf{k'}}/ \partial \mathbf{k'}  \rangle.
\end{equation}

We construct the Berry connection in the Wannier basis [Eq.~\eqref{eq:WannierGauge}] on the coarse $\mathbf{k}$-point grid and compute the derivative using finite differences as~\cite{Wang2006}:
\begin{equation}
\mathbf{A}_{nm\mathbf{k}}^{\text{W}} = i \sum_{\mathbf{b}} w_b \mathbf{b} ( \langle u_{n\mathbf{k}}^{\text{W}} | u_{m\mathbf{k+b}}^{\text{W}} \rangle - \delta_{nm} ),
\end{equation}
where $\mathbf{b}$ are the vectors connecting $\mathbf{k}$ to its nearest neighbors and $w_b$ a weight associated with each shell of neighbours $|\mathbf{b}|$.
These vectors are constructed in such a way as to satisfy~\cite{marzari1997}:
\begin{equation}
\sum_{\mathbf{b}} w_b b_\alpha b_\beta = \delta_{\alpha\beta}
\end{equation}
using the smallest number of neighbours.
The rotated overlap matrices (from Bloch to Wannier basis) are obtained as:
\begin{equation}
\langle u_{n\mathbf{k}}^{\text{W}} | u_{m\mathbf{k+b}}^{\text{W}} \rangle = \sum_{n'm'} U_{nn'\mathbf{k}}^{\dagger} M_{n'm'}(\mathbf{k,b})U_{m'm\mathbf{k+b}},
\end{equation}
where $M_{nm}(\mathbf{k,b})=\langle u_{n\mathbf{k}} | u_{m\mathbf{k+b}} \rangle $ is the overlap matrix between the cell-periodic Bloch eigenstates at neighboring points \textbf{k} and \textbf{k}+\textbf{b}.

We first Fourier transform the position matrix into real space following Eq.~(43) of Ref.~\onlinecite{Wang2006}:
\begin{equation}
\langle \Psi_{n\mathbf{0}} | \hat{\mathbf{r}} | \Psi_{m\mathbf{R}} \rangle  = \frac{1}{N_p} \sum_{\mathbf{k}} e^{-i\mathbf{k}\cdot \mathbf{R}} \mathbf{A}_{nm\mathbf{k}}^{\text{W}},
\end{equation}
where $\mathbf{k}$ belongs to the same grid employed for the Wannier interpolation of the band structure.
The quantity on the left hand side decreases rapidly with $\mathbf{R}$ owing to the localization of MLWFs,
therefore we can use these real-space quantities to interpolate back to arbitrary momenta $\mathbf{k'}$:
\begin{equation}
A_{nm\mathbf{k'},\alpha}^{\text{W}} = \sum_{\mathbf{R}} e^{i\mathbf{k'}\cdot \mathbf{R}} \langle \Psi_{n\mathbf{0}} | \hat{\mathbf{r}} | \Psi_{m\mathbf{R}} \rangle.
\end{equation}
We do the same for the $\mathbf{k'}$ derivative of the Hamiltonian following Eq.~(38) of Ref.~\onlinecite{Wang2006}:
\begin{equation}
\frac{\partial}{\partial \mathbf{k'}} H_{nm\mathbf{k'}}^{\text{W}} = \sum_{\mathbf{R}} e^{i\mathbf{k'}\cdot \mathbf{R}} i\mathbf{R} \langle \Psi_{n\mathbf{0}} | \hat{H} | \Psi_{m\mathbf{R}} \rangle.
\end{equation}

Finally we un-rotate from the Wannier basis to the Bloch state basis using the diagonalizers of the Hamiltonian:
\begin{align}
\frac{\partial}{\partial \mathbf{k'}}H_{nm\mathbf{k'}} &= \sum_{n'm'} U_{nn'\mathbf{k'}}^\dagger \frac{\partial}{\partial \mathbf{k'}}H_{n'm'\mathbf{k'}}^{\text{W}} U_{m'm\mathbf{k'}},\\
\mathbf{A}_{nm\mathbf{k'}} &= \sum_{n'm'} U_{nn'\mathbf{k'}}^\dagger \mathbf{A}_{n'm'\mathbf{k'}}^{\text{W}} U_{m'm\mathbf{k'}},
\end{align}
which gives the desired Eq.~\eqref{eq:final_velo}.

We have implemented the interpolation of the velocity matrix elements including the correction for non-local potentials in the EPW software\cite{Ponce2016a}.
Figure~\ref{fig:effects} shows that the local approximation yields significant errors in the calculated mobilities, ranging from $-80$\% to $+40$\% of the correct value.
For this reason, we use the exact velocities of Eq.~\eqref{eq:final_velo} throughout this study.

To evaluate the quality of the Wannier interpolation of velocities, we perform direct DFT calculations using central finite differences, with six neighbouring points spaced by $5\cdot 10^{-4}2\pi/a$.

\subsection{Long range corrections: dipoles and quadrupoles}\label{sec:dynam_quad}

The analytic properties of the long wavelength limit of the electron-phonon matrix elements in bulk polar materials, resulting from the Fr\"ohlich interaction, have been know for a long time~\cite{Vogl1976}, but the generalization to first-principles calculations using Wannier-Fourier interpolation~\cite{Giustino2007} appeared only recently~\cite{Verdi2015,Sjakste2015}.
Following a line of thinking closely related to the Fourier interpolation of the phonon frequencies~\cite{Giannozzi1991,Gonze1997a,Baroni2001}, one can decompose the matrix elements into a short- ($\mathcal{S}$) and long-range ($\mathcal{L}$) contribution~\cite{Verdi2015,Sjakste2015}:
\begin{equation}\label{geqfirst}
g_{mn\nu}(\mathbf{k},\mathbf{q}) = g_{mn\nu}^{\mathcal{S}}(\mathbf{k},\mathbf{q}) + g_{mn\nu}^{\mathcal{L}}(\mathbf{k},\mathbf{q}).
\end{equation}
In Eq.~\eqref{geqfirst}, the long-range part is given by the infinite multipole expansion:
\begin{equation}\label{eq:multipole}
g_{mn\nu}^{\mathcal{L}}(\mathbf{k},\mathbf{q}) = g_{mn\nu}^{\mathcal{L},\rm{D}}(\mathbf{k},\mathbf{q}) + g_{mn\nu}^{\mathcal{L},\rm{Q}}(\mathbf{k},\mathbf{q}) + g_{mn\nu}^{\mathcal{L},\rm{O}}(\mathbf{k},\mathbf{q}) + \cdots,
\end{equation}
where the first term comes from dipole contribution, then quadrupole, octopoles and higher.
The analytic form of the long-range part of the dipole contribution is given as~\cite{Verdi2015}:
\begin{multline}\label{gl3d2}
g_{mn\nu}^{\mathcal{L},\rm{D}}(\mathbf{k},\mathbf{q}) = i\frac{4\pi}{\Omega}\frac{e^2}{4\pi\varepsilon^0}  \sum_{\kappa} \bigg[ \frac{\hbar}{2 N_{p'} M_\kappa \omega_{\mathbf{q}\nu}}\bigg]^{\frac{1}{2}} \sum_{\mathbf{G} \neq -\mathbf{q}}  \\
\times  \frac{ (\mathbf{G} + \mathbf{q}) \cdot \mathbf{Z}_{\kappa}^* \cdot \mathbf{e}_{\kappa\mathbf{q}\nu}}{(\mathbf{G} + \mathbf{q}) \cdot \boldsymbol{\varepsilon}^\infty \cdot (\mathbf{G} + \mathbf{q})} e^{-i (\mathbf{G} + \mathbf{q}) \cdot \boldsymbol{\tau}_{\kappa}}  \\
\times \langle \Psi_{m \mathbf{k+q}} | e^{i(\mathbf{q+G})\cdot \mathbf{r}} | \Psi_{n\mathbf{k}} \rangle,
\end{multline}
where $\mathbf{G}$ is a reciprocal lattice vector, $\bm{\tau}_{\kappa}$ is the position of atom $\kappa$, $\mathbf{Z}_\kappa^*$ is the Born effective charge tensor of the atom $\kappa$ of mass $M_\kappa$, $\mathbf{e}_{\kappa\mathbf{q}\nu}$ is the vibrational eigendisplacement vector normalized in the unit cell, $\varepsilon^{0}$ is the vacuum dielectric constant and $\boldsymbol{\varepsilon}^\infty$ is the high-frequency dielectric tensor of the material.
In the past, a Gaussian filter has been added to Eq.~\eqref{gl3d2}.
Since this filter breaks the periodicity of the matrix element in reciprocal space, we recommend not to include it in future calculations.
One can notice that Eq.~\eqref{gl3d2} diverges as $1/|\mathbf{q}|$ as we approach the zone center.
This is an integrable singularity, so that upon performing integrations in reciprocal space one obtains finite values of physical properties.

The quadrupolar component of the matrix element at long wavelength is given by:
\begin{multline}\label{gldq}
g_{mn\nu}^{\mathcal{L},\rm{Q}}(\mathbf{k},\mathbf{q}) = \frac{4\pi}{\Omega}\frac{e^2}{4\pi\varepsilon^0}   \sum_{\kappa} \bigg[ \frac{\hbar}{2N_{p'}M_\kappa \omega_{\mathbf{q}\nu}}\bigg]^{\frac{1}{2}} \!\!  \sum_{\mathbf{G} \neq -\mathbf{q}} \\
\times \frac{ (\mathbf{G+q}) \cdot (\mathbf{G+q}) \cdot  \mathbf{e}_{\kappa \mathbf{q}\nu} \cdot \tilde{\mathbf{Q}}_{mn\kappa}(\mathbf{k,q}) }{(\mathbf{G+q})\cdot  \boldsymbol{\varepsilon}^\infty \cdot (\mathbf{G+q})} e^{-i (\mathbf{G} + \mathbf{q}) \cdot \boldsymbol{\tau}_{\kappa} },
\end{multline}
where $\tilde{\mathbf{Q}}_{mn\kappa}(\mathbf{k,q})$ is a third-rank effective quadrupole tensor defined as:
\begin{multline}\label{eq:effectivequad}
\tilde{Q}_{mn\kappa\alpha\beta\gamma}(\mathbf{k,q}) = \frac{1}{2}Q_{\kappa\alpha\beta\gamma} \langle \Psi_{m \mathbf{k+q}} | e^{i(\mathbf{q+G})\cdot \mathbf{r}} | \Psi_{n\mathbf{k}} \rangle \\
 - e Z_{\kappa\alpha,\beta}^* \langle \Psi_{m \mathbf{k+q}} | e^{i(\mathbf{q+G})\cdot \mathbf{r}} V^{\rm Hxc, \mathcal{E}_\gamma}(\mathbf{r}) | \Psi_{n\mathbf{k}} \rangle,
\end{multline}
and $Q_{\kappa\alpha\beta\gamma}$ is the dynamic quadrupole tensor~\cite{Royo2019}.
The third-rank quadrupole tensor should obey the same symmetry rules as other third-rank tensors
 such as the piezoelectric tensor or the second-order magnetoelectric tensor~\cite{Resta2010}.
One can therefore use tools such as the Bilbao Crystallographic Server~\cite{Aroyo2011} to determine the non-zero components of the quadrupole tensor for a given space group, and their relations.
In the case of all the polar materials investigated in this work, the quadrupole tensor takes the form:
\begin{equation}\label{eq:quad2}
Q_{\kappa\alpha\beta\gamma} = Q_{\kappa}|\varepsilon_{\alpha\beta\gamma}|,
\end{equation}
where $\varepsilon_{\alpha\beta\gamma}$ is the Levi-Civita symbol and $Q_{\kappa}$ is an atom-dependent scalar.
In the case of silicon and diamond, the effective tensor in Eq.~\eqref{eq:effectivequad} is completely specified by a single scalar~\cite{Royo2019}:
\begin{multline}\label{eq:quad1}
\tilde{Q}_{mn\kappa\alpha\beta\gamma}(\mathbf{k,q}) = \\
\frac{1}{2} (-1)^{\kappa+1}Q|\varepsilon_{\alpha\beta\gamma}| \langle \Psi_{m \mathbf{k+q}} | e^{i(\mathbf{q+G})\cdot \mathbf{r}} | \Psi_{n\mathbf{k}} \rangle .
\end{multline}

The quantity $V^{\rm Hxc, \mathcal{E}_\gamma}$ in Eq.~\eqref{eq:effectivequad} is the change of the Hartree and exchange-correlation potential with respect to the electric field $\mathcal{E}_\gamma$~\cite{Brunin2020}.
This term is null in the case of materials with vanishing Born effective charges and has been shown to be small in the case of polar materials such as GaAs and GaP~\cite{Brunin2020,Brunin2020a}.
Finally, we note that in the $\mathbf{q+G} \rightarrow 0$ limit, the wavefunction overlap in Eqs.~\eqref{gl3d2},\eqref{eq:effectivequad} can be written in terms of the Wannier rotation matrices $U_{nm\mathbf{k}}$ as~\cite{Verdi2015}:
\begin{equation}
\langle \Psi_{m \mathbf{k+q}} | e^{i(\mathbf{q+G})\cdot \mathbf{r}} |  \Psi_{n\mathbf{k}} \rangle = [U_{\mathbf{k+q}} U_{\mathbf{k}}^\dagger  ]_{mn}.
\end{equation}

A similar approach can be used to enforce the correct behavior of the dynamical matrices at long wavelength~\cite{Gonze1997a,Royo2020}:
\begin{equation}
D_{\kappa\alpha, \kappa'\beta}(\mathbf{q}) = D_{\kappa\alpha, \kappa'\beta}^{\mathcal{S}}(\mathbf{q}) + D_{\kappa\alpha, \kappa'\beta}^{\mathcal{L}}(\mathbf{q}),
\end{equation}
where the nonanalytical, direction-dependent term, is given by the contribution of dipole-dipole, dipole-quadrupole and quadrupole-quadrupole terms as:
\begin{multline}\label{eq:longrange_dyn}
D_{\kappa\alpha, \kappa'\beta}^{\mathcal{L}}(\mathbf{q}) = \frac{4\pi e^2}{\Omega} \Bigg[ \sum_{\mathbf{G}\neq -\mathbf{q}} D_{\kappa\alpha, \kappa'\beta}^{\mathcal{L},\rm{D+Q}}(\mathbf{G+q}) \\
- \delta_{\kappa \kappa'} \sum_{\kappa''} \sum_{\mathbf{G}\neq \mathbf{0}} D_{\kappa\alpha, \kappa''\beta}^{\mathcal{L},\rm{D+Q}}(\mathbf{G}) \Bigg]
\end{multline}
with
\begin{multline}\label{eq:longrange_dyn2}
D_{\kappa\alpha, \kappa'\beta}^{\mathcal{L},\rm{D+Q}}(\mathbf{q}) = \frac{e^{i \mathbf{q}\cdot(\boldsymbol{\tau_{\kappa}} - \boldsymbol{\tau}_{\kappa'})} e^{\frac{- \mathbf{q}\cdot \boldsymbol{\varepsilon}^\infty \cdot \mathbf{q}}{4 \Lambda^2}}}{\mathbf{q} \cdot \boldsymbol{\varepsilon}^{\infty} \cdot \mathbf{q}}
 \bigg[\mathbf{q} \cdot \mathbf{Z}_{\kappa \alpha}^* \cdot \mathbf{q} \cdot \mathbf{Z}_{\kappa' \beta}^* \\
+\frac{1}{4}  \mathbf{q} \cdot \mathbf{q} \cdot \mathbf{Q}_{\kappa\alpha} \cdot \mathbf{q} \cdot \mathbf{q} \cdot  \mathbf{Q}_{\kappa'\beta}
+\frac{i}{2} \mathbf{q} \cdot \mathbf{Z}_{\kappa \alpha}^* \cdot \mathbf{q} \cdot \mathbf{q} \cdot \mathbf{Q}_{\kappa'\beta} \\
- \mathbf{q} \cdot \mathbf{q} \cdot \mathbf{Q}_{\kappa\alpha} \cdot \mathbf{q} \cdot \mathbf{Z}_{\kappa'\beta}^* \bigg].
\end{multline}

In this expression we neglected octopoles and higher-orders as well as fourth-order and higher dielectric functions.
The sum over the lattice of charges is performed using the Ewald technique, with a parameter $\Lambda$.
In our calculations we use $\Lambda=1$~bohr$^{-1}$ so that the real-space contribution in the Ewald summation can be neglected~\cite{Gonze1997a}.

\subsection{Adaptive broadening}

The numerical evaluation of Eqs.~\eqref{eq:iter} and \eqref{eq:scattering_rate} requires to approximate Dirac delta functions by Lorentzian or Gaussian functions of finite broadening.
This means that the resulting mobility will depend on the value of the broadening used, and careful convergence tests are required~\cite{Ponce2018}.
To circumvent this difficulty, various schemes have been proposed including adaptive broadening~\cite{Li2014b} and linear tetrahedron methods \cite{Bloechl1994,Brunin2020}.

We implemented the adaptive broadening approach in the EPW software following Eq.~(18) from Ref.~\onlinecite{Li2014b}, where the state-dependent broadening is given by:
\begin{equation}\label{eq:eta_para}
\eta_{n\mathbf{k}}(\mathbf{q} \nu) = \frac{\hbar}{\sqrt{12}} \sqrt{ \sum_{\alpha}  \bigg[ \Big(\mathbf{v}_{\mathbf{q}\nu\nu}- \mathbf{v}_{nn\mathbf{k+q}}\Big)\cdot \frac{\mathbf{G}_\alpha}{N_\alpha}   \bigg]^{2}},
\end{equation}
where $\mathbf{G}_{\alpha}$ are the three primitive reciprocal lattice vectors, $\alpha$ the three crystalline directions,
$\mathbf{v}_{nn\mathbf{k+q}}$ the electronic velocities defined by Eq.~\eqref{eq:final_velo}, $N_{\alpha}$ the $\mathbf{q}$-point grid density in the three crystalline directions.
Eq.~\eqref{eq:eta_para} can be scaled arbitrarily, and for this work we choose empirically a factor 1/2 after a number of numerical tests.
In addition, the phonon velocity $\mathbf{v}_{\mathbf{q}\nu\nu}$ is obtained from the momentum derivative of the dynamical matrix, following an approach similar to Eq. (38) of Ref.~\onlinecite{Wang2006}:
\begin{equation}
v_{\mathbf{q}\mu\nu\beta} = \frac{1}{2\omega_{\mathbf{q}\nu}} \frac{\partial D_{\mu\nu}(\mathbf{q})}{\partial q_\beta}
                                      = \frac{1}{2\omega_{\mathbf{q}\nu}} \sum_{\mathbf{R}} i R_\beta  e^{i\mathbf{q}\cdot \mathbf{R}}  D_{\mu\nu}(\mathbf{R}).
\end{equation}

In the case of polar materials, the momentum derivative of the long-range dipole part is obtained by taking the derivative of Eq.~\eqref{eq:longrange_dyn} as $\frac{\partial D_{\mu\nu}(\mathbf{q})}{\partial q_\beta} = \sum_{\mathbf{G}\neq \mathbf{-q}} \frac{\partial D_{\kappa\alpha,\kappa'\beta}^{\mathcal{L},\rm{D}}(\mathbf{G+q})}{\partial q_l} $ with:
\begin{multline}\label{eq:adaptderiv}
\frac{\partial D_{\kappa\alpha,\kappa'\beta}^{\mathcal{L},\rm{D}}(\mathbf{q})}{\partial q_l} = \frac{e^{i \mathbf{q}\cdot(\boldsymbol{\tau}_{\kappa}-\boldsymbol{\tau}_{\kappa'})} e^{-\frac{\mathbf{q}\cdot\boldsymbol{\epsilon}^{\infty} \cdot \mathbf{q}}{4\Lambda^2}} }{\mathbf{q} \cdot \boldsymbol{\epsilon}^{\infty} \cdot \mathbf{q}} \times \\
 \bigg[ Z^*_{\kappa\alpha, l} \mathbf{Z}^*_{\kappa'\beta} \cdot \mathbf{q} + \mathbf{Z}^*_{\kappa\alpha}\cdot \mathbf{q} Z^*_{\kappa'\beta, l}
-  \frac{\mathbf{Z}^*_{\kappa\alpha} \cdot \mathbf{Z}^*_{\kappa'\beta} \cdot \mathbf{q} \cdot \mathbf{q} \cdot \boldsymbol{\epsilon}_{l}^{\infty} \cdot \mathbf{q}}{ \mathbf{q} \cdot \boldsymbol{\epsilon}^{\infty} \cdot \mathbf{q}}  \\
-  \frac{\mathbf{Z}^*_{\kappa\alpha} \cdot \mathbf{Z}^*_{\kappa'\beta} \cdot \mathbf{q} \cdot \mathbf{q} \cdot \mathbf{q} \cdot \boldsymbol{\epsilon}_{l}^{\infty}} { \mathbf{q} \cdot \boldsymbol{\epsilon}^{\infty} \cdot \mathbf{q}}
+\mathbf{Z}^*_{\kappa\alpha} \cdot \mathbf{Z}^*_{\kappa'\beta} \cdot \mathbf{q} \cdot \mathbf{q} i (\tau_{\kappa l}- \tau_{\kappa' l}) \\
- \mathbf{Z}^*_{\kappa\alpha} \cdot \mathbf{Z}^*_{\kappa'\beta} \cdot \mathbf{q} \cdot \mathbf{q} \frac{ \boldsymbol{\epsilon}_{l}^{\infty} \cdot \mathbf{q}}{4\Lambda^2}
- \mathbf{Z}^*_{\kappa\alpha} \cdot \mathbf{Z}^*_{\kappa'\beta} \cdot \mathbf{q} \cdot \mathbf{q} \frac{ \mathbf{q} \cdot\boldsymbol{\epsilon}_l^{\infty}}{4\Lambda^2}\bigg],
\end{multline}
where we have considered only the dipole-dipole interaction term for simplicity.
This choice does not change any of the results presented in this work, because it only affects the broadening parameter.
This point will be discussed further in Section~\ref{sec:fine}.
All results described below were obtained using this adaptive broadening scheme.

\section{Computational methodology}\label{computsection}

In this section we describe the software and computational parameters employed.
We discuss the interpolation of electron-phonon matrix elements using the long-range dipole and quadrupole corrections.
We look at the convergence rate with respect to the coarse and fine grid interpolation, and show that the drift mobility, Hall mobility, and Hall factor can be extrapolated to the limit of continuous Brillouin zone sampling.
Finally, we propose in Appendix~\ref{app.0} a way to construct an optimal Wigner-Seitz cell for the double three dimensional Fourier interpolation of the electron-phonon matrix elements.

\subsection{Density functional theory and density functional perturbation theory calculations}\label{sec:DFT}
We use the Quantum ESPRESSO software distribution~\cite{Giannozzi2009,Giannozzi2017} with the relativistic Perdew-Burke-Ernzerhof (PBE) parametrization~\cite{Perdew1996} of the generalized gradient approximation (GGA) to density-functional theory.
All the pseudopotentials are norm-conserving, generated using the ONCVPSP code~\cite{Hamann2013}, and optimized via the PseudoDojo initiative~\cite{Setten2018}.
To assess the effect of the exchange-correlation functional, we also regenerated the pseudopotentials using the ONCVPSP code, and changed the exchange-correlation functional from PBE to the Perdew-Zunger parametrization~\cite{Perdew1981} of the local density approximation (LDA)~\cite{Ceperley1980} but keeping all other parameters unchanged.
To assess the effect of pseudization parameters, we also used PBE pseudopotentials from the SG15 ONCV library~\cite{Schlipf2015} that were extended to fully relativistic pseudopotentials~\cite{Scherpelz2016}.

The electron wave functions are expanded in a plane-wave basis set with the kinetic energy cutoff reported in Table~\ref{table:lattice}, such that the total energy is converged to less than 1~mRy per atom.
The Brillouin zone is sampled with a homogeneous $\Gamma$-centered grid of at least 20$\times$20$\times$20 \textbf{k}-points to converge linear-response quantities such as dielectric tensor, Born effective charges and phonon frequencies using DFPT~\cite{Gonze1997a,Baroni2001}.
The relative threshold for the self-consistent solution of the Sternheimer equations was set to 10$^{-17}$ or lower to ensure accurate first-order perturbed wavefunctions.
All the DFPT data were produced in binary or XML format (as opposed to the default text format) to preserve machine precision.
We include SOC in all our calculations.
To obtain accurate results for properties such as the spin-orbit splitting, semicore states have been used for the pseudopotential of Ga, As, and Sb.

\begin{table}[t]
  \begin{tabular}{
      r
      r
      S[table-format=-3.3]
      S[table-format=-3.3]
      r}
  \toprule
         &    &   \multicolumn{2}{c}{{Lattice (bohr)}}  & {ecut} \\
\cline{3-4}
         &    {Valence electrons} &  {~~~~~~Exp.} & {~~~~PBE}   &  {(Ry)}        \\
\hline
\\[-1.0em]
C       & 2s$^2$2p$^2$          & 6.740~\cite{Gildenblat1996} &  6.751 & 100 \\
Si      & 3s$^2$3p$^2$          & 10.262~\cite{Sze2007}       & 10.336 &  40 \\
GaAs    & 3d$^{10}$4s$^2$4p$^1$ - 3d$^{10}$4s$^2$4p$^3$ & 10.683~\cite{Lee1996}       & 10.865 & 130 \\
SiC  & 3s$^2$3p$^2$ - 2s$^2$2p$^2$ & 8.238~\cite{Taylor1960}     &  8.276 & 80 \\
AlP     & 3s$^2$3p$^1$ - 3s$^2$3p$^3$ & 10.318~\cite{Singh1993}     & 10.406 & 80 \\
GaP     & 3d$^{10}$4s$^2$4p$^1$ - 3s$^2$3p$^3$ & 10.299~\cite{Addamiano1960} & 10.408 &  80  \\
BN    & 2s$^2$2p$^1$ - 2s$^2$2p$^3$ & 6.832~\cite{Madelung1991} & 6.848  & 100 \\
AlAs    & 3s$^2$3p$^1$ - 3d$^{10}$4s$^2$4p$^3$ & 10.696~\cite{Pearson1967} & 10.825  & 150 \\
AlSb    & 3s$^2$3p$^1$ - 4s$^2$4p$^6$4d$^{10}$5s$^2$5p$^3$  & 11.595~\cite{Sirota1962}    &  11.767 & 150 \\
SrO     & 4s$^2$4p$^6$5s$^2$ - 2s$^2$2p$^4$ &  9.754~\cite{Bashir2002} & 9.813 & 100  \\
  \botrule
  \end{tabular}
  \caption{\label{table:lattice} Valence electronic configuration, lattice parameter, and planewaves kinetic energy cutoff used in our calculations.
The first lattice parameter for each compound with a bibliographical reference is the experimental value, while the second parameter is obtained by performing a structural relaxation with the PBE exchange-correlation functional.}
\end{table}
For definiteness we used experimental lattice parameters in all cases, as reported in Table~\ref{table:lattice}.
The relaxed lattice parameters are also given in Table~\ref{table:lattice}, and tend to overestimate experiment by an average of 1.3\%, which is typical for PBE calculations.
As shown in Fig.~\ref{fig:effects}, the lattice parameter can have a significant impact on the calculated mobility, leading to differences of up to 20\% when comparing calculations with the optimized or the experimental parameter.

To ensure reproducibility and follow the FAIR principles (findable, accessible, interoperable, and reusable)~\cite{Wilkinson2016}, all pseudopotentials are made available with the manuscript as well as a tagged release of the EPW software that can be accessed at [DOI: 10.24435/materialscloud:b2-j5].

\subsection{Construction of Wannier functions and spatial localization}

\begin{table}[hb]
  \begin{tabular}{r r r r r r r r }
  \toprule\\
Material & \#  & Initial  & Wind. & Froz.  & k    &  \#      & Spread  \\
          &  WF   &  proj.  & eV    & eV    & grid & iter.   & (\AA$^2$) \\
\hline
C-e      &  8 & C1:sp3      &  13.25 &  4.65 & 20$^3$ &  210 &  2.40 \\
C-h      &  8 & C1:sp3      &      - &     - & 20$^3$ &  200 &  0.79 \\
Si-e     & 12 & Si1:d+Si1:s &  14.20 &  4.00 & 16$^3$ & 1320 &  5.67 \\
Si-h     &  8 & Si1:sp3     &      - &    -  & 20$^3$ &  210 &  2.22 \\
GaAs-e   &  8 & Ga:sp3      &   9.88 &  5.75 & 14$^3$ & 3750 &  8.85 \\
GaAs-h   & 16 & Ga+As:sp3   &   9.88 &  5.75 & 16$^3$ &  170 &  3.04 \\
3C-SiC-e &  8 & Si:sp3      &  13.15 &  5.85 & 22$^3$ &  390 &  4.36 \\
3C-SiC-h &  8 & Si:sp3      &      - &    -  & 16$^3$ &  160 &  1.21 \\
AlP-e    &  8 & Al:sp3      &   9.69 &  2.79 & 18$^3$ &  480 &  8.37 \\
AlP-h    &  6 &  P:p        &      - &    -  & 16$^3$ &  170 &  3.00 \\
GaP-e    &  8 & Ga:sp3      &   9.19 &  3.79 & 14$^3$ & 2950 &  8.21 \\
GaP-h    &  6 &  P:p        &      - &    -  & 16$^3$ &  205 &  3.65 \\
c-BN-e   &  8 &  B:sp3      &      - &    -  & 18$^3$ &  260 &  2.51 \\
c-BN-h   &  6 &  N:p        &      - &    -  & 14$^3$ &  205 &  1.22 \\
AlAs-e   &  8 & Al:sp3      &   9.05 &  2.34 & 16$^3$ &  240 &  8.51 \\
AlAs-h   &  6 & As:p        &      - &    -  & 14$^3$ &  105 &  3.36 \\
AlSb-e   &  8 & Al:sp3      &   7.90 &  1.36 & 16$^3$ & 4160 &  9.80 \\
AlSb-h   &  6 & Sb:p        &      - &    -  & 12$^3$ & 2260 &  4.46 \\
SrO-e    & 12 & Sr:d+Sr:s   &  13.26 &  5.06 & 12$^3$ & 7830 &  3.26 \\
SrO-h    &  6 & O:p         &      - &    -  & 12$^3$ &   30 &  1.38 \\
\hline
  \botrule
  \end{tabular}
  \caption{\label{table:wan}
Data related to the Wannierization of all band structures considered in this work.
The valence and conduction manifold are computed separately.
We indicate the initial projections and the number of iterations to reach a relative convergence of 10$^{-12}$ in the spread, the total spread, and the average spread per Wannier function for the converged \textbf{k}-point grids.
For the initial projection, a number next to the atom means that the initial projection is applied to only one of the two equivalent atoms in the primitive cell.
The disentanglement window as well as the frozen window are given with respect to the band edge.
All calculations include SOC.
 }
\end{table}

When performing calculations of carrier mobility, and especially in the presence of magnetic fields,
it is crucial to have symmetric and highly localized Wannier functions.
In order to provide a benchmark for future work, here we describe the procedure followed to generate MLWFs and the properties of the resulting matrix elements.

For diamond we choose 4 $sp^3$ orbitals (8 Wannier functions) located on each atom of the inversion-symmetric unit cell as initial projection for the valence and conduction manifold.
This choice leads to a Wannier function spread $\sigma^2$ of 2.40~\AA$^2$ per function for the converged coarse \textbf{k}-point grid, and 0.79~\AA$^2$ for the valence bands, see Table~\ref{table:wan}.
The strong localization of these MLWFs leads to a fast spatial decay of the matrix elements in the Wannier representation of the electronic Hamiltonian, the electron velocity, the interatomic force constants, and the electron-phonon matrix elements, down to nine orders of magnitude, as seen in Fig.~\ref{fig:wan}.
In all the tests reported here, we find as expected that Wannier functions describing the valence bands are more localized than for the conduction bands.

\begin{table}[t]
  \begin{tabular}{
      r
      r
      S[table-format=-1.2]
      S[table-format=-1.2]
      S[table-format=-1.2]
      S[table-format=-1.2]
      S[table-format=-1.2]
      }
  \toprule
         &    {Excluded}   &  \multicolumn{5}{c}{{Decay length h$^{-1}$~(\AA)}} \\
\cline{3-7}
         &    {bands} &  {H} & {v} & {D} & {g(R$_e$,0)} & {g(0,R$_p$)}      \\
\hline
\\[-1.0em]
C-e     & 1-8  & 1.057 & 1.784 & 0.859 & 1.524 & 1.487 \\
C-h     &   -  & 1.028 & 1.224 & 0.859 & 1.172 & 1.156 \\
Si-e    & 1-8  & 1.478 & 2.101 & 1.290 & 2.133 & 2.034 \\
Si-h    &   -  & 1.710 & 1.617 & 1.290 & 1.547 & 1.644 \\
GaAs-e  & 1-28 & 1.465 & 2.850 & 1.601 & 2.459 & 2.278 \\
GaAs-h  & 1-20 & 1.533 & 1.906 & 1.603 & 2.171 & 2.038 \\
SiC-e   & 1-8  & 1.187 & 1.909 & 1.214 & 1.751 & 1.765 \\
SiC-h   &  -   & 1.245 & 1.362 & 1.215 & 1.381 & 1.719 \\
AlP-e   &  1-8 & 1.517 & 2.785 & 1.373 & 2.511 & 2.272 \\
AlP-h   & 1-2 $\vert$ 9-40   & 1.328 & 1.328 & 1.373 & 1.736 & 2.060 \\
GaP-e   & 1-18 $\vert$ 35-40 & 1.434 & 2.716 & 1.397 & 2.360 & 2.205 \\
GaP-h   & 1-12 $\vert$ 19-40 & 1.293 & 1.490 & 1.397 & 1.903 & 1.985 \\
BN-e    & 1-8 $\vert$ 17-40  & 1.057 & 2.057 & 0.994 & 1.532 & 1.507 \\
BN-h    & 1-2 $\vert$ 9-40   & 0.976 & 1.091 & 1.000 & 1.254 & 1.539 \\
AlAs-e  & 1-18               & 1.509 & 2.647 & 1.450 & 2.465 & 2.267 \\
AlAs-h  & 1-12 $\vert$ 19-40 & 1.247 & 1.375 & 1.450 & 1.720 & 2.057 \\
AlSb-e  & 1-26               & 1.642 & 2.712 & 1.536 & 2.751 & 2.437 \\
AlSb-h  & 1-20 $\vert$ 27-50 & 1.305 & 1.638 & 1.537 & 2.041 & 2.182 \\
SrO-e   & 1-16               & 1.284 & 1.813 & 1.505 & 2.409 & 1.732 \\
SrO-h   & 1-10 $\vert$ 17-40 & 0.952 & 1.035 & 1.505 & 1.326 & 2.076 \\
  \botrule
  \end{tabular}
  \caption{\label{table:decayval}
Number of bands excluded from the Wannierization step, and the exponential decay length of the Hamiltonian, velocity matrix, dynamical matrix, and electron-phonon vertex in the limiting case of $\mathbf{R}=0$ or $\mathbf{R'}=0$.
The decays have been determined from a least-squares fitting of the form $\exp(-hx)$ in a 10~\AA\ range.
}
\end{table}

In the case of silicon, the Wannierization of the conduction band manifold is especially challenging.
Using the wannier90 software we need 1320 iterations to converge to a relative accuracy of 10$^{-12}$~\AA$^2$ in the spread (we note in passing the usefulness of the conjugate gradients algorithm such that we changed the default of
resetting the search direction to steepest descents every 100 iterations, rather than 5, as default).
For the conduction bands, we find that adding one $d$ and one $s$ orbital on one of the silicon atoms as initial projection works best, and yields a spread of 5.67~\AA$^2$.
The comparison between the electronic bandstructure of the conduction bands of silicon using this projection and an
$sp^3$ one is shown in Appendix, Fig.~\ref{fig:sibandswannier}.
The challenge in obtaining carefully converged effective masses lies in the fact that the conduction band minimum of silicon does not lie at a high symmetry point and
is therefore not included in the $\mathbf{k}$-point grid used for interpolation.
The calculation of the conduction band manifold is accelerated in silicon and diamond by excluding the valence bands.
The corresponding Wannier functions are shown in Fig.~\ref{fig:wan}.
They display a relatively complex shape that deviates from the simple chemical picture found for the other materials.
The difficulty in generating Wannier functions for the conduction bands of silicon
has already been discussed in earlier work~\cite{Agapito2013}.
For the conduction band of SrO, we find that using a similar combination of $d$ and $s$ orbitals on the Sr atom works better than $sp^3$, but it also results in a more complicated and entangled set of Wannier functions, as shown in Fig.~\ref{fig:wan}.

For GaAs we used 16 Wannier functions (8 Wannier functions times two due to spin-orbit coupling); on the Ga and As atoms, with $sp^3$ character.
These span both the valence and conduction manifolds, and are used to calculate hole mobilities.
To reduce the computational cost for the electron mobility, we used only 8 Wannier functions centered on the Ga atom, with $sp^3$ character, to describe the conduction manifold.
Since we used semicore states for the pseudopotentials, we excluded the first 20 bands from the Wannierization in the case of valence bands, and 28 bands in the case of conduction bands.
The spread of the maximally localized Wannier functions associated with Ga is 8.85~\AA$^2$.
The case of the valence band of GaAs is unique among the studied compounds.
We find that Wannierizing both the valence and conduction bands yielded a smaller spread of 3.04~\AA$^2$ on the valence band manifold than using 6 or 8 Wannier functions of $p$ or $sp^3$ character on the As atom with a spread of 4.12~\AA$^2$ or 3.29~\AA$^2$, respectively.
We attribute the decrease in spread with increasing number of Wannier functions in the case of GaAs to an increase in the flexibility of the Wannier function basis set and some hybridization with the conduction  manifold due to the small bandgap in DFT.
For SiC we used $sp^3$ orbitals located on the Si atom, with 8 Wannier functions for both valence and conduction, and in the conduction band calculations we excluded 8 bands.
After optimization, the spreads are 4.36~\AA$^2$ and 1.21~\AA$^2$ for all orbitals, respectively.

For the other six materials, we report in Table~\ref{table:wan} all the chosen initial projections, disentanglement energy windows, and  frozen windows~\cite{Souza2001}.
For completeness, we mention in Table~\ref{table:decayval} the bands that were excluded from the Wannierization.
In crystals with inversion symmetry such as diamond or silicon, states with opposite spin orientations are degenerate.
In zinc-blend structures however, the lack of inversion symmetry causes a spin splitting along all directions except [100]~\cite{Theodorou1999}.
We verify in all cases that our choice of initial Wannier projections preserves the expected crystal symmetries in the spread and in the electronic band structure.

\afterpage{%
  \clearpage
\begin{turnpage}
\begin{figure}[ht]
  \centering
  \includegraphics[width=0.95\linewidth]{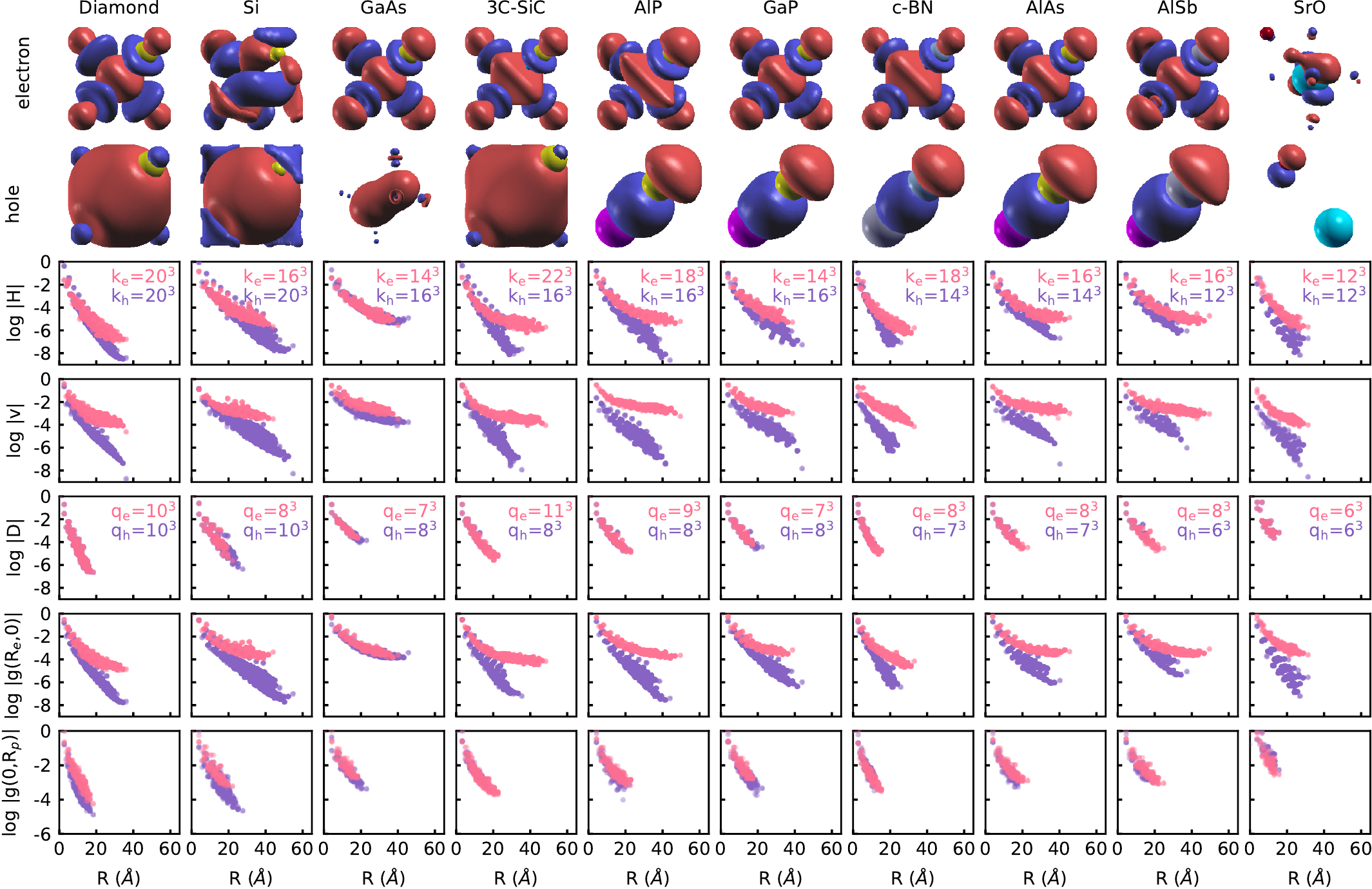}
  \caption{\label{fig:wan}
The first two rows present the Wannier functions in real space for the conduction and valence bands, respectively; generated using the XCrySDen visualization program~\cite{Kokalj1999}.
In the case of conduction bands, we show the $\pm$1.0\AA$^{-3/2}$ isosurface everywhere, except in the case of SrO for which we show the isosurface at $\pm$2.0\AA$^{-3/2}$.
In the case of valence bands, we show the isosurfaces at $\pm$0.2\AA$^{-3/2}$, $\pm$0.5\AA$^{-3/2}$, $\pm$1.0\AA$^{-3/2}$, and $\pm$1.0\AA$^{-3/2}$ for Si, diamond, 3C-SiC and GaAs, respectively; and the isosurface at $\pm$2.0\AA$^{-3/2}$ for all the other materials.
By denoting the spinor components as $[\phi,\psi]$, in the figure we show $|\phi|\times$sign$(\Re\{\phi\})$.
The next five rows show the spatial decays of the electronic Hamiltonian, velocity matrix, dynamical matrix and electron-phonon vertex in the Wannier representation, in the limiting case of $\mathbf{R}=0$ or $\mathbf{R'}=0$, as a function of $R=|\mathbf{R}-\mathbf{R'}|$.
The data points correspond to the largest value taken over the Wannier functions in each of the two unit cells at a distance $R$, and are normalized such that the value is 1 at $R=0$.
The pink and purple colors indicate the Wannierization for the conduction and valence manifold, respectively.
The size of the \textbf{k}- and \textbf{q}-points grids used for each manifold are also indicated in pink and purple colors, respectively.
  }
\end{figure}
\end{turnpage}
\clearpage
}
As seen in Fig.~\ref{fig:wan}, the electronic Hamiltonian, velocity matrix, dynamical matrix and electron-phonon vertex in the limiting case of $\mathbf{R}=0$ or $\mathbf{R'}=0$ in the Wannier representation as a function of $R=|\mathbf{R}-\mathbf{R'}|$ decay very rapidly.
We know that the decay of the Hamiltonian for the valence band manifold must be exponential~\cite{He2001,Brouder2007}.
For the conduction band the exponential decay is not guaranteed, and for the dynamical matrix and the phonon part of the electron-phonon matrix elements we expect a power law decay.
To characterize the decay using a simple unified descriptor, we fit our data with an exponential function for all cases. The resulting decay length is reported in Table~\ref{table:decayval} for all materials.

In the specific case of the valence band manifold of silicon, the Hamiltonian decay length is 1.710~\AA, improving on earlier work
which reported a decay length of 3.2~\AA~with an empirical pseudopotential starting from four bond-centered trial functions~\cite{He2001}.
The worst localization is found for the conduction band manifold of GaAs, GaP, and AlSb ($> 2.7$~\AA), while
the best localization is achieved for the valence manifold of c-BN, diamond, and SrO ($\sim 1$~\AA).
We also note that the spread and number of iterations required to reach convergence systematically increase with $\textbf{k}$-point grid density.
Furthermore, it has been reported that convergence fails for ultra-dense grids (80$\times$80$\times$80 and above)~\cite{Cances2017}.

We also tested two aditional procedures for constructing Wannier functions, namely the selectively-localized Wannier functions (SLWFs)~\cite{Wang2014} and the symmetry-adapted Wannier functions (SAWFs)~\cite{Sakuma2013}.
With SLWFs one constrains the Wannier center $\mathbf{r}_n$ to be located at the position $\mathbf{r}_{0n}$ using a Lagrange multiplier $\lambda_{\rm c}$ as $\lambda_{\rm c} \sum_{n}(\mathbf{r}_n - \mathbf{r}_{0n})^2$, where the summation can be restricted to selected Wannier functions.
In this case we consider the entire manifold.
The multiplier $\lambda_{\rm c}$ can be chosen arbitrarily; the larger the value, the more difficult the convergence, so that in practice the Wannier centers are close to but not exactly positioned at the target positions.
The SAWFs guarantee that the Wannier functions respect the point group symmetry of the crystal.
We note that the existing implementation in the wannier90~\cite{Pizzi2020} code does not currently support band disentanglement with a frozen window nor SOC.
In both cases, we find that the minimum spread is larger than those reported in Table~\ref{table:wan}.
Therefore, we did not purse these avenues further.

The MLWFs obtained as described in this section were used to calculate interpolated band structures, phonon dispersions, and electron-phonon matrix elements in the following sections.

\subsection{Dynamical quadrupoles}\label{sec:quaddyn}

\begin{table}[h]
  \begin{tabular}{r r r r r }
  \toprule\\
          &  \multicolumn{2}{c}{This work} & \multicolumn{2}{c}{Previous work~\cite{Brunin2020}} \\
Compounds &  $Q_{\kappa_1}$ & $Q_{\kappa_2}$ & $Q_{\kappa_1}$ & $Q_{\kappa_2}$ \\
  \hline
Diamond   & \multicolumn{2}{c}{ 3.2372} & \multicolumn{2}{c}{-} \\
     Si   & \multicolumn{2}{c}{11.8307} & \multicolumn{2}{c}{13.67} \\
   GaAs   & 28.88 & -20.41 & 16.54 & -8.57 \\
 3C-SiC   &  7.41 &  -2.63 & - & - \\
   AlP    &  9.73 &  -4.03 & - & - \\
   GaP    & 13.72 &  -6.92 & 12.73 & -5.79\\
   c-BN   &  4.50 &  -0.82 & - & - \\
   AlAs   & 12.25 &  -6.18 & - & - \\
   AlSb   & 14.61 &  -8.70 & - & - \\
   SrO    &  0.00 &   0.00 & - & -\\
  \botrule
  \end{tabular}
  \caption{\label{table:quad_values}
Effective quadrupole tensor ($e\,$bohr) computed in this work.
The complete tensors are given by $\tilde{Q}_{\kappa\alpha}^{\beta\gamma} = (-1)^{\kappa+1}Q|\varepsilon_{\alpha\beta\gamma}|$ for Si and diamond and  $\tilde{Q}_{\kappa\alpha}^{\beta\gamma} = (Q_{\kappa_1} + Q_{\kappa_2} )|\varepsilon_{\alpha\beta\gamma}|$ for all other materials.
We compare our values with previous calculations based on perturbation theory from Ref.~\onlinecite{Brunin2020}.
  }
\end{table}

For practical calculations, we need a procedure for evaluating the quadrupole tensor introduced in Sec.~\ref{sec:dynam_quad}.
An efficient way to compute dynamical quadrupoles rests on perturbation theory and has been implemented in the Abinit software~\cite{Royo2019}.
However, this implementation is currently limited to the LDA exchange and correlation functional, and to pseudopotentials without non-linear core corrections.
To circumvent these limitations, we propose an alternative strategy for estimating the quadrupole tensor, which can be used with any exchange and correlation functional and pseudopotential type.
The general idea is to determine the quadrupole tensor by matching Eqs.~\eqref{eq:multipole}-\eqref{eq:effectivequad} to explicit DFPT calculations at small \textbf{q}.
The minimum number of calculations required is equal to the number of independent components of the tensor.
In practice, to minimize numerical noise we compute the DFPT matrix elements for several points on a sphere with $|\mathbf{q}|=$~constant.
\begin{figure}[b]
  \centering
  \includegraphics[width=0.95\linewidth]{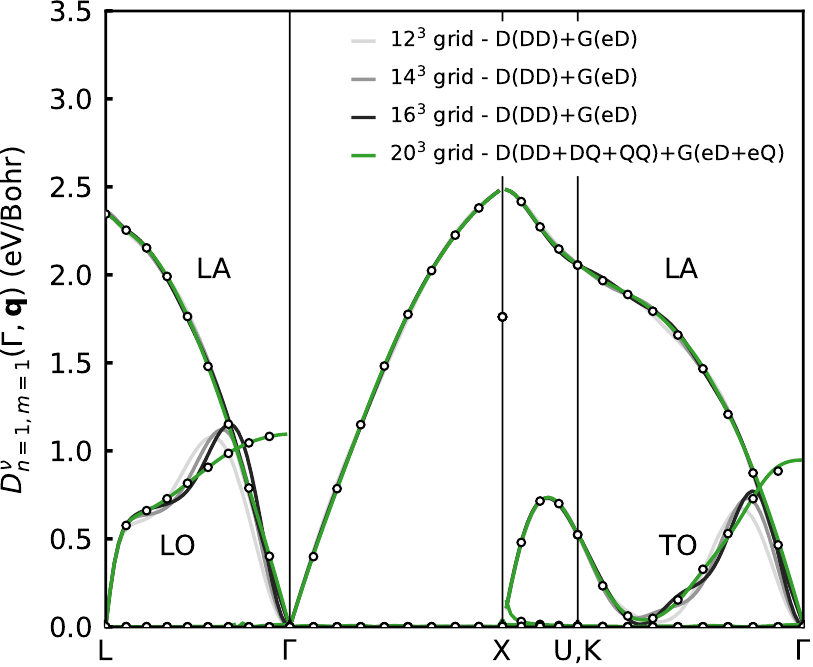}
  \caption{\label{fig:gwithquad}
Comparison between the deformation potential of the $\Gamma_1$ valence band of silicon from the direct DFPT (dots) and the Wannier interpolation with and without quadrupoles, for different \textbf{k}-point grids (the \textbf{q}-point grid is half of that).
The components included in the dynamical matrix and in the electron-phonon matrix elements are labelled as ``D" and ``G", respectively.
The labels ``DD", ``DQ", and ``QQ" indicate the dipole-dipole, dipole-quadrupole and quadrupole-quadrupole contributions, while ``eD" and ``eQ" denote the monopole-dipole and monopole-quadrupole interactions, respectively.
  }
\end{figure}
\begin{figure*}[ht]
  \centering
  \includegraphics[width=0.92\linewidth]{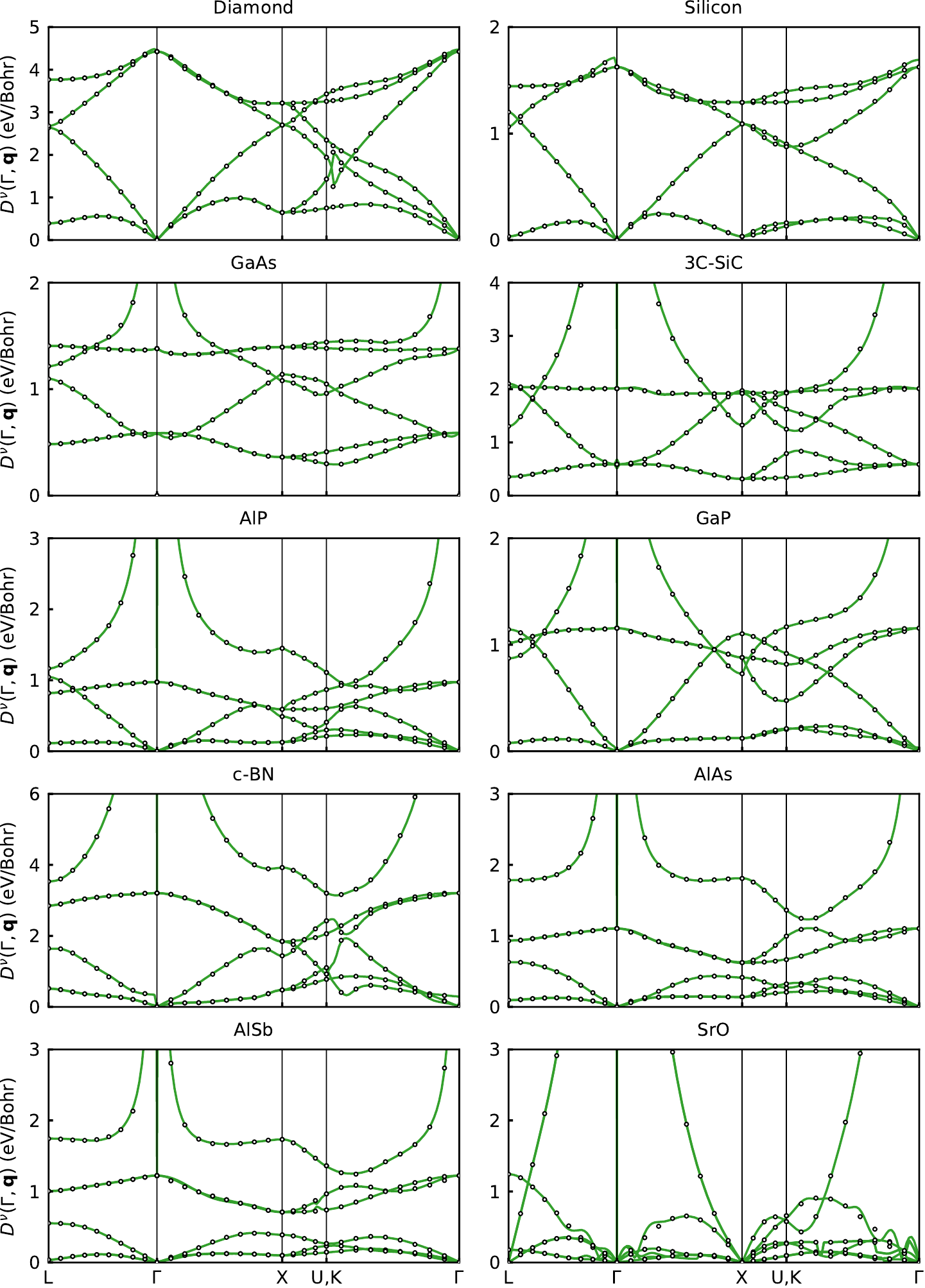}
  \caption{\label{fig:deformation_pot}
Comparison between the deformation potential of the valence band manifold from the direct DFPT calculations (dots) and the Wannier interpolation including dipoles and quadrupoles (green lines).
The $\mathbf{k}$-point for the deformation potential is located at the valence band maximum.
  }
\end{figure*}
\clearpage
To this aim, we use a Python script with the long-range analytic electron-phonon matrix elements from Eq.~\eqref{eq:multipole}, and the phonon frequencies, Born effective charges, dielectric constant, eigendisplacements and wavefunction overlaps from the DFPT calculations.
We then perform least squares optimization with respect to the components of the quandrupole tensor, and include a mode-dependent offset as an additional minimization parameter to capture the $\mathcal{O}(|\mathbf{q}|^0)$ term of the matrix elements, following Eq.~(3.17) of Ref.~\onlinecite{Vogl1976}.
For the compounds considered in this work, we constrain the quadrupole tensor to have the form of Eq.~\eqref{eq:quad2} or Eq.~\eqref{eq:quad1} during the minimization, although the procedure holds for tensors of any symmetry.

We emphasize that this procedure for calculating the quadrupole tensor is inexpensive, as it requires the calculation of only a dozen $\mathbf{q}$-points,
see Fig.~\ref{fig:quad_fit}.

As a sanity check, we computed the quadrupoles of 3C-SiC using the Abinit software~\cite{Gonze2016,Gonze2019}.
To use the Abinit implementation we considered LDA pseudopotentials without non-linear core correction and without SOC.
We obtained $Q_{\rm Si} = 7.025$~$e\,$bohr and $Q_{\rm C} = -1.416$~$e\,$bohr.
Using the same pseudopotentials and the Quantum Espresso software~\cite{Giannozzi2017} we obtained
$\tilde{Q}_{\rm Si} = 7.392$~$e\,$bohr and $\tilde{Q}_{\rm C} = -2.483$~$e\,$bohr.
%
%
We attribute the small difference to the fact that the implementation of Ref.~\onlinecite{Brunin2020} does not include the second term of Eq.~\eqref{eq:effectivequad}, and that our procedure does not capture the $\bf{q}\rightarrow 0$ limit with the same precision as in DFPT.

Using the above procedure, we computed the effective quadrupole tensor for all ten compounds.
The results are reported in Table~\ref{table:quad_values}.
Among the compounds considered here, GaAs has the quadrupole tensor with the largest elements, while this tensor vanishes in SrO due to the cubic $Fm\overline{3}m$ space group symmetry.
In Table~\ref{table:quad_values} we also report the calculated quadrupole terms from Refs.~\onlinecite{Brunin2020,Brunin2020a} for comparison, as obtained using perturbation theory, LDA without non-linear core correction, and neglecting SOC.
Our calculations compare well with Refs.~\onlinecite{Brunin2020,Brunin2020a}, although close agreement is not expected due to the aforementioned differences.

\subsection{Interpolation of the electron-phonon matrix elements}

To assess the quality of the Wannier interpolation, we compare the interpolated electron-phonon matrix element with those obtained from a direct DFPT calculation.
For an easier comparison, we compute the total deformation potential~\cite{Zollner1990,Sjakste2015}:
\begin{equation}\label{eq:deformation}
D^{\nu}(\Gamma, \mathbf{q}) =
 \frac{1}{\hbar N_{\rm w}}\bigg[ 2 \rho \Omega_{\rm} \hbar\omega_{\mathbf{q}\nu} \! \sum_{nm}  |g_{mn\nu}(\Gamma,\mathbf{q})|^2  \bigg]^{1/2},
\end{equation}
where the $\mathbf{k}=\Gamma$ point was chosen, the sum over bands is carried over the $N_{\rm w}$ states of the Wannier manifold, and $\rho$ is mass density of the crystal.
The advantage of examining the deformation potential rather than the matrix elements directly is that electronic degeneracies are traced out, and the frequency that enters the zero-point amplitude is removed, so that one can concentrate on the linear variation of the potential.
To investigate specific states, we also define the band-resolved deformation potential $D_{mn}^{\nu}(\Gamma, \mathbf{q})$ where the band summation in Eq.~\eqref{eq:deformation} has been removed.

In Fig.~\ref{fig:gwithquad}, we compare the deformation potential of silicon for the lowest valence band state at $\Gamma$, for the direct DFPT calculation (black dots) and the Wannier interpolation, with and without quadrupoles (lines).

In silicon there is no Fr\"ohlich contribution to the matrix elements due to inversion symmetry.
There are, however, quadrupole contributions.
As can be seen in Fig.~\ref{fig:gwithquad}, the Wannier interpolation without quadrupole correction reproduces the direct DFPT calculation everywhere, except close to $\Gamma$ in the $\Gamma$L and $\Gamma$K directions for the LO and TO modes.
The discontinuity of the LO and TO modes at $\Gamma$ poses a challenge to the interpolation without quadrupole correction, as shown by the solid lines in gray.
Upon including quadrupole corrections, the deformation potential is found to be independent of grid size.
We therefore only report the curve corresponding to the grid used throughout this manuscript (see Sec.~\ref{sec:coarse}).
In contrast, the longitudinal acoustic (LA) mode of silicon shown in Fig.~\ref{fig:gwithquad} does not exhibit any discontinuities at the zone center, and is therefore well described without quadrupole correction.
The deformation potential of the transverse-acoustic modes (TA) vanishes and is not shown.
These data are consistent with Ref.~\onlinecite{Park2020a}.

Figure~\ref{fig:deformation_pot} shows a systematic comparison between the deformation potential computed via explicit DFPT calculations and the results of our Wannier interpolation including dipole, dipole-quadrupole and quadrupole corrections using Eq.~\eqref{eq:deformation}.
We note the sharp change in deformation potential occurring near the K point in diamond, and to a smaller extent in c-BN.
This effect is due to the avoided crossing in the phonon bandstructure of diamond and c-BN at the same point in momentum space.
Quantitatively, diamond and c-BN show the largest deformation potential, which is consistent with these compounds being the hardest materials in our set.

Overall, the agreement between the direct calculation and the Wannier interpolation with quadrupole correction is excellent, guaranteeing a good interpolation.
The only exception is SrO which is the only material investigated here for which the quadrupole contribution vanishes by symmetry.
This finding suggests that a very accurate interpolation for SrO might require the inclusion of octopole contributions.
We leave the investigation of the effect of octopoles for further studies.


\subsection{Convergece of the mobility with the coarse Brillouin zone grid}\label{sec:coarse}

\begin{figure*}[ht]
  \centering
  \includegraphics[width=0.99\linewidth]{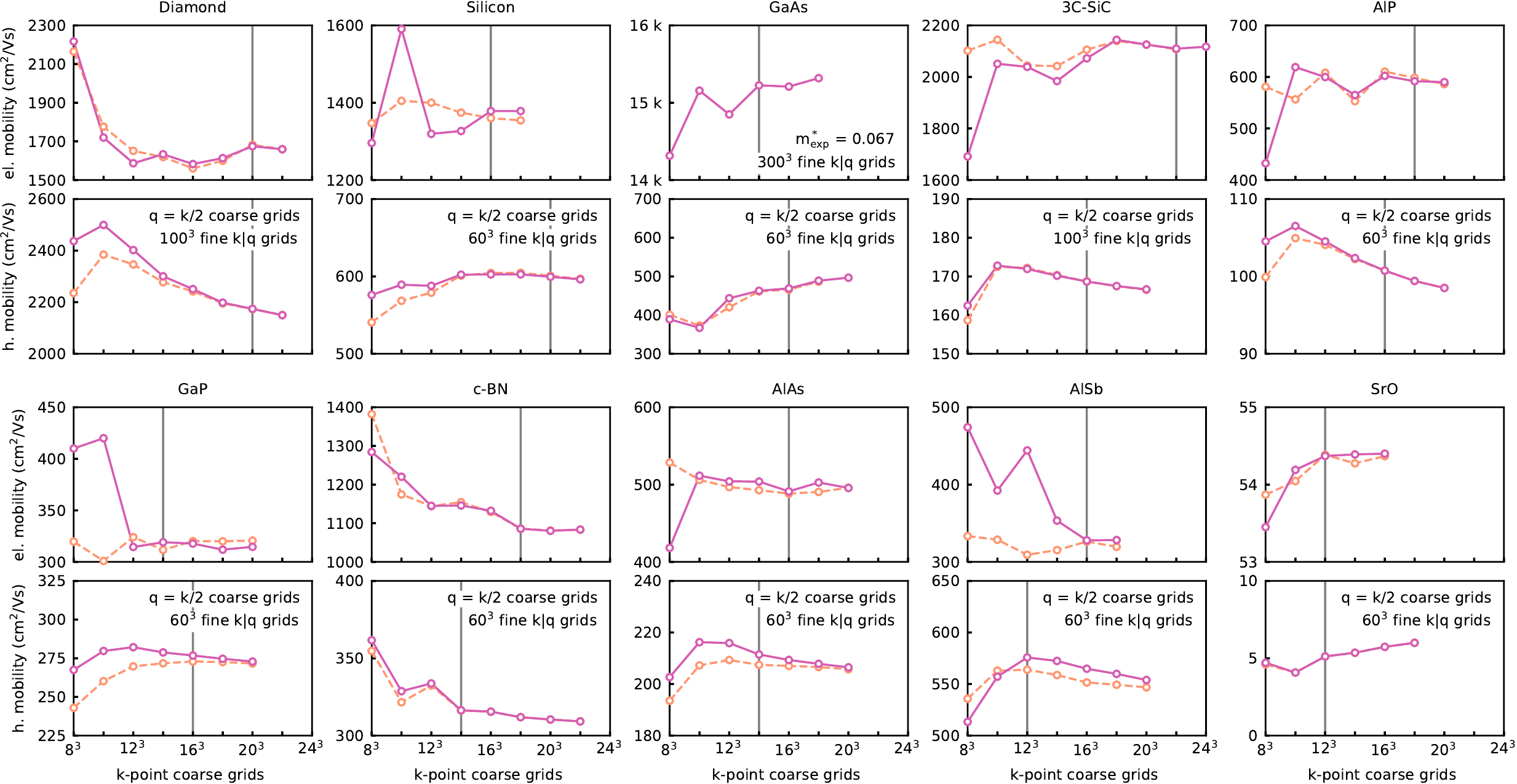}
  \caption{\label{fig:coarseconv}
Convergence of BTE mobility as a function of the coarse \textbf{k}-point grid size (magenta dots).
The coarse \textbf{q}-point grid and the fine \textbf{k} and \textbf{q}-point grids are given in the inset.
The vertical gray line indicates that convergence has been achieved, and represents the coarse grid employed in the remainder of this work.
The orange dots represent the BTE mobility \textit{after} the rescaling defined by Eq.~\eqref{eq:carta}.
}
\end{figure*}

\begin{figure*}[t]
  \centering
  \includegraphics[width=0.99\linewidth]{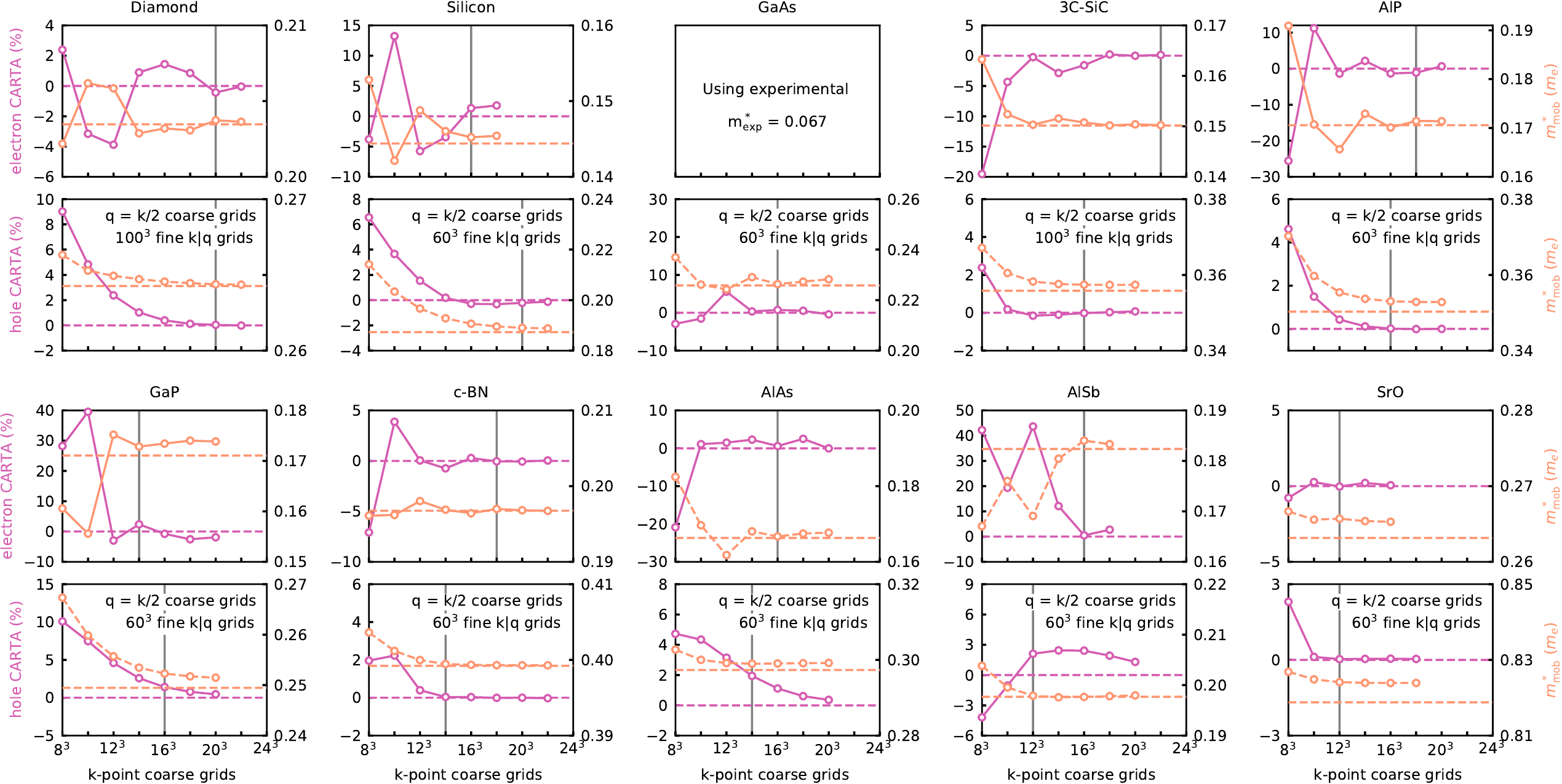}
  \caption{\label{fig:carta}
Convergence of the CARTA mobility as a function of the coarse \textbf{k}-point grid size (magenta dots).
We only show the relative convergence since the CARTA mobility contains the arbitrary parameter $g^2$, see Eq.~\eqref{eq:carta}.
The coarse \textbf{q}-point grid and the fine \textbf{k} and \textbf{q}-point grids are given in the inset.
The vertical gray line indicates that convergence has been achieved, and represents the coarse grid that is used in the remainder of this work.
The orange dots are the mobility effective masses, and are referred to the vertical axis on the right.
The horizontal orange and magenta dashed lines are the DFT reference values.
  }
\end{figure*}

\begin{figure}[h]
  \centering
  \includegraphics[width=0.95\linewidth]{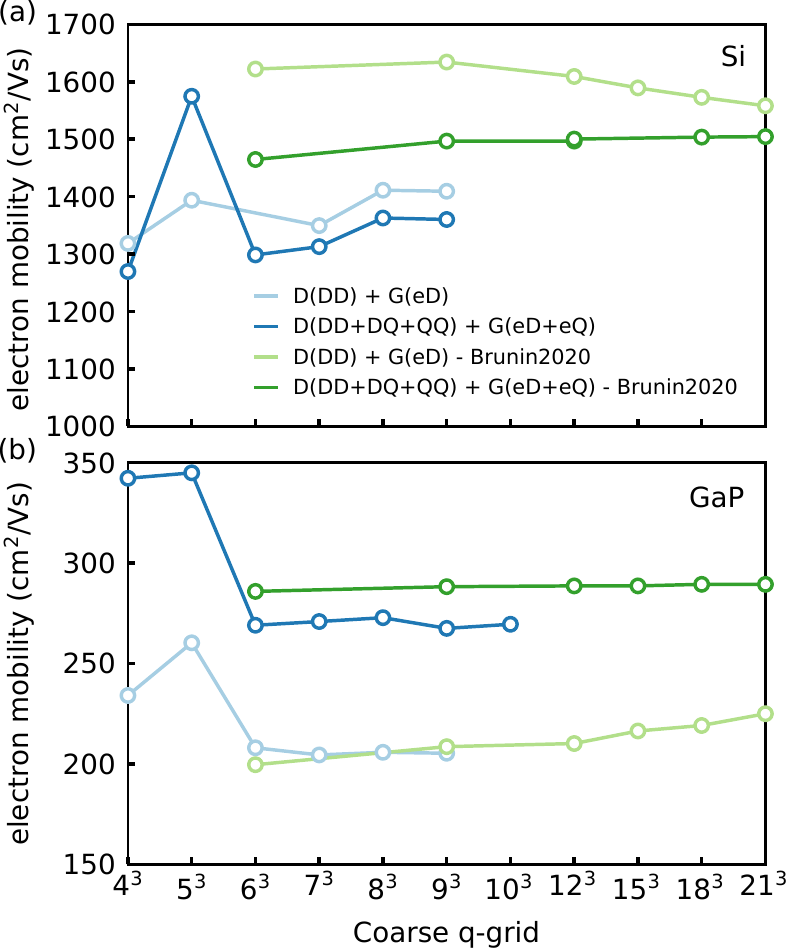}
  \caption{\label{fig:compaBrunin}
Comparison between the coarse-grid convergence in the present work and in Ref.~\onlinecite{Brunin2020a},
for the electron mobility of (a) Si and (b) GaP as a function of the coarse \textbf{q}-point grid.
All calculations are performed within SERTA to be consistent with Ref.~\onlinecite{Brunin2020a}.
The coarse $\mathbf{k}$-point grid used in Ref.~\onlinecite{Brunin2020a} is fixed to 18$^3$, while we used
a coarse $\mathbf{k}$-point grid that is twice the one reported in this figure for the \textbf{q}-point grid.
We used equal fine \textbf{k}- and \textbf{q}-point grids of 150$^3$ for Si and  120$^3$ for GaP, while Ref.~\cite{Brunin2020a} used a fine 72$^3$ $\mathbf{k}$-point and a 144$^3$ $\mathbf{q}$-point grid for Si and a 78$^3$ $\mathbf{k}$-point and a 156$^3$ $\mathbf{q}$-point grid for GaP.
}
\end{figure}

We present in Fig.~\ref{fig:coarseconv} the convergence rate of the electron and hole carrier mobility with increasing coarse grid sizes.
Overall, we find that using the same coarse \textbf{k} and \textbf{q} grid leads to the fastest convergence.
On the other hand, using a coarse \textbf{k}-grid three times denser than the \textbf{q}-grid leads to similar results but at a much higher computational cost, as shown in Fig.~\ref{fig:comparisoncoarse} in the Appendix for c-BN.
Therefore it is recommended that calculations be performed using the same \textbf{k}- and \textbf{q}- coarse grids.
We note that the calculations presented below were performed using a \textbf{k}-grid twice as dense as the \textbf{q}-grid for historical reasons, but in future work equal grids will be employed.

To perform the convergence study, we use either a 60$^3$ or a 100$^3$ fine grids as in indicated in Fig.~\ref{fig:coarseconv}.
We verify that the cross convergence between coarse and fine grids is weak.
As can be seen in Fig.~\ref{fig:coarseconv}, the interpolated mobility converges slowly with respect to the coarse grid size, despite dipole and quadrupole corrections being already included in the calculations.
The origin of this slow convergence has not been investigated so far.
To shed light on this effect, we compute the mobility effective mass defined in Eq.~\eqref{eq:effectivemassmob} and report the convergence of this purely electronic quantity with orange dots in Fig.~\ref{fig:carta}.
We note that the mobility effective mass reflects the quality of the Wannier interpolation of the band structures.
Its value should converge to the mobility effective mass computed directly from DFT bands without interpolation.
In Fig.~\ref{fig:carta} the reference DFT value is shown as horizontal dashed orange lines.
Figure~\ref{fig:carta} shows that the mobility effective mass plays a significant role in the slow convergence of the mobility as a function of coarse grid size.
However, the convergence of the mobility effective mass is a necessary but not sufficient condition for the overall convergence of the mobility with coarse grid size.
Since computing the mobility effective mass is much cheaper than computing the mobility, this quantity can be used to estimate a minimum coarse grid size.
In the cases of diamond, Si, 3C-SiC, AlP, GaP, AlAs, AlSb and SrO in Fig.~\ref{fig:carta}, we see that the electron mobility effective mass does not behave consistently with the convergence of the mobility.
It is also the case for the hole mobility effective mass of GaAs and AlSb.
The reason for this deviation is that the effective mass does not fully capture the scattering phase space, since it neglects the energy conservation connected with the electron-phonon scattering amplitudes.

To incorporate these effects, we employ the CARTA approximation defined in Eq.~\eqref{eq:carta}.
In this case, we have to compute the sum over \textbf{q}-points explicitly, making it more computationally involved, but still far cheaper than the SERTA since we do not have to interpolate the dynamical matrix and electron-phonon matrix elements.
For the cases considered here the CARTA calculations are at least two orders of magnitude faster than the SERTA.
We note that in CARTA calculations we cannot employ adaptive broadening, therefore we use a constant 10~meV Gaussian broadening for the Dirac deltas.
The CARTA results are shown as magenta dots in Fig.~\ref{fig:carta}, and show how the electronic degrees of freedom influence the convergence with respect to the coarse Brillouin zone sampling.
Since these calculations do not carry information about the electron-phonon matrix element, we present the data as a deviation from the CARTA mobility computed without Wannier interpolation, by using directly DFT eigenvalues.

By comparing Figs.~\ref{fig:coarseconv} and \ref{fig:carta}, we observe that the CARTA mobility converges at a similar rate as the BTE mobility.
This observation suggests to speed up the coarse grid convergence by factoring out the electronic degrees of freedom.
To this aim, we rescale our BTE mobilities $\mu_{\alpha\beta}$ using the following expression:
\begin{equation}\label{eq:scaledCARTA}
\mu_{\alpha\beta}^{\rm scaled} = \mu_{\alpha\beta}\frac{\mu_{\alpha\beta}^{\rm CARTA, DFT}}{\mu_{\alpha\beta}^{\rm CARTA}}.
\end{equation}
where $\mu_{\alpha\beta}^{\rm CARTA, DFT}$ is the mobility obtained in the CARTA with DFT eigenvalues and velocities.
Eq.~\eqref{eq:scaledCARTA} allows one to correct for the small off-set observed in some materials for the effective mass mobility.
Moreover, computing $\mu_{\alpha\beta}^{\rm CARTA}$ is over two orders of magnitude faster than $\mu_{\alpha \beta}$ since we only need to interpolate the eigenvalues and velocities.
The results for the scaled BTE mobility are shown as orange dots in Fig.~\ref{fig:coarseconv}.
Importantly, not only $\mu_{\alpha\beta}^{\rm scaled}$ converges faster, but it also does so in a smoother and more systematic way.

The vertical thin gray line in Fig.~\ref{fig:coarseconv} is the coarse grid that we consider converged and that is used in all subsequent calculations, unless stated otherwise.
For the electron mobility of GaAs we do not report the mobility effective mass since we used the experimental value to describe the bands and velocities.
From Fig.~\ref{fig:coarseconv}, we can note that the hole mobility converges in a smooth and systematic way while this is not necessarily the case for the electron mobility.
This behavior reflects the better real-space localization of Wannier functions obtained for the valence manifold than the conduction manifold, see Fig.~\ref{fig:wan}.

For the case of silicon, the electron mobility is significantly improved upon using the scaling of Eq.~\eqref{eq:scaledCARTA}.
However, we can notice that for our chosen converged grid, the calculated mobility does not converge to the correct limit without scaling (see Fig.~\ref{fig:coarseconv}).
Specifically, the interpolated CARTA result overestimates the CARTA/DFT result by 1~\%.
We expect a similar error to occur for the BTE mobility.
We anticipate that the slight difference will disappear upon using denser grids or by using a more extended Wannier functions basis set, but we have not investigated this further due to the computational cost.
An overestimation of 2~\% is also observed for the hole mobility of AlSb.
In all other cases, the interpolated CARTA values converge smoothly to the CARTA/DFT result.

\begin{figure*}[ht]
  \centering
  \includegraphics[width=0.99\linewidth]{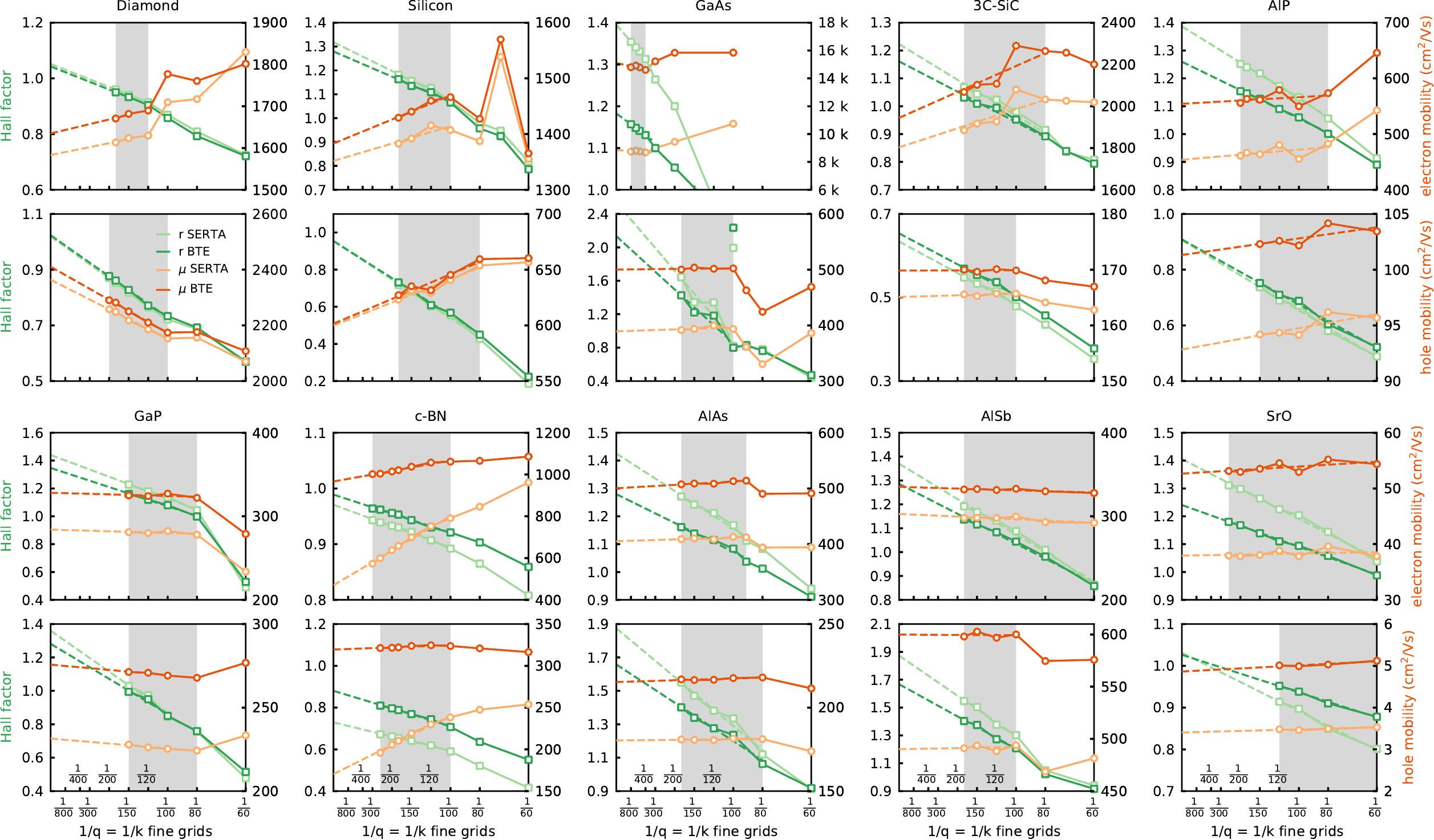}
  \caption{\label{fig:fineconvergence}
Convergence of the SERTA and BTE mobility and Hall factor as a function of the fine \textbf{k}-and \textbf{q}-point grid size, using the converged coarse grids from Fig.~\ref{fig:coarseconv}.
Both quantities converge linearly with the fine grid density, and can therefore be extrapolated to infinity.
The shaded area denotes the interpolation range for the linear least squares fit of the Hall factor and mobility.
}
\end{figure*}

We are now in a position to answer the question on why is the convergence of the carrier mobility slow with respect to coarse grid size even when dipole and quadrupole corrections are included.
The answer is material dependent.
In the case of the electron mobility of Si, 3C-SiC, AlP, GaP, AlAs, AlSb, and SrO, the slow convergence is mostly due to purely electronic effects as demonstrated by the CARTA results.
Specifically, the Wannier-interpolated electron bands converge slowly with grid size.
A possible way to accelerate convergence would be to increase the number of Wannier functions in order to improve the flexibility of the basis functions.
In general, for the valence bands the convergence is smoother and more systematic, and, therefore, less can be gained by using the CARTA rescaling.
Nevertheless, we noticed some improvements for diamond, AlP, GaP, c-BN, AlAs, and AlSb.

In the case of diamond and c-BN, very little improvement is noticed using the CARTA rescaling.
This indicates that for these two materials the slow convergence can be attributed to a slow convergence of the phonon dispersions and electron-phonon matrix elements with grid size.

In Fig.~\ref{fig:compaBrunin} we compare our coarse grid convergence results with previous work for Si and GaP in Ref.~\onlinecite{Brunin2020a}.
Since only the SERTA mobility is reported in Ref.~\onlinecite{Brunin2020a}, we compare with our results at the same approximation level.
Overall our results agree quite well given that there are a number of differences in the calculations.
The coarse $\mathbf{k}$-point grid in Ref.~\onlinecite{Brunin2020a} was fixed to 18$^3$, while we used
a coarse $\mathbf{k}$-point grid that is twice the one reported in Fig.~\ref{fig:compaBrunin} for the $\mathbf{q}$-point grid.
We used an equal fine \textbf{k}- and \textbf{q}-point grids of 150$^3$ for Si and 120$^3$ for GaP, while Ref.~\onlinecite{Brunin2020a} used a fine 72$^3$ $\mathbf{k}$-point grid and a 144$^3$ $\mathbf{q}$-point grid for Si and a 78$^3$ $\mathbf{k}$-point grid and a 156$^3$ $\mathbf{q}$-point grid for GaP.
Different pseudopotentials and exchange correlation functionals were used as well.
In addition, we included SOC, while Ref.~\onlinecite{Brunin2020a} neglects it, but as shown in Fig.~\ref{fig:effects}, the effect on electron mobility is negligible.
The main difference between our results and those of Ref.~\onlinecite{Brunin2020a} is that our calculations including dipole but not quadrupole corrections seem to converge rapidly, while in the earlier work there is a slow drift of in the SERTA mobility calculated with dipole correction.
However, in agreement with Ref.~\onlinecite{Brunin2020a}, we observe that the inclusion of quadrupole corrections leads to a shift of the calculated mobility even at convergence.

\subsection{Convergence of the mobility with the fine Brillouin zone grid}\label{sec:fine}

Using the converged coarse grids determined in the previous section, we can proceed to testing the convergence of the mobility with respect to the fine grid.
The presence of the $\partial f_{n\mathbf{k}}^0 / \partial \varepsilon_{n\mathbf{k}}$ term in the mobility, see for example Eq.~\eqref{eq:serta}, implies that only those states close to the valence band or conduction band edge will contribute.
To reduce the computational cost of these calculations we use the following scheme.
We construct a fine \textbf{k}-point grid composed of only the points within a small energy window around the band edges, and rely on crystal symmetry to further reduce the number of points.
For example in the case of the hole mobility of c-BN, using an energy window of 0.3~eV and an homogeneous fine grid of 250$^3$ points, we only have 12,390 irreducible \textbf{k}-points.
Importantly, at no point we store the entire uniform grid in memory. 
This reduction allowed us to use extremely dense grids for the electron mobility of GaAs, up to 800$^3$ points in the first Brillouin zone.

Since all \textbf{q}-points can contribute (for example through inter-valley scattering), we do not use symmetry reduction, and keep all the \textbf{q}-points of the uniform grid that lie with the small energy window.
For example, in the case of c-BN and a grid of 250$^3$ points, we have to compute 814,981~\textbf{q}-points.
To this aim we generate the \textbf{q}-points on the fly, so that we do not have to store the full grid in memory.

Finally, we interpolate the electron-phonon matrix elements for all these \textbf{k} and \textbf{q} points and store on disk only the \textbf{k}- and \textbf{q}-points identified above that contribute to the mobility.
To retain only the information that will contribute to the mobility calculation, for each
$\mathbf{q}$-point we store the matrix elements for the $n\mathbf{k}$ states that fulfil the following condition:
\begin{equation}\label{eq:threshold}
\hbar \frac{\partial f_{n\mathbf{k}}^0}{\partial \varepsilon_{n\mathbf{k}}} \tau_{n\mathbf{k}}^{-1}(\mathbf{q}) > \frac{10^{-16}}{N_{k} N_{q} N_b^2},
\end{equation}
where $N_{k}$, $N_{q}$ and $N_{b}$ are the total number of \textbf{k}-points, \textbf{q}-points, and Wannierized bands, respectively.
The threshold in Eq.~\eqref{eq:threshold} guarantees that the cumulative absolute error made by neglecting very small matrix elements will be lower than machine precision, or $10^{-16}$.
In the c-BN example mentioned above, this procedure required the storage of a binary file of approximately 40~GB instead of multiple TBs required to store all the matrix elements.
Once all the scattering rates are written to disk, we read the file in parallel using the message passing interface input-output (MPI-IO) and solve the iterative BTE.
The first step of the iterative solution yields the SERTA mobility.
The iterative solution of the BTE is extremely fast (a few minutes) and converges typically within a few tens of iterations.
For the compounds considered in this work, we determined that the optimal energy window is between 0.2 and 0.3~eV, as shown in Table~\ref{table:energywindow} of the Appendix.

The results for the SERTA and BTE mobility and Hall factor are shown for all materials in Fig.~\ref{fig:fineconvergence}.
For similar reasons as in Ref.~\onlinecite{Ponce2015}, we stress that we are showing the results as a function of inverse grid density and the fine \textbf{k}-point and \textbf{q}-point grids are always the same.
For example, the label 1/100 on the horizontal axis corresponds to a 100$^3$ fine grid, such that the left hand side of each figure corresponds to infinitely dense sampling.

Interestingly, both the mobility and the Hall factor appear to converge linearly with inverse grid density and can therefore be extrapolated.
The linear interpolation range is depicted in Fig.~\ref{fig:fineconvergence} with a gray shaded region.
We perform the linear extrapolation as soon as we enter the linear regime.
This linearity comes from the fact that we compute the momentum integrals using the basic rectangular integration, therefore the error scales linearly with the grid spacing.

\begin{figure}[ht]
  \centering
  \includegraphics[width=0.95\linewidth]{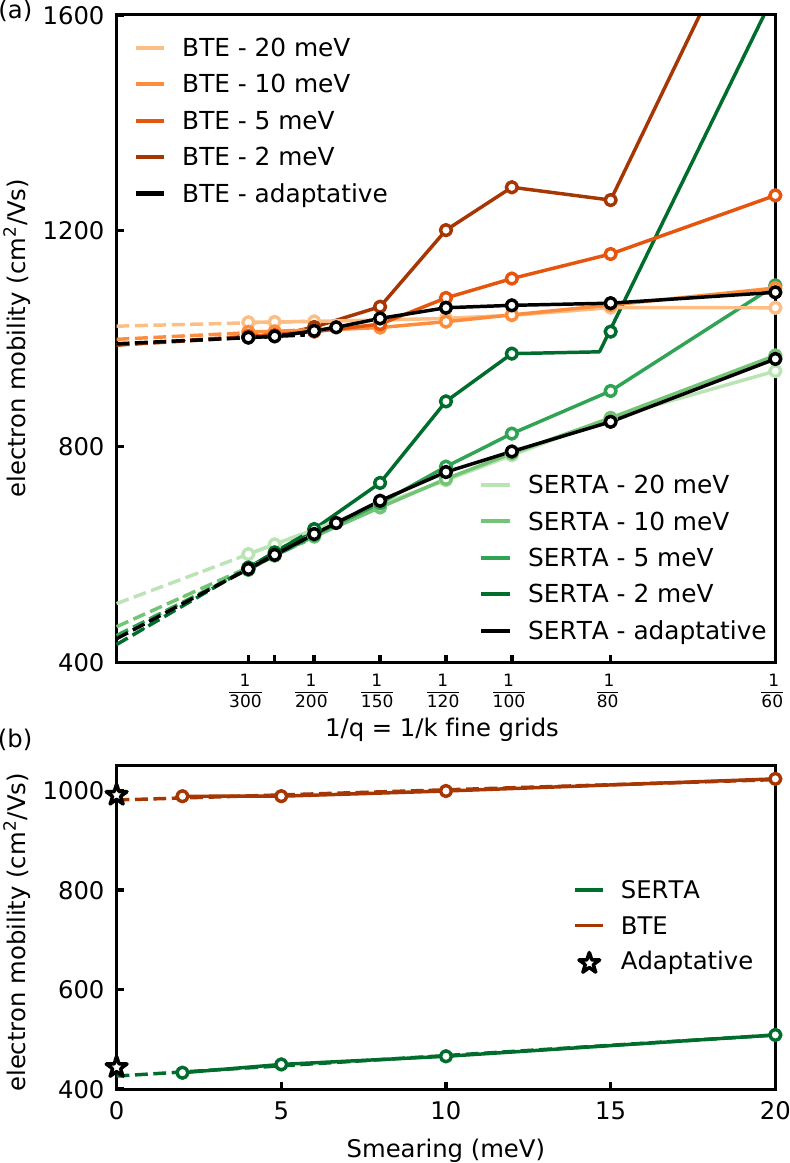}
  \caption{\label{fig:fineconvBN}
(a) Convergence of the SERTA and BTE electron mobility of c-BN as a function of the fine \textbf{k}-and \textbf{q}-point grid size.
The coarse grid employed in these calculations is given in Fig.~\ref{fig:coarseconv}.
We test various fixed smearing parameters as well as the adaptative smearing.
(b) Electron mobility of c-BN at extrapolated infinite grid density, plotted as a function of smearing parameter.
The dashed lines are the extrapolation to zero smearing.
The use of adaptative smearing (stars) yields results close to the zero-smearing extrapolation.
}
\end{figure}

We notice that the electron and hole mobility of diamond, Si, c-BN as well as the electron mobility of GaAs and 3C-SiC show a relatively steep slope, therefore it is important to extrapolate the values for accurate results.
In contrast, all the other cases considered here exhibit a milder slope, therefore the extrapolation is not necessary in these cases.

The linear behavior of the mobility is closely linked to the smearing used for the energy conserving Dirac deltas in Eq.~\eqref{eq:iter}.
In Fig.~\ref{fig:fineconvBN} we focus on the electron mobility of c-BN since the linear trend is very pronounced, and
we show how the carrier mobility depends on the grid density for various Gaussian smearing parameters.
We see that, the smaller the smearing, the more pronounced are the fluctuations, and finer grids are needed before the linear regime is reached.
The extrapolated value depends on the smearing used.

The above discussion does not take into account the finite lifetimes of carriers in the calculations.
In reality, a finite smearing in the range of 10-50~meV is to be expected as a result of the electronic linewidths.
This effect could be incorporated by setting the smearing to $\hbar/\tau_{n\bf{k}}$  with $\tau_{n\bf{k}}$ given by Eq.~\eqref{eq:scattering_rate}, but we have not explored this direction.

Since the linear extrapolation at fixed smearing is rather tedious, in practice we proceed by using the adaptive smearing introduced in Eq.~\eqref{eq:eta_para}.
As seen in Fig.~\ref{fig:fineconvBN}, this approach offers a very good approximation to the zero-smearing extrapolation.
All data presented in Fig.~\ref{fig:fineconvergence} have been generated using adaptive smearing.
We also show a similar analysis for the case of the electron mobility of GaAs in Fig.~\ref{fig:fineconvGaAs} of the Appendix.

In Fig.~\ref{fig:fineconvergence} we also present the convergence of the SERTA and BTE Hall factor defined in Eq.~\eqref{eq:hallfactor}.
In this case, the Hall factor systematically converges linearly with grid density and can be extrapolated to infinite grid density.
Also in this case we can either extrapolate the results to zero smearing, or use adaptive smearing.
The results in either case are extremely close, therefore we proceed with adaptive smearing in the following.

\section{Results}\label{resultsection}

In this section we discuss the main results of our calculations for all the compounds considered in our work.
We first describe the electronic and vibrational properties, then we discuss the Hall factor and its temperature dependence, as well as drift and Hall carrier mobilities.
We also analyze in details the microscopic mechanisms responsible for limiting the carrier mobilities.
Lastly, we compare our results to available experimental data.

\subsection{Wannier interpolation of electron bands and phonon dispersions}

\begin{figure*}[ht]
  \centering
  \includegraphics[width=0.99\linewidth]{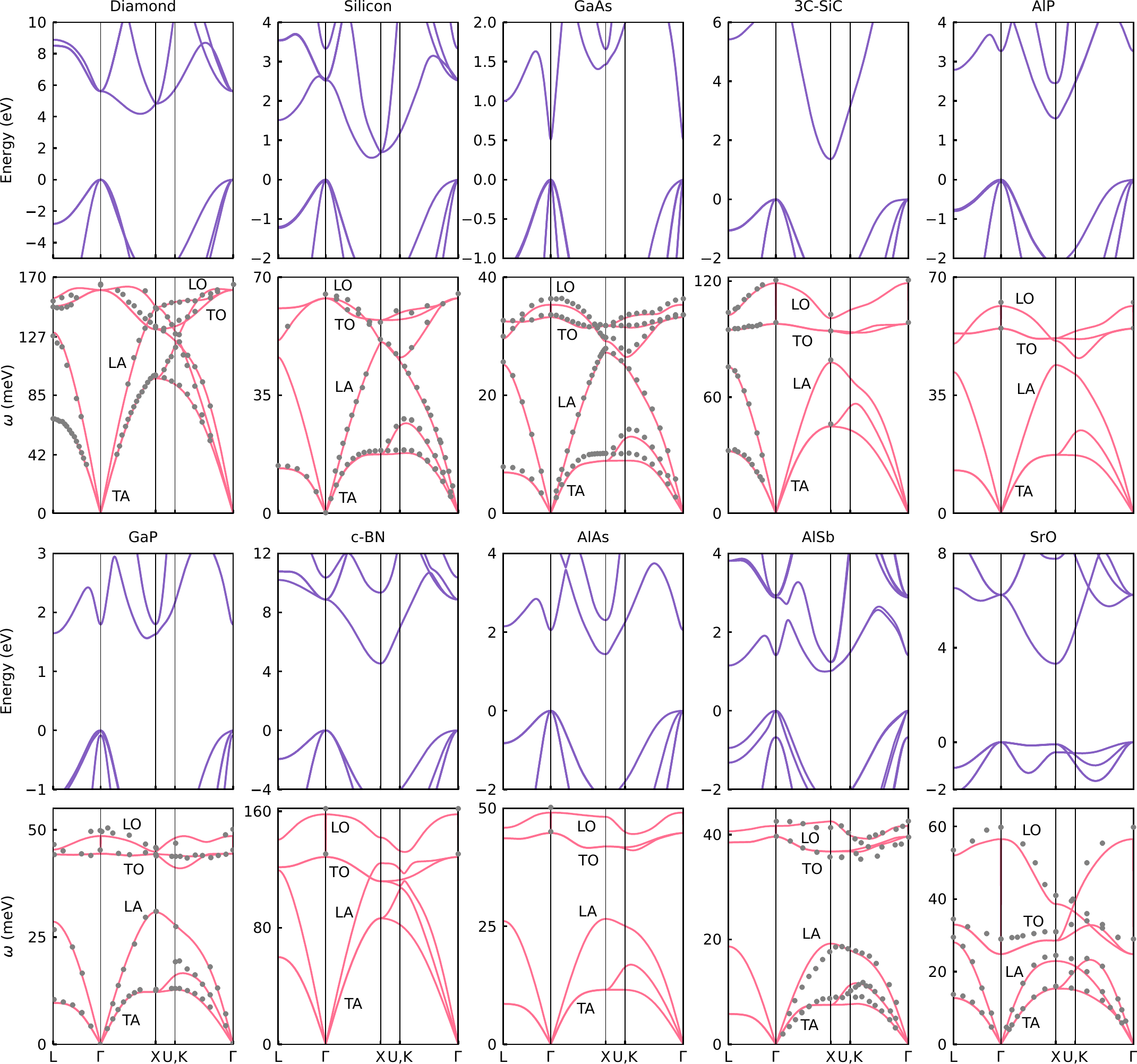}
  \caption{\label{fig:bandstructures}
Wannier-interpolated band structures and phonon dispersions of the compounds considered in this work.
The gray dots are experimental neutron scattering data for diamond~\cite{Warren1967}, silicon~\cite{Nilsson1972,Kulda1994}, GaAs~\cite{Strauch1990}, 3C-SiC~\cite{Feldman1968}, AlP~\cite{Onton1970}, GaP~\cite{Borcherds1979}, c-BN~\cite{Karch1997}, AlAs~\cite{Strauch1990}, AlSb~\cite{Jha1997}, and SrO~\cite{Rieder1975}.
}
\end{figure*}

The interpolated electronic band structures and phonon dispersions for the ten compounds considered in this work are presented in Fig.~\ref{fig:bandstructures}.

The phonon dispersions are compared to neutron scattering data.
Close agreement between theory and experiment is found in all cases except for SrO, where the theory underestimates measured frequencies.
The associated Born effective charges, high-frequency dielectric constants and maximum phonon frequencies are reported in Table~\ref{table:bandstructures}.
The average deviation for the highest phonon frequency between calculations and experiments is 1.6\%.

Since DFT/PBE systematically underestimates the electronic bandgap, the dielectric constant is systematically overestimated with respect to experiment.
For the systems considered here, the average overestimation of the dielectric constant is 10\%, but the largest discrepancy is as large as 30\%.

As expected, the band gaps are severely underestimated.
The average deviation from experiments is of 41\%.
On the other hand, the spin-orbit splitting is in slightly better agreement with experiment, the average deviation being 18\%.
Given these data, we expect that the electron-phonon matrix elements will be overscreened, and we anticipate that our calculated mobilities will tend to overestimate experimental data.
We will come back to this point in Sec.~\ref{eq:exp_comp}.

\begin{table}[t]
\centering
  \begin{tabular}{@{}l l@{\hskip 0.02in} l@{\hskip 0.02in} l@{\hskip -0.02in} l l l l@{\hskip -0.02in}}
  \toprule
           & Z$^*$ & $\varepsilon^{\infty}$ & $\omega_{\textrm{max}}$ & &  $E_g$ & $\Delta_{\rm so}$ &  \\
           &       &                        & (meV)                   & & (eV)  & (meV)             &  \\
\hline
\\[-1.0em]
      C & 0 & 5.77 & 163.20 & & 4.18 & 13 &   \\
        & - & 5.5~\cite{Sze2007} & 165~\cite{Warren1967} & & 5.46~\cite{Clark1964} & 12~\cite{Cardona2001a} &   \\
     Si & 0 & 12.89 &  63.71 & & 0.55  & 48 &   \\
        & - & 11.94~\cite{Sze2007}  &  62.74~\cite{Dolling1963} & & 1.17~\cite{Kittel2004} & 44~\cite{Bona1985} &   \\
 GaAs   & 2.11 & 14.30 &  35.35 & & 0.41 & 335 &    \\
        &  -   & 10.86~\cite{Lockwood2005} & 36.20~\cite{Lockwood2005}  & & 1.42~\cite{IOFF} & 340~\cite{IOFF} &   \\
 SiC & 2.70 &  6.92 & 119.00 & & 1.36 & 14 &    \\
        & -    &  6.52~\cite{IOFF} & 120.54~\cite{IOFF}  & & 2.36~\cite{IOFF}  & 10~\cite{IOFF} &   \\
 AlP    & 2.21 &  8.08 &  61.45 & & 1.56 & 60 &    \\
        &  -   & 7.5~\cite{Christensen1996} &  62.55~\cite{Onton1970a} & & 2.51~\cite{Madelung1996} & 65~\cite{Vurgaftman2001} &  \\
  GaP   & 2.14 & 10.43 &  47.29 & & 1.55 & 84 &   \\
        &  -   &  9.2~\cite{Lockwood2005} & 49.90~\cite{Lockwood2005} & & 2.35~\cite{Haynes2020} & 130~\cite{Haynes2020} &   \\
  c-BN  & 1.91 &  4.54 & 157.92 & & 4.53 & 21 &   \\
        &  -   & 4.46~\cite{IOFF} & 158.82~\cite{IOFF} & & 6.4~\cite{IOFF} & 9~\cite{IOFF} &    \\
 AlAs   & 2.11 &  9.24 &  49.09 & & 1.34 & 299 &     \\
        &    - & 8.16~\cite{Lockwood2005} & 49.54~\cite{Lockwood2005} & & 2.25~\cite{Guzzi1992} & 275~\cite{Onton1970} &  \\
 AlSb   & 1.79 & 11.48 & 39.59  & & 0.99 & 670 &     \\
        &   -  & 10.9~\cite{Seeger1991} & 39.18~\cite{Turner1962} & & 1.7~\cite{Haynes2020} & 645~\cite{Rustagi1976} &     \\
  SrO   &  2.45 &  3.78  &  56.52 & & 3.31 & 61 &    \\
        &    -  & 3.5~\cite{Duan2008} &  59.8~\cite{Rieder1975} & & 5.22~\cite{Duan2008} & - &    \\ [0.5em]
 & \multicolumn{4}{c}{hole mass (m$_{\rm e}$)} &  \multicolumn{3}{c}{electron mass (m$_{\rm e}$)} \\
 & $m_{\rm hh}^{*}$ & $m_{\rm lh}^{*}$ &  $m_{\rm so}^{*}$ & $m_{\rm mob}^{*,\rm h}$ & $m_{\rm \parallel}^{*}$ & $m_{\rm \perp}^{*}$ &  $m_{\rm mob}^{*,\rm el}$  \\
\hline
     C  & 0.413 & 0.294 & 0.343 & 0.264 & 1.645 & 0.290 & 0.204 \\
        & 0.588~\cite{Willatzen1994} & 0.303~\cite{Willatzen1994} & 0.394~\cite{Willatzen1994} & - & 1.4~\cite{Nava1980} &  0.36~\cite{Nava1980} & - \\
     Si & 0.260 & 0.189 & 0.225 & 0.187 & 0.957 & 0.193 &  0.144 \\
        & 0.49~\cite{IOFF} & 0.16~\cite{Dexter1954} & 0.234~\cite{Barber1967} &  - & 0.92~\cite{Hensel1965} & 0.19~\cite{Hensel1965} & - \\
   GaAs & 0.324 & 0.034 & 0.107 & 0.226 & \multicolumn{2}{c}{-} & 0.034 \\
        & 0.51~\cite{IOFF} & 0.08~\cite{IOFF} & 0.15~\cite{IOFF}  & - & \multicolumn{2}{c}{0.067~\cite{Vurgaftman2001}}  & - \\
   SiC  & 0.592 & 0.421 & 0.492 & 0.356 & 0.672 & 0.260 & 0.150 \\
        &   -   & 0.45~\cite{Kono1993} & - & - & 0.68~\cite{Kaplan1985} & 0.25~\cite{Kaplan1985}  & - \\
   AlP  & 0.545 & 0.253 & 0.354 & 0.350 & 0.787 & 0.277 & 0.171 \\
   GaP  & 0.376 & 0.141 & 0.214 & 0.249 & 1.051 & 0.410 & 0.171 \\
        & 0.54~\cite{Bradley1973} & 0.16~\cite{Bradley1973} & - & - & 0.87~\cite{Baranskii1976} & 0.252~\cite{Baranskii1976} & - \\
    BN  & 0.53 & 0.52 & 0.52 & 0.399 & 0.914 & 0.303 & 0.197 \\
        & 0.38~\cite{Madelung1996} & 0.15~\cite{Madelung1996} & - & - & 1.2~\cite{IOFF} &  0.26~\cite{IOFF}  & - \\
  AlAs  & 0.463 & 0.151 & 0.264 & 0.297 & 0.851 & 0.240 & 0.166  \\
        & 0.81~\cite{Adachi1994} & 0.16~\cite{Adachi1994} & 0.30~\cite{Adachi1994} & - & 1.1~\cite{Lay1993} & 0.19~\cite{Lay1993} & - \\
  AlSb  & 0.322 & 0.105 & 0.236 & 0.198 & 1.41 & 0.474 & 0.182 \\
        & 0.5~\cite{Cardona1966} & 0.11~\cite{Cardona1966} & 0.29~\cite{Lawaetz1971} & - & 1.0~\cite{Glinskii1979} & 0.26~\cite{Glinskii1979} & - \\
   SrO  & 4.324 & 0.464 & 0.871 & 0.819 & 1.222 & 0.403 & 0.263 \\
  \botrule
  \end{tabular}
  \caption{\label{table:bandstructures} Computed Born effective charges, high frequency dielectric constants, maximum phonon frequencies, electronic band gaps, spin-orbit splitting, and hole and electron effective masses (as well as mobility effective mass), compared to experiments.
The experimental data are reported in the row below each compound name.
The mobility effective mass is evaluated for the temperature $T=300$~K.
The electron effective mass of GaAs is not reported as we used the experimental mass, as discussed in the text.
}
\end{table}

The DFT mobility effective masses obtained without Wannier interpolation  are reported in Table~\ref{table:bandstructures}.
The values range from 0.1 to 0.8~$m_{\rm e}$.
Direct comparison of such mobility effective mass with experiment is not possible, however we can compare the effective masses calculated using the standard definition with experiments.
We report all the experimental and theoretical electron and hole effective masses in Table~\ref{table:bandstructures}.
Whenever the experimental values were reported along the principal axis, we transform them to the density of state effective mass using for example $m_{\rm hh}^{*} = [(m_{\rm hh}^{110})^2 m_{\rm hh}^{100}]^{1/3}$, and similarly for the light-hole effective masses.

For diamond, the experimental longitudinal and transverse electron effective masses are 1.4 and 0.36~$m_{\rm e}$~\cite{Nava1980}, respectively.
For the hole, we used the values from Willatzen and Cardona~\cite{Willatzen1994} which are obtained from linear muffin-tin orbital calculations corrected with measured energies, and are considered most reliable
(experimental values obtained from the field dependence of the hole mobility show significant dispersion, with 1.1~\cite{Reggiani1981}, 0.3~\cite{Reggiani1981}, and 1.06~\cite{Rauch1962} for the heavy-hole, light-hold and spin-orbit band, respectively).

For silicon, the experimental longitudinal and transverse electron effective masses are 0.92 and 0.19~$m_{\rm e}$~\cite{Hensel1965}, respectively, while the light and heavy hole masses are 0.16 and 0.5~$m_{\rm e}$~\cite{Dexter1954}, respectively.
%

For GaAs, the electron effective mass is isotropic, with a recommended experimental value of 0.067~$m_{\rm e}$~\cite{Vurgaftman2001}.
The measured spin-orbit, light hole, and heavy hole effective masses of GaAs are 0.15, 0.082, and 0.51~$m_{\rm e}$~\cite{IOFF}, respectively.
%

For 3C-SiC, the measured longitudinal and transverse electron masses are 0.68 and 0.25~$m_{\rm e}$~\cite{IOFF}, respectively; for the holes we could find the density of states effective mass, 0.45~$m_{\rm e}$~\cite{Kono1993}.

For AlP there are no experimental effective masses available, but we note a $\mathbf{k}\cdot \mathbf{p}$ study in relatively good agreement with our results which reports 2.68~\cite{Vurgaftman2001} and 0.16~$m_{\rm e}$ ~\cite{Vurgaftman2001} for the transverse and perpendicular electron effective masses, respectively.
For the hole, the same study obtains 0.73, 0.19, and 0.3~$m_{\rm e}$~\cite{Vurgaftman2001} for the heavy-hole, light-hole, and spin orbit hole, respectively.

In the case of GaP, the low-temperature electron transverse and longitudinal effective masses were measured to be 0.252 and 0.87~$m_{\rm e}$~\cite{Baranskii1976}, respectively.
The heavy and light hole effective masses were obtained by cyclotron resonance at 1.6~K, and are 0.54 and 0.16~$m_{\rm e}$~\cite{Bradley1973}, respectively.

In the case of c-BN, the longitudinal and transverse electron effective masses are 1.2 and 0.26~$m_{\rm e}$~\cite{IOFF}; the heavy hole masses were reported in the range 0.38-0.96~$m_{\rm e}$, and the light-hole masses were reported in the range 0.11-0.15~$m_{\rm e}$~\cite{Madelung1996}.
In the case of AlAs, a compilation of experimental data from Ref.~\onlinecite{Nakwaski1995} shows that the electron transverse effective mass ranges between 0.124 (from optical absorption) and 0.19~$m_{\rm e}$ (from quantum-well spectroscopic measurements).
%

For AlSb, the room temperature transverse and longitudinal electron effective masses were measured to be 0.26 and 1.0~$m_{\rm e}$~\cite{Glinskii1979} using indirect exciton absorption measurements.
%
For SrO we could not find reliable measurements of the effective masses.

Overall, the agreement between DFT effective masses and experiment is reasonable (within a factor 2-3) but there are a few large discrepancies.
For the electron effective masses, the transverse mass of GaP and AlSb are overestimated with respect to the experimental values by 63~\% and 82~\%, respectively.
In the case of silicon, the computed electron effective masses are very close to the experimental values.
We note that our results are in line with previous calculations employing perturbation theory~\cite{LaflammeJanssen2016}.
Since DFT strongly underestimates the effective mass of GaAs, with values ranging from 0.03~$m_{\rm e}$~\cite{Kim2010} to 0.053~$m_{\rm e}$~\cite{Brunin2020}, we decided to use the experimental effective mass for the calculation of the electron mobility.
To this aim we employed a parabolic band approximation to evaluate analytically the eigenenergies and electron velocities from the experimental effective mass.
This semi-empirical procedure was used only for the electron mobility of GaAs.

In the case of hole effective masses, the largest discrepancies are found in the cases of c-BN (the mass is underestimated by almost a factor of three), GaAs, Si, and AlAs (all underestimating experimental values by 40-60\%).
The impact of these discrepancies on the calculated mobilities will be analyzed in Sec.~\ref{eq:exp_comp}.

\subsection{Impact of various approximations on the drift mobility}\label{sec:effects}

Having introduced all the required concepts and results, we are now able to discuss Fig.~\ref{fig:effects}.
For each of the approximations tested, we only change a single parameter, keeping everything else the same as in our most accurate calculations.
The reference calculations have been performed using converged coarse and fine grids, velocities including non-local contributions, dipole and quadrupole corrections, SOC, BTE, adaptive broadening, and experimental lattice parameters.
The following considerations also apply to the Hall factor.

We discuss the impact of the various approximations in order of importance.
We do not discuss the constant relaxation time approximation (CRTA) which depends on the empirical choice of the scattering rates and whose predictive power is unclear~\cite{Ganose2020}.

The most severe approximation concerns the long-range treatment of the dynamical matrix and the electron-phonon matrix element.
In Fig.~\ref{fig:effects} we break down the long-range treatment into four levels: (i) no long-range treatment; (ii) including Fr\"ohlich
 dipoles; (iii) including dipole and quadrupole corrections in the dynamical matrix; and (iv) including dipole and quadrupole corrections in the electron-phonon matrix elements.

Completely neglecting the long-range behavior has been recognized in 2015 to be a severe limitation~\cite{Verdi2015,Sjakste2015} but its impact on the mobility has not yet been quantified.
The consequence of neglecting long-range contributions during Fourier transformation of the matrix elements is that the Fr\"ohlich divergence when approaching the zone centre is spuriously suppressed.
As a result, the overall electron-phonon coupling, and therefore the scattering rates, will be strongly reduced.
This leads to a strong overestimation of the carrier mobility, up to 220\% in the case of the electron mobility of SrO.
Surprisingly, the hole mobility of 3C-SiC is only slightly underestimated (8\%) when ignoring long-range couplings.
This error cancellation is due to incorrect eigendisplacement vectors and will be discussed shortly.

Next in order of significance, with a standard deviation of 30\%, is the neglect of dynamical quadrupoles during the interpolation of the dynamical matrix and the electron-phonon matrix element.
Neglecting quadrupoles typically leads to an underestimation of the mobility by an average of -5.6\%, and up to -45\% in the case of 3C-SiC.
A notable exception is the case of c-BN, for which the neglect of quadrupole corrections leads to an overestimation of the mobility by 71\%.
This important effect has been discovered recently, and to our knowledge it has been taken into account only in a handful of works~\cite{Brunin2020,Brunin2020a,Jhalani2020,Park2020a}.
In the light of these developments it will be necessary to revisit prior mobility calculations.

In the case of silicon, neglecting quadrupoles for the LO and TO modes in the $\Gamma-L$ and $\Gamma-K$ directions
shown in Fig.~\ref{fig:gwithquad} produces spurious oscillations, leading to a 4\%/5\%  overestimation of the electron/hole mobility.
In the case of diamond, the neglect of quadrupole corrections yields an overestimation of less than 1\% of the mobilities.
Data for all compounds are collected in Fig.~\ref{fig:effects}, and their magnitude is in line with prior work~\cite{Agapito2018,Brunin2020}.

\begin{figure}[t]
  \centering
  \includegraphics[width=0.9\linewidth]{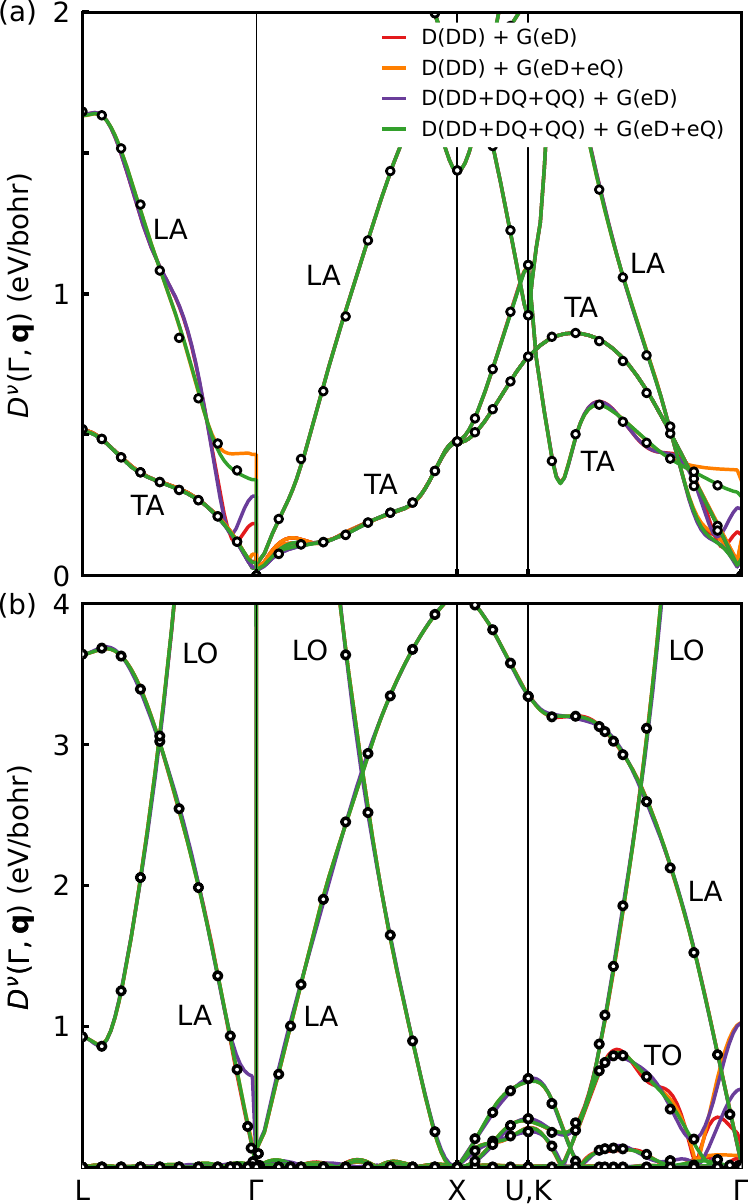}
  \caption{\label{fig:deformationBN}
(a) Valence band deformation potential corresponding to c-BN from 0 to 2 eV/bohr.
The label ``D" refers to the dynamical matrix, the label ``G" refers to electron-phonon matrix elements.
The labels ``DD", ``DQ" and ``QQ" indicate that dipole-dipole, dipole-quadrupole, or quadrupole-quadrupole contributions are included.
The labels ``eD" and ``eQ" indicate monopole-dipole and monopole-quadrupole terms.
The black circles are the reference DFPT calculations.
(b) Same as in (a), but for the $\Gamma_1$ valence band of 3C-SiC.
}
\end{figure}

The effect of quadrupole correction can be substantial in some compounds.
For example, as seen in Fig.~\ref{fig:effects}, the hole mobility of c-BN is strongly overestimated.
To better understand why this is the case, in Fig.~\ref{fig:deformationBN}(a) we compare the deformation potential obtained by including or neglecting the quadrupole term.
We can see that the high-energy modes are unaffected by quadrupoles in this case.
Conversely, the quadrupole correction becomes important for low-energy acoustic modes.
The red lines in Fig.~\ref{fig:deformationBN}(a) shows the result of the interpolation performed by including only the dipole long-range correction.
This result clearly deviates from the direct DFPT calculations, shown as the black dots close to the zone center.
In particular, the deformation potential corresponding to the LA mode in the $L-\Gamma$ and $K-\Gamma$ directions is strongly underestimated, leading to the observed large overestimation of the hole mobility of c-BN (in Fig.~\ref{fig:effects}).

Interestingly, we found that including quadrupole corrections to the electron-phonon matrix elements only was not sufficient to correctly describe the deformation potential, see the orange lines in Fig.~\ref{fig:deformationBN}.
Indeed, also in this case the deformation potential of the LA mode is overestimated, albeit the overestimation is less significant, and the hole mobility of c-BN is underestimated accordingly (23\%, see fourth column of Fig.~\ref{fig:effects}).
This result is counter-intuitive since the phonon frequencies are extremely well described by including dipole-only terms in the dynamical matrix, see Fig.~\ref{fig:wsphonons} for example (in certain classes of materials such as piezoelectrics this is often not the case~\cite{Royo2020}).
A careful investigation shows that the root cause which explains the difference between the orange and green curves in Fig.~\ref{fig:deformationBN} are the eigendisplacement vectors obtained by diagonalization of the dynamical matrix.
The eigendisplacement vectors $e_{\kappa\alpha,\mathbf{q}\nu}$ enter in Eq.~\eqref{eq:gkk} during the Fourier transformation of the electron-phonon matrix elements and within the long-range term in Eq.~\eqref{gldq}.
It is therefore crucial to include at least dipole-dipole, dipole-quadrupole and quadrupole-quadrupole in the dynamical matrix, see Eq.~\eqref{eq:longrange_dyn} as well as dipole-dipole and quadrupole-quadrupole effects in the electron-phonon matrix elements, see Eq.~\eqref{gldq} to achieve a reliable interpolation of the deformation potential.
We also emphasize that this correction must be applied both to the dynamical matrix and the scattering matrix elements: a partial correction of the dynamical matrix only is not sufficient, as shown by the purple line in Fig.~\ref{fig:deformationBN} and the third column of Fig.~\ref{fig:effects}.

Another interesting situation is found for 3C-SiC.
We show the deformation potential for the lowest energy band in Fig.~\ref{fig:deformationBN}(b).
Again we can see that the deformation potential close to $\Gamma$ along the $K-\Gamma$ direction has a spurious finite value at the zone center.
Including quadrupoles in the dynamical matrix and electron-phonon matrix elements fixes the problem.
Among all materials, we note that including quadrupole correction in the electron-phonon matrix element but not in the dynamical matrix, yields an average underestimation of the mobility by 5\%.

\begin{figure*}[t]
  \centering
  \includegraphics[width=0.99\linewidth]{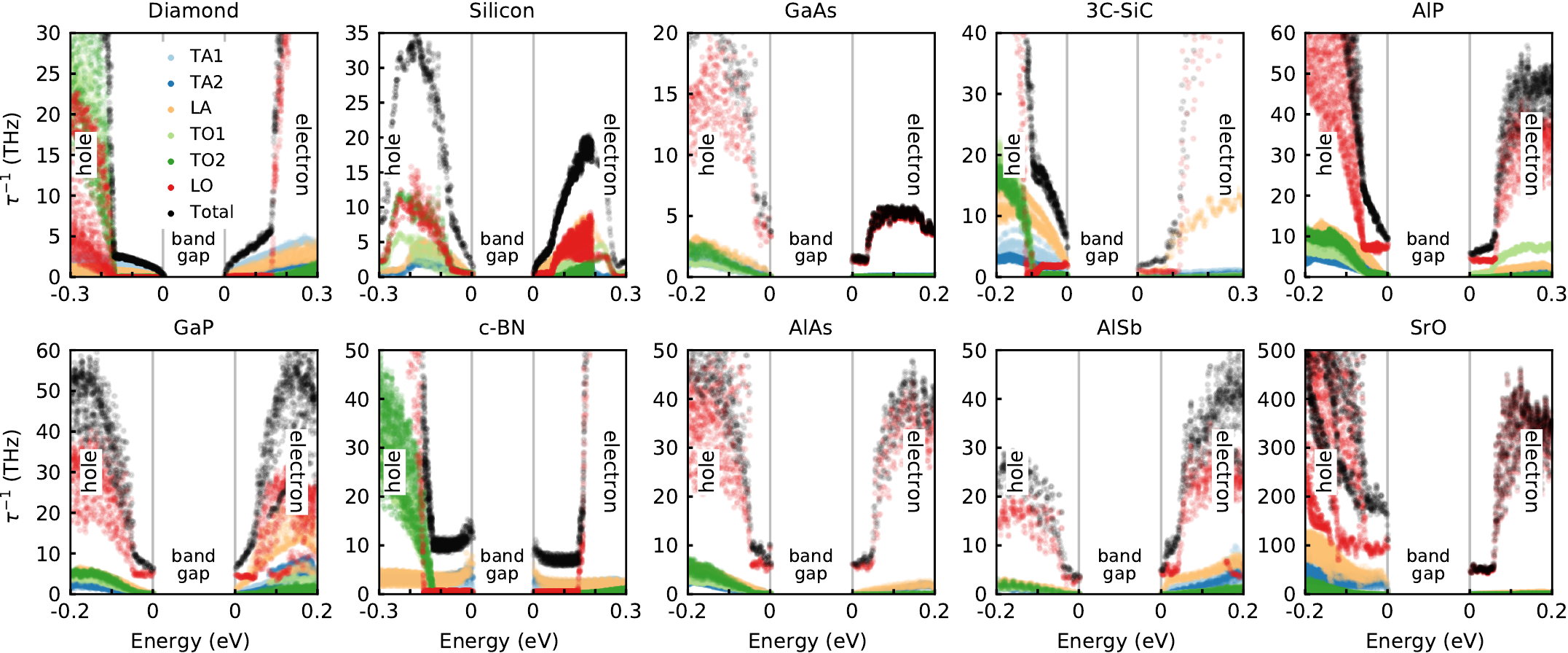}
  \caption{\label{fig:scatdecomposition}
Mode decomposition and total scattering rates at room temperature as a function of energy from the band edges.
The energy range covers the energy window used in the calculations.
  }
\end{figure*}

Beyond long-range dipole and quadrupole corrections, the most significant source of error in mobility calculations is the local velocity approximation, whereby the non-local part of the pseudopotential is neglected.
This approximation yields a large standard deviation of 28\%  with values ranging from -71\% in c-BN to +37\% for 3C-SiC.
We note that the local approximation was used by some of us for the calculation of the mobility of silicon~\cite{Ponce2018}.
We are now able to state that this approximation yields an overestimation of the mobility by +17\% and +7\% for the electron and hole mobility of silicon, respectively.
It is worth mentioning that the velocity including non-local contributions was implemented in the EPW software shortly after the publication of Ref.~\onlinecite{Ponce2018}, so that subsequent work is not impacted by this approximation.

The next source of error in order of importance is the neglect of SOC.
Without SOC the mobility is underestimated by up to 62\%.
For the compounds investigated here, this effect is almost negligible for electron mobility, while it is significant for hole mobility.
The strongest underestimation occurs in AlSb with a -62\% decrease in mobility.
These results are in line with earlier findings~\cite{Ponce2018,Ma2018}.
The reason for this underestimation is that SOC only affects valence bands in the semiconductors investigated here, because the valence band maximum is predominantly of $p$ character.
The effect of SOC is to lower the spin-split band, thereby reducing the phase space available for scattering, which increases the mobility.
For systems with $p$ states at the conduction band bottom we expect a similar effect, for example in halide perovskites~\cite{Ponce2019}.

Less significant than the above but still important is the error that one makes by using the popular SERTA method.
This approximation systematically underestimates the mobility within the range 1\% to 40\%.
At an intuitive level, the fact that the SERTA mobility underestimates the BTE mobility can be understood as follows: in the SERTA all scattering processes reduce the electron current in the same way, while in the BTE one takes into account the fact that forward scattering does not reduce the current as strongly as backward scattering.
Therefore the BTE is expected to yield a higher mobility.

\begin{figure*}[ht]
  \centering
  \includegraphics[width=0.99\linewidth]{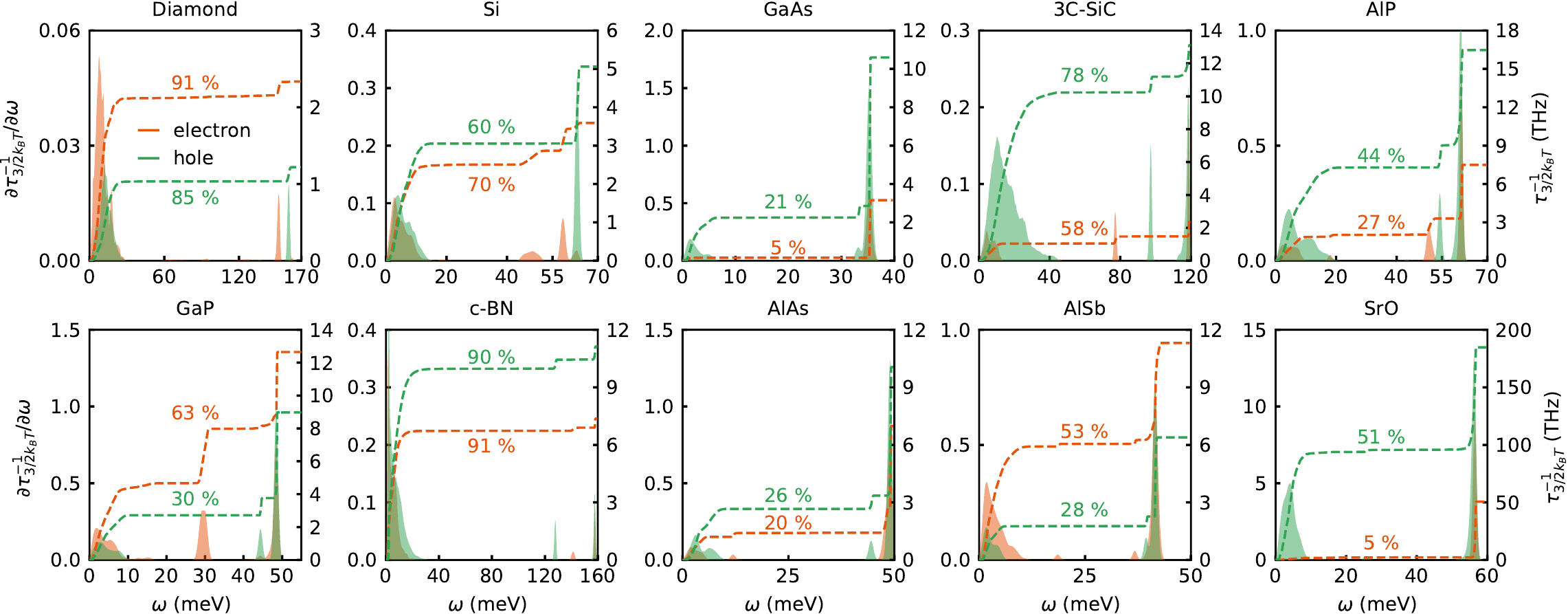}
  \caption{\label{fig:spectraldecomposition}
Spectral decomposition of the electron (orange) and hole (green) scattering rates as a function of phonon energy at 300~K.
The rates are calculated as angular averages for carriers at an energy of $3k_{\rm B}T/2$ = 39~meV away from the band edges.
The dashed lines represent the cumulative integrals of the calculated rates, and add up to the carrier scattering rate $\tau_{3/2 k_{\rm B} T}^{-1}$.
The percentage indicates the contributions from acoustic modes.
  }
\end{figure*}

The last three effects shown in Fig.~\ref{fig:effects} have a smaller impact on mobility, with errors up to 20\%.
We go over these effects in order of appearance in the figure.
First, we investigated the sensitivity of the mobility to the lattice parameter.
To this aim we re-calculated all mobilities using the relaxed lattice parameters.
As discussed in Sec.~\ref{sec:DFT}, the PBE lattice parameter is overestimated with respect to experiment by an average of +1.3\%.
This effect yields an average decrease of mobility by -6.5\%.
We then considered the effect of the exchange-correlation functional by using LDA instead of PBE but keeping all other parameters the same.
In this case we find only small changes, with an average of -1.5\%.
Lastly, we explored the effect of pseudization parameters.
To this aim we repeated the calculations using the SG15 ONCV library~\cite{Schlipf2015} that was extended to fully relativistic potentials~\cite{Scherpelz2016} instead of the PseudoDojo library~\cite{Setten2018}.
In this case we find an average deviation of the order of 2\%.
The small impact of these effects on the calculated mobilities highlights the robustness of our approach.

We note that many other effects can impact the calculation of carrier mobilities, including electron-two-phonon interactions~\cite{Lee2020}, thermal lattice expansion~\cite{Ponce2018}, and band structure renormalization due to electron-phonon coupling~\cite{Giustino2010,Ponce2015}.
We expect these effects to be small in most materials, but it will be important in the future to perform a detailed analysis.
The effect of quasiparticle corrections~\cite{Ponce2018,Ponce2019c} and many-body renormalization of the electron-phonon matrix elements~\cite{Antonius2014,Li2019a} will be discussed in Sec.~\ref{eq:exp_comp} in relation with experimental data.

\begin{figure}[t]
  \centering
  \includegraphics[width=0.99\linewidth]{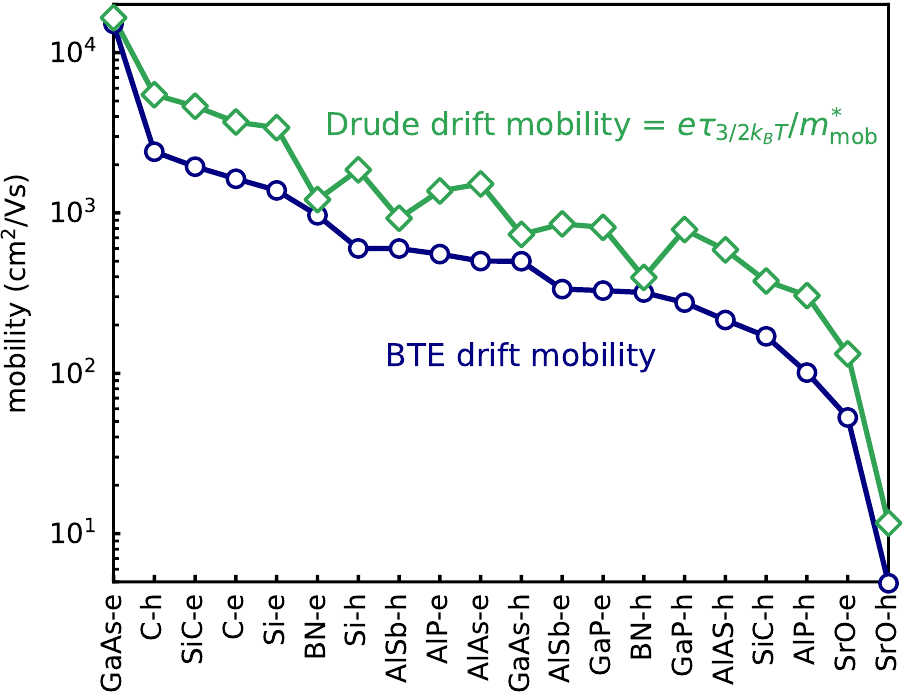}
  \caption{\label{fig:mobilitycomparison}
Comparison  between the room temperature first-principles BTE drift mobility (blue) and the mobility obtained with the Drude formula from Eq.~\eqref{eq:Drudeformula} (green).
The scattering lifetimes $\tau_{3/2 k_{\rm B} T}$ are calculated as angular averages for carriers at an energy of $3k_{\rm B}T/2$ = 39 meV away from the band edge.
The materials are arranged in decreasing order of their BTE drift mobility value.
  }
\end{figure}

\subsection{Dominant scattering mechanisms}

To gain a better understanding of the microscopic phenomena responsible for limiting the intrinsic transport properties of the compounds investigated here, we show in the Appendix Fig.~\ref{fig:scattemp} the electron
and hole scattering rates $\tau_{n\mathbf{k}}$ computed via Eq.~\eqref{eq:scattering_rate} at 150~K, 300~K, and 500~K.
Without surprises, the higher the temperature, the higher the scattering rates and the lower the mobility.

As expected, in most cases we find a mild increase of the scattering rates as we move away from the band edges, which corresponds to acoustic-phonon scattering.
At the onset for the emission or absorption of optical phonons, the rates show a marked increase.
For example in diamond this sharp increase occurs at around 160~meV, when optical phonon emission/absorption processes become allowed.
We note a couple of exceptions to this general trend: the scattering rates of electrons in GaAs and SrO, and both the electron and hole scattering rates of c-BN first decrease when moving away from the band edges which is attributed
to LO phonon absorption and was investigated in Ref.~\onlinecite{Liu2017}.

In Fig.~\ref{fig:scatdecomposition} we present the mode decomposition of the room-temperature scattering rates, as well as the total scattering rates.
Starting from the holes:
For diamond, the onset of scattering is mostly related to transverse acoustic mode scattering, while for silicon the longitudinal acoustic mode also plays an important role.
In the case of GaAs, AlP, GaP, AlAs, AlSb, and SrO, the hole scattering near the band edges is dominated by  the longitudinal optical mode.
For the electrons:
The scattering of electrons near the band edge in GaAs, AlP, GaP, AlAs, AlSb, and SrO is dominated by
longitudinal optical modes, while diamond, silicon, 3C-SiC and c-BN show significant contributions from LA and TA modes.

In 1966, Berman, Lax and Loudon~\cite{Birman1966} have derived intervalley-scattering selection rules for III-V semiconductors.
For example, in the case of electrons in GaP, $X$-$X$ intervalley scattering is expected.
The rule is that, if the mass of the group-V element is heavier than that of the group-III element, then LO scattering should be allowed by selection rules, otherwise the LA scattering would be allowed.
Therefore, for GaP one would expect LA-phonon scattering, while for AlP, AlSb, and AlAs one would expect LO-phonon scattering to dominate the intervalley scattering.
Looking at Fig.~\ref{fig:scatdecomposition} confirms the dominance of LO scattering in AlP, AlSb, and AlAs.
For the case of GaP, it seems the scattering at an energy $3/2 k_{\rm B} T$ away from the band edge has about equal contribution from LA and LO phonon modes.

A complementary viewpoint on the scattering rates is obtained by considering only carriers at a given energy, and resolving the rates by phonon energy.
This analysis is shown in Fig.~\ref{fig:spectraldecomposition}.
The calculations correspond to carriers at an energy of $3k_{\rm B}T/2$ away from the band edge and room temperature, following previous work~\cite{Ponce2019}.
From this analysis, we can see that the scattering rate is overwhelmingly dominated by acoustic scattering in diamond and c-BN.
In contrast, optical scattering dominates for electrons in GaAs, SrO, AlAs, AlP, and for holes in GaAs, AlAs, AlSb, GaP, and AlP.

We also mention a few counter examples to the common wisdom whereby if a material is dominated by deformation potential acoustic scattering, then the temperature exponent of the mobility is $(T^{-3/2})$~\cite{Li2006}.
Diamond, silicon, c-BN and 3C-SiC are all dominated by acoustic-phonon scattering, but the temperature exponents evaluated on the 150~K-500~K range are -1.81, -2.05, -1.33, and -2.86 for the electrons and -2.06, 2.93, -1.51, and -1.97 for the holes, respectively.
These data suggest that one should exert some caution when applying simplified models based on the electron gas to real materials.

For completeness we also examine the predictive power of the classic Drude formula [Eq.~\eqref{eq:Drudeformula}].
To this aim we compare our BTE calculations for all the compounds considered in this work with the Drude prediction.
In the Drude formula we employ the mobility effective masses reported in Table~\ref{table:bandstructures} and the total
scattering rates evaluated as in Fig.~\ref{fig:spectraldecomposition}.
The resulting data are shown in Fig.~\ref{fig:mobilitycomparison}.
The Drude formula reproduces the trend of \textit{ab initio} BTE calculations very closely, although it tends to overestimate mobilities by a factor of 1.1 to 3.1.
This favourable comparison implies that the materials descriptors that we introduced, namely the mobility effective mass and the scattering rates of carriers at $3k_{\rm B}T/2$, provide a meaningful description of carrier transport in the systems considered here.

\subsection{Temperature dependence of the Hall factor}\label{sec:hallfact}

\begin{figure*}[t]
  \centering
  \includegraphics[width=0.99\linewidth]{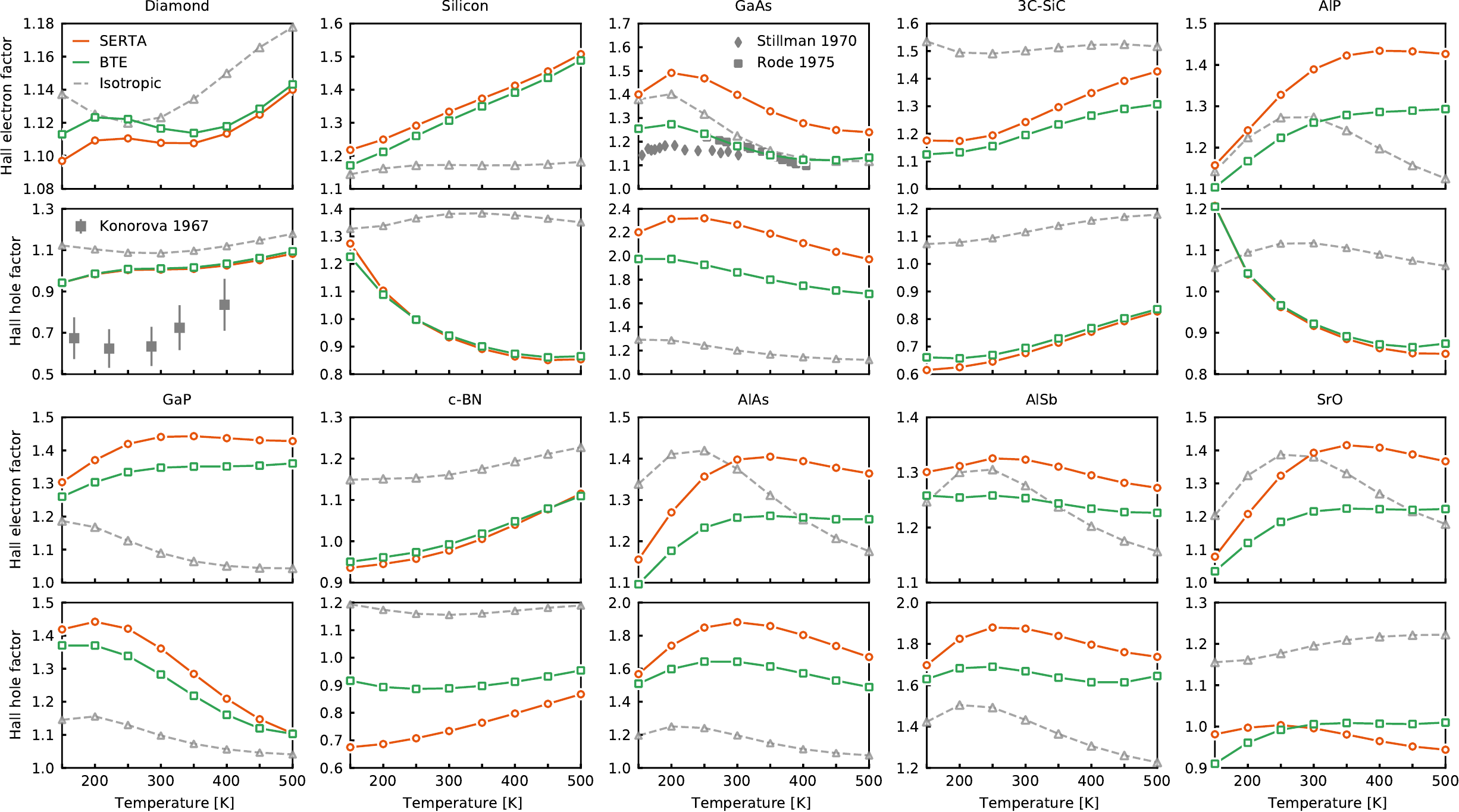}
  \caption{\label{fig:hallfigfact}
Temperature dependence of the calculated Hall factor.
We show results within the SERTA (orange) and BTE (green).
The calculations have been extrapolated to infinitely dense Brillouin zone grids.
The gray dashed lines correspond to the isotropic approximation introduced in Eq.~\eqref{eq65}.
Experimental data for diamond are taken from Konorova and Shevchenko~\cite{Konorova1967}.
Experimental data for GaAs are taken from Rode~\cite{Rode1975a} and Stillman \textit{et al.}~\cite{Stillman1970}, for  a low magnetic field of 0.5~kG.
  }
\end{figure*}

In this section we investigate the temperature dependence of the Hall factor.
We use Eqs.~\eqref{eq:iterwithbimpl}-\eqref{eq:hallfactor} to compute the Hall factors within both the SERTA and full BTE.
For each temperature we extrapolate the fine grids as discussed in Sec.~\ref{sec:fine}, and we present the results in Fig.~\ref{fig:hallfigfact}.
We find that the Hall factor can vary substantially with temperature, ranging from 0.7 to 1.9 at room temperature.
These results suggest that caution should be used when comparing \textit{ab initio} calculations of carrier mobilities to experiments, since most reported experimental data refer to Hall mobilities, not drift mobilities.
As a technical note, we mention that below 150~K we could not converge the Brillouin zone grids due to excessive computational cost.
Accordingly these data are not shown in Fig.~\ref{fig:hallfigfact}.

Our calculations indicate that the Hall factor exhibits rather complex temperature trends, which reflects the variety of phonon energy scales involved.

Comparison with experimental data for the Hall factor is challenging given the scarcity of available data.
We were only able to find experimental values for the hole Hall factor of diamond and for the electron Hall factor of GaAs, which we report in Fig.~\ref{fig:hallfigfact}.
In both cases we find reasonable agreement between our calculations and experiments, with the largest deviation being of the order of 30\% (note the magnified scales in Fig.~\ref{fig:hallfigfact}).

In a recent preprint~\cite{Desai2021}, the authors calculated the electron Hall factor of silicon to be 1.15 at room temperature.
These results are consistent with our Fig.~\ref{fig:fineconvergence} for the same fine grid where we have an electron Hall factor of 1.12.
The authors of Ref.~\onlinecite{Desai2021} also computed the electron Hall factor of GaAs, finding a BTE value ranging from 1.22 to 1.14 in the 250-400~K temperature range.
Their results compare well with our calculations, yielding a Hall factor of 1.28 to 1.13 in the same temperature range.

In Fig.~\ref{fig:hallfigfact} we also show the Hall factor as computed using the standard isotropic approximation given by Eqs.~\eqref{eq65}-\eqref{eq66} (gray dashed lines).
For these calculations we evaluated the Dirac deltas using a Gaussian of width 1~meV.
This approximation reproduces the BTE results quite well for diamond, GaAs, AlP(e), GaP(h), c-BN(e), AlAs, AlSb, and SrO.
However, it fails even qualitatively in the case of silicon, AlP(h), and GaP(e).
One important limitation of the isotropic approximation is that it cannot lead to a Hall factor smaller than unity.

\begin{figure*}[ht]
  \centering
  \includegraphics[width=0.99\linewidth]{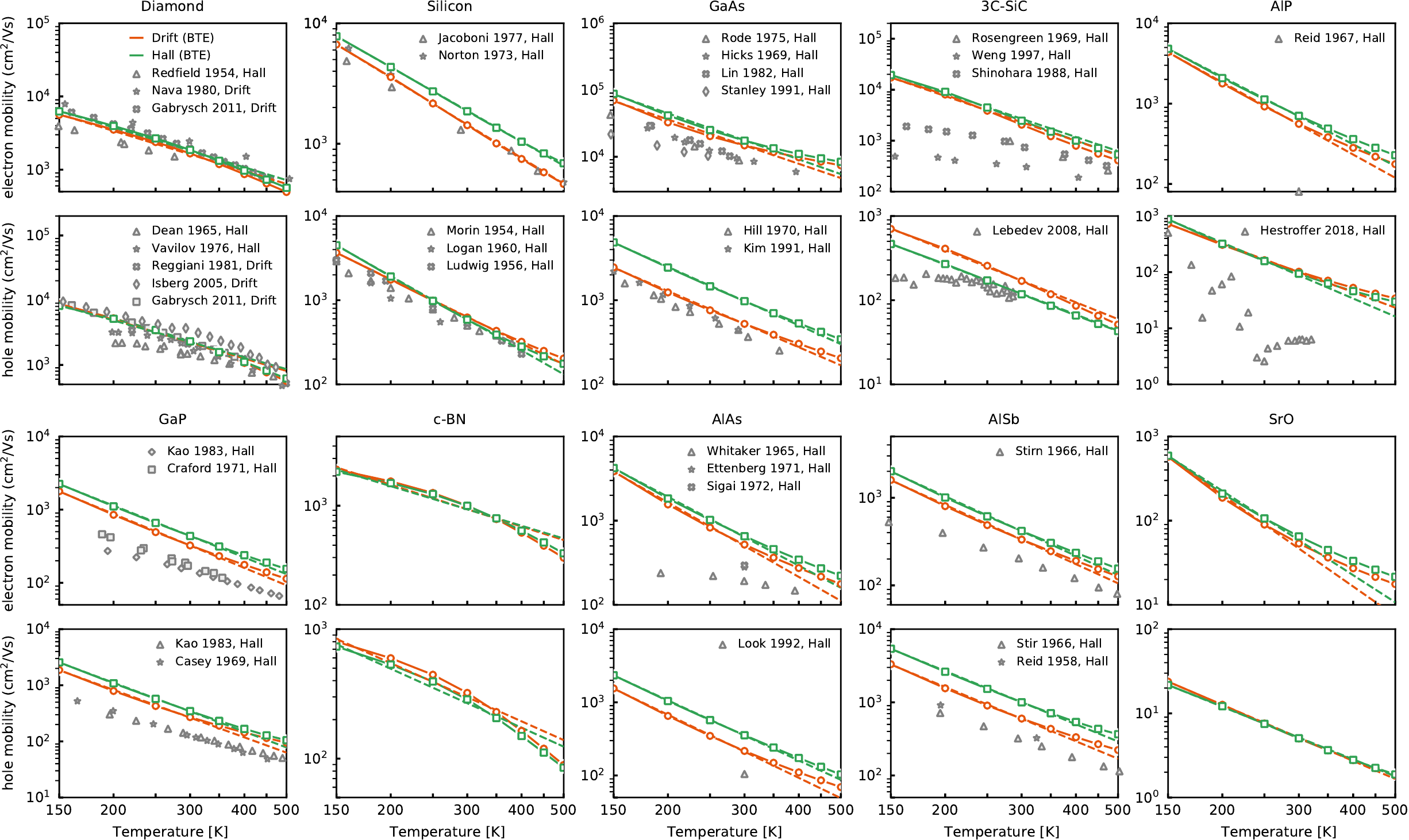}
  \caption{\label{fig:experimentaltempmob}
Calculated temperature dependence of mobilities, and comparison to experimental data.
The calculated drift mobilities are in red, the Hall mobilities are in green, and the experimental datapoints are in gray.
The dashed lines indicate the power law fits with exponents reported in Table~\ref{table:mobility_val} and obtained
on the 150~K to 500~K range.
The references for the experimental data are as follows:
diamond~\cite{Redfield1954, Dean1965, Vavilov1976, Nava1980,
Isberg2005, Gabrysch2011},
silicon~\cite{Morin1954,Ludwig1956,Logan1960,Norton1973,Jacoboni1977},
GaAs~\cite{Hicks1969,Hill1970,Rode1975a,Lin1982,Stanley1991,Kim1991},
3C-SiC~\cite{Rosegreen1969,Shinohara1988,Weng1997,Lebedev2008},
AlP~\cite{Reid1967,Hestroffer2018},
GaP~\cite{Casey1969,Craford1971,Kao1983},
AlAs~\cite{Whitaker1965,Ettenberg1971,Sigai1972,Look1992}, and
AlSb~\cite{Reid1958, Stirn1966a, Stirn1966}.
  }
\end{figure*}

\subsection{Temperature dependence of the drift and Hall mobility and comparison to experiment}\label{eq:exp_comp}

In this section we examine the accuracy of the various methods presented here in reproducing experimental carrier mobility.
In Fig.~\ref{fig:experimentaltempmob} we show the dependence of the BTE drift and Hall mobility on temperature, and compare our calculations to experimental data whenever available.
The calculations are presented in a log-log scale.
We indicate with dashed lines the linear fits used to determine the temperature exponents in Table~\ref{table:mobility_val} of the Appendix.
Our results span a large range of power law decays, from $T^{-1.18}$ for c-BN to $T^{-3.15}$ for AlP.
In the temperature range considered here (150-500~K), we find that a single exponent captures fairly well the calculated trends for all materials.
Some deviations are found to occur above 400~K in the case of GaAs, AlP, c-BN, AlAs, and  SrO.
These deviations are indicative of high-frequency optical phonons becoming increasingly important in carrier scattering at these temperatures.
We note that this fitting procedure is done in order to extract a single exponent which can be compared to experimental data fitted with the same procedure and in the same temperature range, whenever available.

\begin{figure}[ht]
  \centering
  \includegraphics[width=0.95\linewidth]{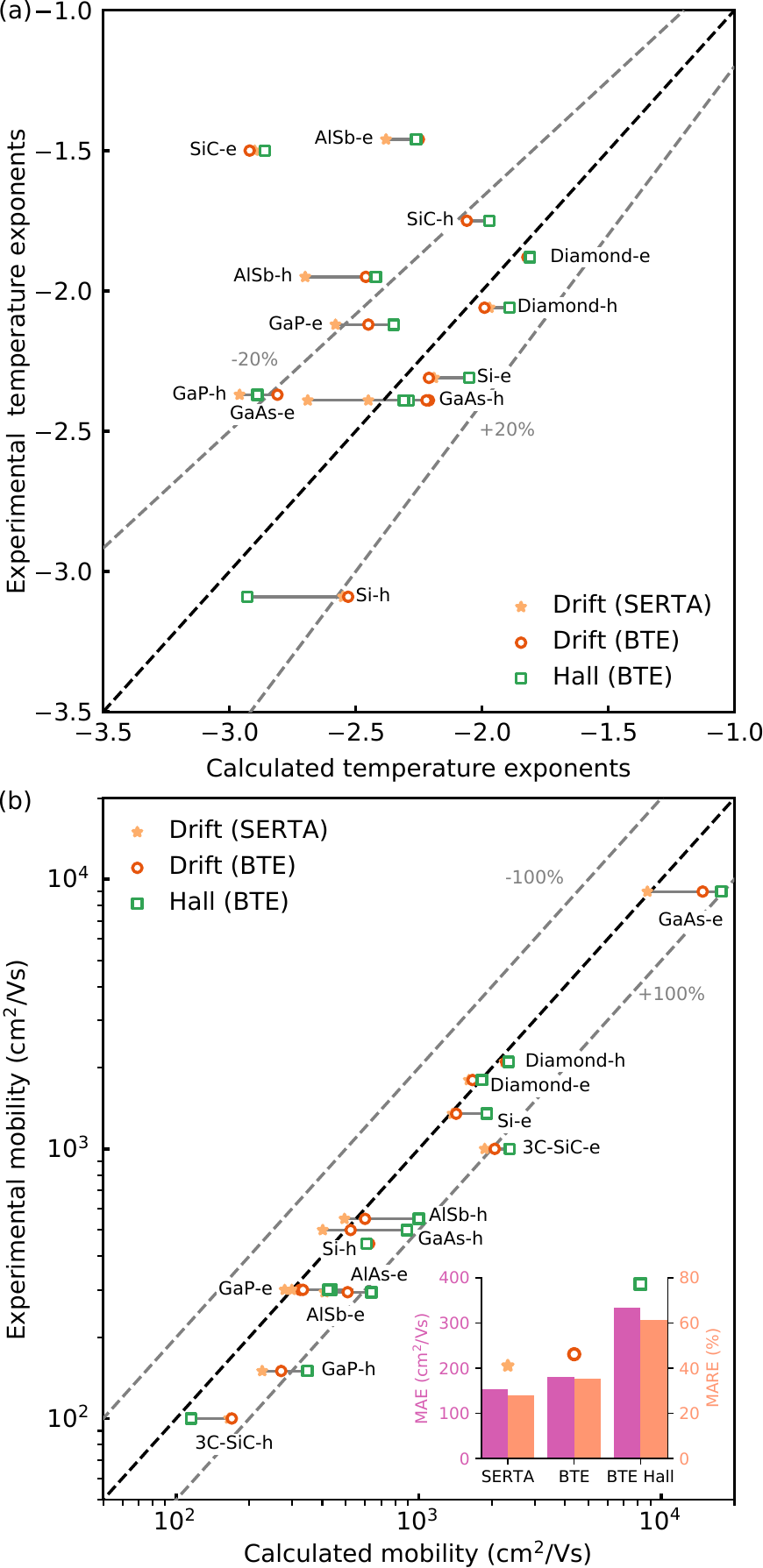}
  \caption{\label{fig:expmob}
Comparison between theoretical and experimental (a) temperature exponents and (b) mobilities.
The values are reported in Table~\ref{fig:experimentaltempmob}.
The mean absolute error (MAE) and mean absolute relative error (MARE) are indicated as inset in the bottom figure and do
not include the electron mobility of GaAs.
  }
\end{figure}

In Fig.~\ref{fig:expmob}(a) we compare the calculated and measured temperature exponents.
We find that in general the measured exponents are well reproduced by the calculations, with an error of less than 20\%.
The only exceptions are found for SiC and AlSb, for which very few experimental studies exist.
Given the overall good agreement found for the other compounds, we expect that the agreement between theory and experiment for SiC and AlSb will improve as new experimental data will be produced in the future.

In Fig.~\ref{fig:expmob}(b) we show the comparison between calculated and measured mobilities, in a logarithmic scale.
In all cases the experimental data lay below the predicted phonon-limited Hall mobility.
The difference can be as severe as 100\% in some cases.
The calculated drift mobility is found to be closer to the experimental data, but we have to emphasize that most experimental data reported in this figure are for the Hall mobility, as indicated in Table~\ref{table:mobility_val}.
The SERTA drift mobility appears to be closer to experiments, and in some cases below the experimental data (for example in diamond, AlSb and GaAs).
This behavior is probably due to a cancellation of errors, and should not be taken as an endorsement of the SERTA method.
Whenever possible we recommend to perform complete BTE calculations.
We note that, in calculations of phonon-limited carrier mobilities, it is not meaningful to validate computational methods by direct comparison with experiments, since experimental data incorporate additional scattering mechanisms.

\subsection{Corrections to band structures and matrix elements from experimental data}\label{sec:mb}

In this section we investigate the role of many-body renormalization of the band structures and electron-phonon matrix elements on the mobility.
Since it is not currently possible to perform complete many-body calculations of the BTE, we estimate these effects by using the experimental effective masses and by rescaling the electron-phonon matrix elements via the experimental dielectric permittivities.

To this aim, starting from the experimental masses reported in Table~\ref{table:bandstructures}, for the electrons we determine the conductivity effective mass defined as the harmonic mean of the transverse and longitudinal effective masses:
\begin{equation}
\frac{3}{m_{\rm cond}^{*,\rm el}} = \frac{1}{m_{\parallel}^*} + \frac{2}{m_{\perp}^*}.
\end{equation}
For the holes we average the band contribution with the following weighted average formula which assumes band parabolicity~\cite{Ricci2017}:
\begin{equation}
m_{\rm cond}^{*,\rm h} = \frac{(m_{\rm hh}^*)^{5/2} + (m_{\rm lh}^*)^{5/2}}{(m_{\rm hh}^*)^{3/2} + (m_{\rm lh}^*)^{3/2}}.
\end{equation}
Using these experimentally-derived quantities, we rescale our calculated electron and hole Hall mobilities using:
\begin{equation}
\mu^{\rm mass} = \mu (m_{\rm cond}^{*,\rm DFT}/m_{\rm cond}^{*,\rm exp}).
\end{equation}
The relative change in mobility obtained by this procedure is reported in Table~\ref{table:massandg}.
In all materials considered here this rescaling decreases the hole mobility, with the exception of SiC and BN.
For the electron mobility, in all cases but diamond the mobility increases upon rescaling.

\begin{table}[t]
  \begin{tabular}{r r r r r}
  \toprule
         & Hall mobility & Effective mass & Screening &     Both  \\
         &    cm$^2$/Vs  &    rescaling   & rescaling & rescaling \\
\hline
\\[-1.0em]
C-e      &  1705 & -16\% &  -9\% & -24\% \\
C-h      &  2467 & -28\% &  -9\% & -35\% \\
Si-e     &  1770 &  +2\% & -14\% & -13\% \\
Si-h     &   574 & -47\% & -14\% & -54\% \\
GaAs-e   & 17860 &     - & -42\% & -42\% \\
GaAs-h   &  1068 & -32\% & -42\% & -63\% \\
SiC-e    &  2257 &  +3\% & -11\% &  -8\% \\
SiC-h    &   111 & +17\% & -11\% &  +4\% \\
AlP-e    &   698 &     - & -14\% & -14\% \\
AlP-h    &    92 &     - & -14\% & -14\% \\
GaP-e    &   442 & +56\% & -22\% & +21\% \\
GaP-h    &   353 & -32\% & -22\% & -47\% \\
BN-e     &   956 & +11\% &  -3\% &  +7\% \\
BN-h     &   281 & +57\% &  -3\% & +52\% \\
AlAs-e   &   640 & +20\% & -22\% & +16\% \\
AlAs-h   &   357 & -45\% & -22\% & -47\% \\
AlSb-e   &   431 & +76\% & -10\% & +59\% \\
AlSb-h   &  1000 & -38\% & -10\% & -44\% \\
SrO-e    &    65 &     - & -14\% & -14\% \\
SrO-h    &     5 &     - & -14\% & -14\% \\
  \botrule
  \end{tabular}
  \caption{\label{table:massandg} Effect of rescaling the Hall mobilities obtained via the BTE using experimental effective masses and dielectric constants.
}
\end{table}

\begin{figure}[ht]
  \centering
  \includegraphics[width=0.98\linewidth]{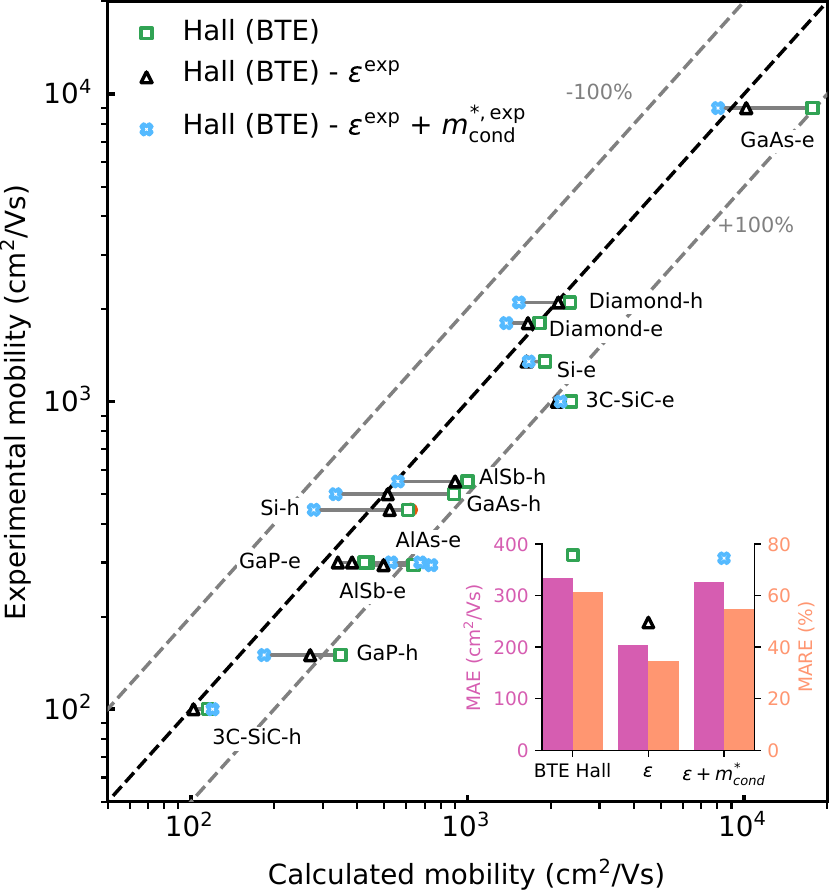}
  \caption{\label{fig:expmobscale}
Comparison between theoretical and experimental mobility with experiment.
Black triangles indicate the impact of rescaling the matrix elements while blue crosses indicate additional effective masses rescaling.
The mean absolute error (MAE) and mean absolute relative error (MARE) are indicated as inset in the bottom figure and do
not include the electron mobility of GaAs.
  }
\end{figure}

Along the same line, we also proceed to renormalize the electron-phonon matrix elements by using the experimental dielectric screening.
This strategy allows us to capture effects similar to the many-body renormalization of the electron-phonon coupling~\cite{Antonius2014, Li2019b}.
Since the mobility contains the square moduli of the matrix elements, we rescale the mobility using:
\begin{equation}
\mu^{\rm screen} = \mu (\epsilon^{\rm DFT}/\epsilon^{\rm exp})^2.
\end{equation}
In this operation we employ the dielectric constants reported in Table~\ref{table:bandstructures}.

In Fig.~\ref{fig:expmobscale} we show the impact of rescaling the matrix elements (black triangles) on the mobility together with
the rescaling of the matrix elements and effective masses (blue crosses).
For all cases we also report the mean absolute error (MAE) and the mean absolute relative error (MARE) between the computed and experimental mobilities.

Our best first-principles BTE Hall mobilities have a large absolute error around 1000~cm$^2$/Vs due to the strong overestimation of the high electron mobility of GaAs and with a MARE of 64~\%.
If we remove GaAs from the dataset, which is the compound with the highest mobility and hence it weighs disproportionately in the calculation of the average, we obtain a maximum absolute error of 335~cm$^2$/Vs, and a MARE of 62\%.
When we proceed to renormalize the electron-phonon matrix elements using the experimental dielectric constant, the results improve drastically with a MAE below 205~cm$^2$/Vs and MARE of 35~\%.
However, this improvement is partially cancelled when we also use the experimental effective masses, leading to a MARE of 55~\%.

These results indicate that improving the accuracy of the band structures and electron-phonon matrix elements, for example by using hybrid functionals~\cite{Yin2013} or many-body perturbation theory~\cite{Antonius2014, Li2019b} will be important to achieve agreement with experiment.
Whether these improvements will lead to close agreement with experiment, or else a more powerful theory beyond the BTE formalism will be needed in the future, remains an important open question.

Finally, we note that regardless of the level of approximation or re-scaling applied, the electron mobility of 3C-SiC and AlAs, and the hole mobility of GaP are significantly higher than experiment.
This suggests that higher mobilities could be achieved in these materials, paving the way for improved electronic devices.

\section{Conclusion}\label{sec:conclusion}

In this work we have carried out an extensive investigation of the carrier transport properties of 10 simple semiconductors including diamond, silicon, GaAs, 3C-SiC, AlP, GaP, c-BN, AlAs, AlSb, and SrO.
Exploiting a precise interpolation approach and an efficient procedure for numerical convergence, we have been able to compute from first-principles transport coefficients with high accuracy.
This key step has allowed us to directly quantify the impact of a wide range of approximations often used in transport calculations.
In addition, we have been able to assess the role of a number of physical factors such as the contribution of long-range dipole and quadrupole interactions, the SOC, effective masses and lattice parameters as well as the effect of temperature and magnetic field.
One of the most interesting result of this extensive study is that our most accurate Hall mobility calculations seem to systematically overestimate the experimental data.
In a few cases the overestimation is as large as a factor of two.
This is a quite promising outcome as in our calculation we only included electron-phonon scattering and our result should be regarded as an upper limit to the Hall mobility achievable in highly pure samples, suggesting the possibility of further improvement in the transport properties of semiconductors of technological relevance.
Here it is also important to stress that our work has clearly shown that the Hall factor exhibits significant variations across materials and as a function of temperature, ranging between 0.7 and 2.
This is a clear indication that the common practice of comparing computed drift mobilities with experimental Hall mobilities should be done with particular care.
Finally, it is important to stress that in our analysis we have also showed that part of the difference between theory and experiment may be connected with the inaccuracy in the DFT effective masses and electron-phonon matrix elements, but further work using explicit many-body perturbation theory calculations will be needed to confirm this point.

Our work demonstrates that first-principles calculations of carrier transport are making considerable progress.
We believe we are reaching a point where detailed comparison with high-quality experimental data on pure samples will be increasingly important.
For example, we highlight the need for new accurate measurements of the carrier mobility of SiC, AlAs, and GaP, and we hope to reignite interest in the experimental community to generate modern transport datasets for theorists to compare with.
This manuscript describes a comprehensive protocol to perform such delicate calculations, and as a blueprint for reporting first-principles transport data in a way that is accessible, reliable, and reproducible.
In this spirit, we encourage the adoption of the FAIR~\cite{Wilkinson2016} data and software principles in the study first-principles transport.

\begin{acknowledgments}
The authors would like to thank
C. Verdi,
X. Gonze, 
Junfeng Qiao, 
Hyungjun Lee, 
Massimiliano Stengel, 
Miquel Royo, 
for useful discussions.
Computer time was provided by the PRACE-17 and PRACE-21 resources MareNostrum at BSC-CNS, and the Texas Advanced Computing Center (TACC) at the University of Texas at Austin.
S.P. acknowledges support from the European Unions Horizon 2020 Research and Innovation Programme, under the Marie Sk\l{}odowska-Curie Grant Agreement SELPH2D No.~839217.
F.M. and N.B. acknowledge the Cirrus UK National Tier-2 HPC Service at EPCC.
E.R.M. acknowledges support from the National Science Foundation (Award No. OAC-1740263).
N.M. acknowledges support from the Swiss National Science Foundation and the NCCR MARVEL.
F.G.'s contribution to this work was supported as part of the Computational Materials Sciences Program funded by the U.S. Department of Energy, Office of Science, Basic Energy Sciences, under Award DE-SC0020129.
\end{acknowledgments}

\appendix

\section{Optimal Wigner-Seitz construction}\label{app.0}

The interpolation of the Hamiltonian, dynamical matrix, and electron-phonon matrix elements from real space to reciprocal space via Eqs.~\eqref{eq:hamiltonian},\eqref{eq:dynamical} and \eqref{eq:gkk} is usually performed by using a radial cutoff on the direct lattice vectors~\cite{Mostofi2014,Ponce2016a}.
While this is certainly the simplest possible choice, it is not the most efficient, especially in the case of unit cells with high aspect ratios.

In order to improve the accuracy of the interpolation, it is convenient to select the direct lattice vectors so that the distance between two Wannier centers or the distance between two atoms are within a cutoff radius.
For example, in the interpolation of the Hamiltonian we construct a \textit{shifted} Wigner-Seitz cell of lattice vectors $\mathbf{R}_{p}$ with the requirement that this cell be centered on one of the two Wannier functions, say $\mathbf{r}_{n}$.
The radial cutoff is then imposed with respect to the distance $|\mathbf{R}_{p} + \mathbf{r}_{m} - \mathbf{r}_{n}|$, where $\mathbf{r}_{m} $ denotes the center of the second Wannier function.

In practice, we determine the set of Wigner-Seitz vectors to describes the electronic Hamiltonian:
\begin{equation}
H_{mn}(\mathbf{R}_{p}) = (1/N_{p})\!\!\!\!\!\!\!\!\!\!\! \sum_{\mathbf{k} \in (i_{p}
\times j_{p} \times o_{p})} \!\!\!\!\!\!\!\!\!\! U_{mm'\mathbf{k}}^\dagger
H_{m'n'\mathbf{k}} U_{n'n\mathbf{k}} e^{-i\mathbf{k}\cdot \mathbf{R}_{p}},
\end{equation}
 such that $\mathbf{R}_{p} + \mathbf{r}_{m} - \mathbf{r}_{n} \in (i_{p} \times j_{p} \times o_{p})$ where $\mathbf{r}_{m}$ is the origin vector pointing to the position of the $m$th Wannier center.
A similar approach was recently implemented in the Wannier90 software for the interpolation of the Hamiltonian~\cite{Pizzi2020}.

Similarly, in the case of the dynamical matrix we consider pairs of atoms $\boldsymbol{\tau}_{\kappa}$ and $\boldsymbol{\tau}_{\kappa'}$.
For each such pair, we construct a Wigner-Seitz cell centered at $\boldsymbol{\tau}_{\kappa}$, and we impose a truncation based on the distance $|\mathbf{R}_{p'} + \boldsymbol{\tau}_{\kappa'} - \boldsymbol{\tau}_{\kappa}|$.
For the optimal choice of a Wigner-Seitz vector for a discretized Brillouin zone grid $i_{p'} \times j_{p'} \times o_{p'}$ , the short-range part of the dynamical matrix takes the form~\cite{Gonze1997a}:
\begin{multline}\label{eq:Dwstransform}
D_{\kappa\alpha,\mu}^{\mathcal{S}}(\mathbf{R}_{p'})= \\
 (1/N_{p'}) \!\!\!\!\!\!\!\!\!\!\! \sum_{\mathbf{q} \in (i_{p'} \times j_{p'} \times
o_{p'})} \!\!\!\!\!\!\!\!\!\!\!  e^{-i\mathbf{q}\cdot \mathbf{R}_{p'}}
e_{\kappa\alpha,\mathbf{q}\nu}^\dagger  D_{\nu,\kappa'\beta}^{\mathcal{S}}(\mathbf{q})
e_{\kappa'\beta,\mathbf{q}\mu}  ,
\end{multline}
for all direct lattice vectors $\mathbf{R}_{p'}$ such that $\mathbf{R}_{p'} + \boldsymbol{\tau}_{\kappa'} - \boldsymbol{\tau}_{\kappa} \in (i_{p'} \times j_{p'} \times o_{p'})$ and 0 otherwise.
In Eq.~\eqref{eq:Dwstransform} we have made a rigid shift to the central primitive cell such that it only depends on one lattice vector $\mathbf{R_{p'}}$.
As a result one needs to store a set of atom-dependent Wigner-Seitz cells.
In first-principles software such as Quantum ESPRESSO~\cite{Giannozzi2017}, a Fourier-transform grid is constructed with weights centered on pairs of atoms and zeros elsewhere.
In EPW, we slightly optimize the computational cost of the interpolation by only retaining the union of nonzero elements between all the Wigner-Seitz cells.

\begin{figure}[t]
  \centering
  \includegraphics[width=0.96\linewidth]{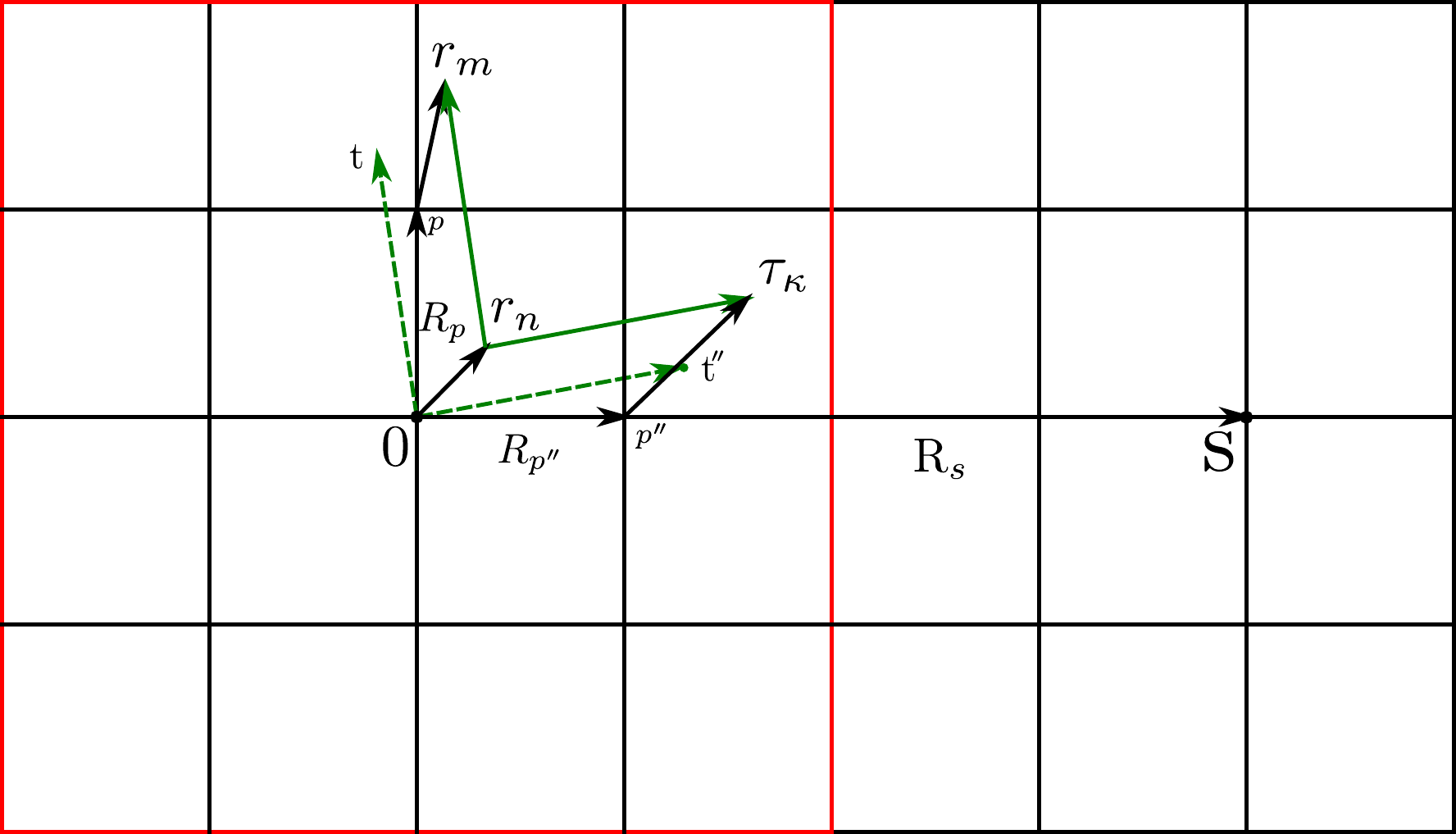}
  \caption{\label{fig:optimizedWS}
Simplified schematic (2 dimensional 2$\times$2 \textbf{k}/\textbf{q} grid) of the electron-phonon matrix elements where $\mathbf{r}_n$ and $\mathbf{r}_m$ indicate the position of the electronic Wannier centres, and $\boldsymbol{\tau}_{\kappa}$ is the position of the atom $\kappa$ within a primitive cell.
The square lattice indicates the unit cells of the crystal and $\mathbf{R}_p$ and $\mathbf{R}_{p''}$ are the lattice vectors for the electronic and phononic grids, respectively.
When performing the electron or phonon Fourier transform, we use the direct lattice point $p$ or $p''$, respectively, if the trial point $\mathbf{t}=\mathbf{R}_p + \mathbf{r}_m - r_n$ or $\mathbf{t}''=\mathbf{R}_{p''} + \boldsymbol{\tau}_\kappa - \mathbf{r}_n$ is closer to $\mathbf{0}$ than any other supercell center
$\mathbf{S}$.
The red square denotes the central supercell.
  }
\end{figure}

In the case of the electron-phonon matrix elements, we exploit the same ideas as above, this time for both the Wannier functions and the atomic positions.
The Wigner-Seitz cells for electrons and phonons are chosen so that both the $|\mathbf{R}_p + \mathbf{r}_m - \mathbf{r}_n|$ and $|\mathbf{R}_{p''} + \boldsymbol{\tau}_\kappa - \mathbf{r}_n|$ distances are minimized:
\begin{multline} \label{eq:gkkWS}
      g_{nm\kappa\alpha}(\mathbf{R}_p, \mathbf{R}_{p''}) = \frac{1}{N_pN_{p''}}
\!\!\!\!\!\! \!\!\sum_{\substack{\mathbf{k} \in (i_{p} \times j_{p} \times o_{p}) \\
\mathbf{q} \in (i_{p''} \times j_{p''} \times o_{p''})}  } \!\!\!\!\!\!\!\!\!\!\!\!\!\!\!
e^{-i(\mathbf{k}\cdot \mathbf{R}_p+\mathbf{q}\cdot\mathbf{R}_{p''})} \!\! \sum_{m'n'\nu}
\\
     \times  \sqrt{\frac{2M_\kappa \omega_{\mathbf{q}\nu}}{\hbar}}
e_{\kappa\alpha,\mathbf{q}\nu}^{\dagger} U_{mm'\mathbf{k+q}}^{\dagger}
g_{m'n'\nu}(\mathbf{k,q})  U_{n'n\mathbf{k}},
\end{multline}
where the lattice vectors are obtained such that  $\mathbf{R}_{p''} + \boldsymbol{\tau}_{\kappa}  - \mathbf{r}_{n} \in (i_{p''} \times j_{p''} \times o_{p''})$ and $\mathbf{R}_{p} +  \mathbf{r}_{m}  - \mathbf{r}_{n} \in (i_{p} \times j_{p} \times
o_{p})$.
We note that the set of lattice vectors $\mathbf{R}_{p''}$ used for the interpolation of electron-phonon matrix elements can in general be different from the set employed for the interpolation of the dynamical matrix.
This procedure used to determine the optimal set of vectors is sketched in Fig.~\ref{fig:optimizedWS}.
The set of direct lattice vectors $\mathbf{R}_{p}$ and  $\mathbf{R}_{p''}$ are selected such that the trial point $\mathbf{t}=\mathbf{R}_p + \mathbf{r}_m - \mathbf{r}_n$, constructed from the distance between two Wannier centers, or $\mathbf{t}^{''}=\mathbf{R}_{p''} + \boldsymbol{\tau}_\kappa - \mathbf{r}_n$ constructed from a Wannier center and an atomic position in the central cell and any other cell, is closer to the supercell center $\mathbf{0}$ than any other supercell centers $\mathbf{S}$.
In the case of a trial point at equal distance between $\mathbf{0}$ and $\mathbf{S}$ (i.e., on the surface of the supercell Wigner-Seitz cell) a weight proportional to the number of degeneracies is given.
The optimal Wigner-Seitz construction is activated with the new \texttt{use\_ws} input variable in EPW.
As a sanity check, we verified that the mobility of 3C-SiC calculated on a 8$\times$8$\times$8 \textbf{k}-point and 4$\times$4$\times$4 \textbf{q}-point grid was the same as the one obtained from a 2$\times$1$\times$1 supercell with a 4$\times$8$\times$8 \textbf{k}-point and 2$\times$4$\times$4 \textbf{q}-point grid.

\begin{figure}[t]
  \centering
  \includegraphics[width=0.98\linewidth]{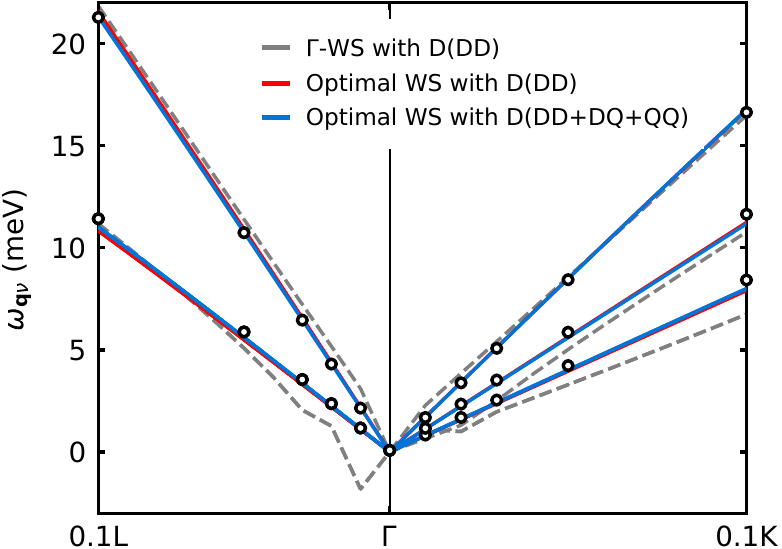}
  \caption{\label{fig:wsphonons}
Acoustic phonon dispersions of 3C-SiC near the zone center, interpolated from a coarse 4$\times$4$\times$4 $\mathbf{k}$-point and $\mathbf{q}$-point grid using a $\Gamma$-centered fixed Wigner-Seitz cell with dipole-dipole correction (dash gray line),
an optimal Wigner-Seitz cell with dipole-dipole correction (red line), and an optimal Wigner-Seitz cell with dipole-dipole, dipole-quadrupole and quadrupole-quadrupole corrections (blue line).
The black dots are reference datapoints obtained via direct DFPT calculations.
  }
\end{figure}

\begin{figure}[t]
  \centering
  \includegraphics[width=0.98\linewidth]{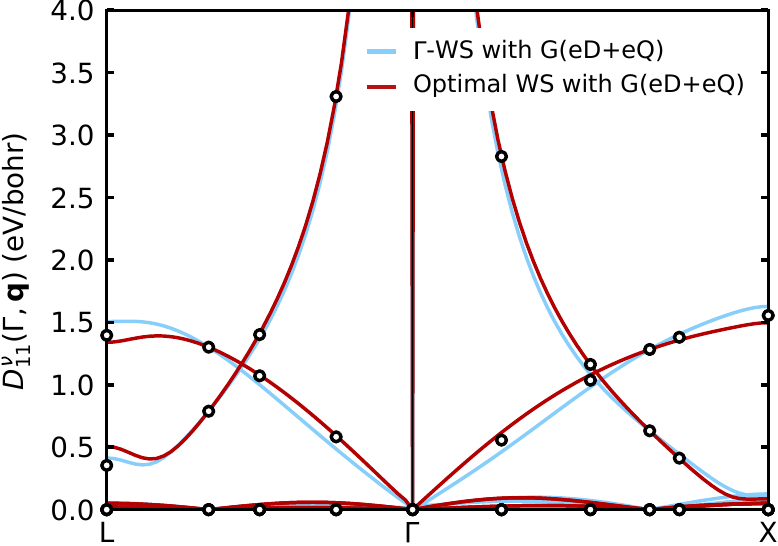}
  \caption{\label{fig:wscomparison}
Deformation potential of 3C-SiC without spin-orbit coupling of the first electronic band ($m=n=1$) on a coarse 3$\times$3$\times$3 $\mathbf{k}$-point and $\mathbf{q}$-point grid including dipole and quadrupole corrections.
The Wigner-Seitz cell is either $\Gamma$-centered or constructed using $\mathbf{R}_{p} + \mathbf{r}_{m}  - \mathbf{r}_{n} \in (i_{p} \times j_{p} \times o_{p})$ for the electrons and $\mathbf{R}_{p''} + \boldsymbol{\tau}_{\kappa}  - \mathbf{r}_{n} \in (i_{p''} \times j_{p''} \times o_{p''})$ for the phonons, respectively.
The black dots are reference datapoints obtained via direct DFPT calculations.
}
\end{figure}

In Fig.~\ref{fig:wsphonons} we report the calculated phonon dispersions of the three acoustic modes of 3C-SiC.
The dashed gray line represents the interpolation from a coarse $\mathbf{q}$-point grid using a $\Gamma$-centered fixed Wigner-Seitz cell  with dipole-dipole correction.
The red line is for an optimal Wigner-Seitz cell with dipole-dipole correction.
The blue line is for an optimal Wigner-Seitz cell with dipole-dipole, dipole-quadrupole and quadrupole-quadrupole corrections.
We can see that the $\Gamma$-centered Wigner-Seitz cell construction yields spurious soft phonon modes along the $\Gamma-L$ direction.
The improved Wigner-Seitz construction eliminates this artifact.

We also show in Fig.~\ref{fig:wscomparison} a comparison between the deformation potential of 3C-SiC computed with a $\Gamma$-centered Wigner-Seitz cell or with the improved construction.
The new construction produces results closer to those obtained directly with DFPT everywhere.
We note that the results shown in Fig.~\ref{fig:wscomparison} are for an unconverged small coarse grid of 3$\times$3$\times$3 for both $\mathbf{k}$- and $\mathbf{q}$-points.
When using denser coarse grids, these difference rapidly disappear (at least for the present case of cubic semiconductors).
Given that we use much denser coarse grids in this work, for simplicity we employed the basic $\Gamma$-centered construction in all calculations reported in this work.

\section{Additional figures and tables}\label{app.1}

\begin{figure}[ht]
  \centering
  \includegraphics[width=0.99\linewidth]{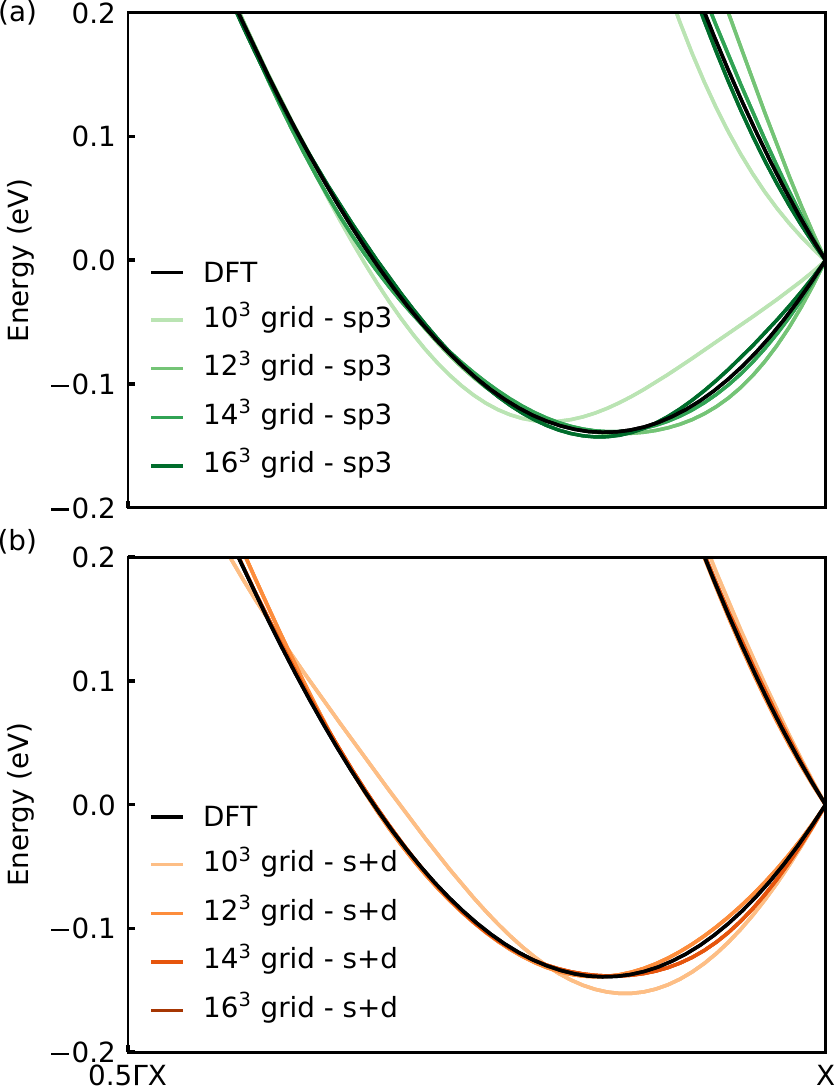}
  \caption{\label{fig:sibandswannier}
Conduction band edge of silicon -- including spin-orbit coupling (SOC) -- where the zero energy has been set at the X high-symmetry point and computed using DFT (black line).
(a) Electronic bands obtained using Wannier interpolation with $\mathbf{k}$-point grids ranging from $10^3$ to $16^3$ and constructed from an initial set of four $sp^3$ projections on each Si atom.
Both the valence and conduction bands have been Wannierized, totalling 16 Wannier functions due to SOC.
(b)  Electronic bands obtained using Wannier interpolation with $\mathbf{k}$-point grids ranging from $10^3$ to $16^3$ and constructed from an initial set of one $s$ and five $d$ projections on one Si atom only.
Only the conduction bands have been Wannierized, totalling 12 Wannier functions due to SOC.
  }
\end{figure}

\begin{figure}[ht]
  \centering
  \includegraphics[width=0.99\linewidth]{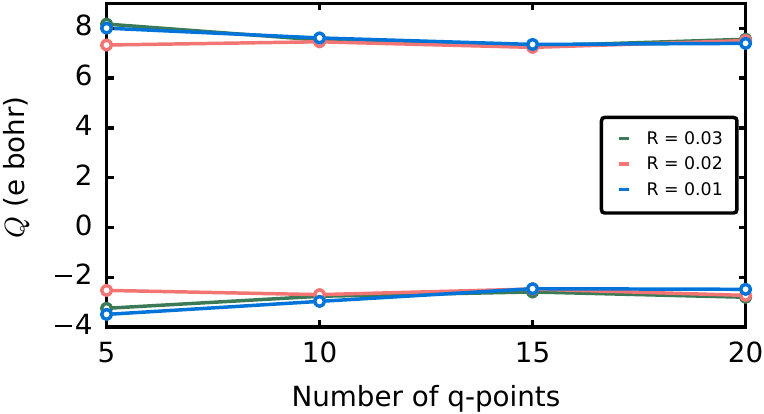}
  \caption{\label{fig:quad_fit}
Effective quadrupole term $\tilde{Q}_{\kappa\alpha}^{\beta\gamma}$ obtained from the least-mean square fit minimization of direct DFPT calculations for 3C-SiC.
In these calculations we used the LDA exchange and correlation functional and without nonlinear core correction, so that we can compare with Ref.~\onlinecite{Brunin2020}.
The two non-equivalent values of the quadrupole tensor are obtained by using the phonon frequencies, Born-effective charges, dielectric constant, eigendisplacement vectors and wavefunction overlaps from explicit DFPT calculations, and by optimizing the quadrupole tensor using Eq.~\eqref{gldq}.
We repeat the fitting for several sets of $\mathbf{q}$-points on spherical surfaces of varying radii $R$.
The radii are in units of $2\pi/a$.
  }
\end{figure}

\begin{table*}
\begin{footnotesize}
  \begin{tabular}{>{\rowmac}r >{\rowmac}r >{\rowmac}r >{\rowmac}r >{\rowmac}r >{\rowmac}r >{\rowmac}r >{\rowmac}r >{\rowmac}r >{\rowmac}r | >{\rowmac}r >{\rowmac}r >{\rowmac}r >{\rowmac}r >{\rowmac}r >{\rowmac}r<{\clearrow}}
  \toprule\\
Compounds & \multicolumn{2}{r}{Coarse}& Temp. & Wind. & Fine & \multicolumn{2}{c}{Hole mobility} & \multicolumn{2}{c}{Hall factor} & Wind. & Fine & \multicolumn{2}{c}{Electron mobility} & \multicolumn{2}{c}{Hall factor}  \\
          &  k & q      & (K)  & eV   & grids &  SERTA & BTE & SERTA & BTE & eV   & grid & SERTA & BTE & SERTA & BTE  \\
\hline
Diamond & 8$^3$ & 4$^3$ & 100 &   0.40 & 100$^3$ & 6694.2 & 6700.1 & 0.721 & 0.749 &   0.40 & 80$^3$ &  11016 &  10675 & 0.656 & 0.662 \\
        &       &       & 300 &        &         & 1048.5 & 1051.3 & 0.755 & 0.792 &        &        & 1713.4 & 1746.3 & 0.809 & 0.795 \\
\bb     & 8$^3$ & 4$^3$ & 100 &   0.30 & 100$^3$ & 6694.4 & 6700.4 & 0.721 & 0.745 &   0.30 & 80$^3$ &  11016 &  10676 & 0.656 & 0.662 \\
\bb     &       &       & 300 &        &         & 1049.6 & 1052.4 & 0.755 & 0.793 &        &        & 1714.2 & 1747.0 & 0.809 & 0.795 \\
        & 8$^3$ & 4$^3$ & 100 &   0.20 & 100$^3$ & 6703.6 & 6709.5 & 0.720 & 0.748 &   0.20 & 80$^3$ &  11035 &  10693 & 0.655 & 0.661 \\
        &       &       & 300 &        &         & 1084.2 & 1087.7 & 0.752 & 0.790 &        &        & 1778.2 & 1817.1 & 0.804 & 0.792 \\
\hline
Silicon & 8$^3$ & 4$^3$ & 100 &   0.30 &  60$^3$ & 7157.9 & 7053.0 & 0.129 & 0.135 &   0.30 &  60$^3$ &  12592 & 13137  & 0.638 & 0.593 \\
        &       &       & 300 &        &         & 691.43 & 708.94 & 0.292 & 0.335 &        &         & 1331.0 & 1358.9 & 0.939 & 0.900 \\
\bb     & 8$^3$ & 4$^3$ & 100 &   0.20 &  60$^3$ & 7175.5 & 7070.4 & 0.129 & 0.135 &   0.20 &  60$^3$ &  12618 & 13163  & 0.637 & 0.592 \\
\bb     &       &       & 300 &        &         & 695.19 & 712.82 & 0.337 & 0.295 &        &         & 1334.1 & 1362.2 & 0.937 & 0.898 \\
        & 8$^3$ & 4$^3$ & 100 &   0.15 &  60$^3$ & 7236.4 & 7130.4 & 0.128 & 0.134 &   0.15 &  60$^3$ &  12740 & 13290  & 0.631 & 0.586 \\
        &       &       & 300 &        &         & 710.44 & 728.95 & 0.297 & 0.339 &        &         & 1352.5 & 1381.2 & 0.927 & 0.889 \\
\hline
GaAs    & 8$^3$ & 4$^3$ & 100 &   0.30 & 100$^3$ & 3118.7 & 3160.3 & 0.518 & 0.535 &   0.30 & 200$^3$ & 529560 & 660550 & 0.711 & 0.801 \\
        &       &       & 300 &        &         & 326.35 & 432.33 & 0.932 & 0.907 &        &         &  23875 &  45296 & 1.060 & 0.971 \\
\bb     & 8$^3$ & 4$^3$ & 100 &   0.20 & 100$^3$ & 3125.2 & 3166.8 & 0.517 & 0.534 &   0.20 & 200$^3$ & 531150 & 662530 & 0.709 & 0.799 \\
\bb     &       &       & 300 &        &         & 326.78 & 433.04 & 0.933 & 0.908 &        &         &  23925 &  45253 & 1.059 & 0.973 \\
        & 8$^3$ & 4$^3$ & 100 &   0.15 & 100$^3$ & 3156.8 & 3198.9 & 0.511 & 0.529 &   0.15 & 200$^3$ & 537830 & 670860 & 0.701 & 0.789 \\
        &       &       & 300 &        &         & 329.13 & 436.55 & 0.953 & 0.936 &        &         &  24216 &  45262 & 1.054 & 0.981 \\
\hline
3C-SiC  & 8$^3$ & 4$^3$ & 100 &   0.30 & 60$^3$ &  274.85 & 308.89 & 0.457 & 0.470 &\bf 0.30 &\bf 60$^3$ &\bf 3418.8 &\bf 3123.1 &\bf 0.964 &\bf 1.111 \\
        &       &       & 300 &        &        &   82.11 &  91.82 & 0.532 & 0.583 & \bf       &\bf        &\bf 875.61 &\bf 923.19 &\bf 0.727 &\bf 0.780 \\
        & 8$^3$ & 4$^3$ & 100 &\bf 0.20 & \bf60$^3$ & \bf 275.03 &\bf 309.10 &\bf 0.457 &\bf 0.469 &   0.20 & 60$^3$ & 3423.1 & 3127.1 & 0.963 & 1.109 \\
        &       &       & 300 &        &        &\bf   82.76 &\bf  92.56 &\bf 0.534 &\bf 0.584 &        &        & 893.84 & 943.62 & 0.731 & 0.784 \\
        & 8$^3$ & 4$^3$ & 100 &   0.15 & 60$^3$ &  276.16 & 310.37 & 0.455 & 0.467 &   0.15 & 60$^3$ & 3443.3 & 3145.5 & 0.958 & 1.103 \\
        &       &       & 300 &        &        &   87.12 &  97.76 & 0.544 & 0.592 &        &        & 1004.1 & 1053.6 & 0.726 & 0.784 \\
\hline
AlP     & 8$^3$ & 4$^3$ & 100 &   0.30 & 60$^3$ & 208.38 & 277.63 & 0.263 & 0.538 &   0.30 & 60$^3$ & 1036.02 & 1276.04 & 0.807 & 0.964 \\
        &       &       & 300 &        &        &  42.74 &  54.74 & 0.396 & 0.589 &        &        &  190.76 &  274.31 & 0.868 & 0.915 \\
\bb     & 8$^3$ & 4$^3$ & 100 &   0.20 & 60$^3$ & 208.67 & 278.02 & 0.262 & 0.538 &   0.20 & 60$^3$ & 1037.23 & 1277.52 & 0.806 & 0.963 \\
\bb     &       &       & 300 &        &        &  42.79 &  54.81 & 0.400 & 0.594 &        &        &  190.87 &  274.36 & 0.867 & 0.914 \\
        & 8$^3$ & 4$^3$ & 100 &   0.15 & 60$^3$ & 210.40 & 280.33 & 0.260 & 0.533 &   0.15 & 60$^3$ & 1042.66 & 1284.22 & 0.802 & 0.958 \\
        &       &       & 300 &        &        &  43.09 &  55.24 & 0.416 & 0.614 &        &        &  191.58 &  274.84 & 0.866 & 0.914 \\
\hline
GaP     & 8$^3$ & 4$^3$ & 100 &   0.30 & 60$^3$ & 846.71 & 937.41 & 0.257 & 0.414 &   0.30 & 60$^3$ & 2379.37 & 2237.48 & 0.619 & 0.722 \\
        &       &       & 300 &        &        & 136.15 & 171.62 & 0.412 & 0.546 &        &        &  225.58 &  275.75 & 0.892 & 0.909 \\
\bb     & 8$^3$ & 4$^3$ & 100 &   0.20 & 60$^3$ & 848.06 & 938.90 & 0.257 & 0.414 &   0.20 & 60$^3$ & 2382.49 & 2240.41 & 0.618 & 0.721 \\
\bb     &       &       & 300 &        &        & 136.30 & 171.88 & 0.419 & 0.555 &        &        &  225.76 &  276.00 & 0.891 & 0.909 \\
        & 8$^3$ & 4$^3$ & 100 &   0.15 & 60$^3$ & 853.61 & 945.04 & 0.255 & 0.411 &   0.15 & 60$^3$ & 2399.07 & 2256.00 & 0.614 & 0.716 \\
        &       &       & 300 &        &        & 137.25 & 173.31 & 0.436 & 0.575 &        &        &  226.73 &  277.23 & 0.889 & 0.906 \\
 \hline
c-BN    & 8$^3$ & 4$^3$ & 100 &   0.40 & 60$^3$ & 985.49 & 1213.15 & 0.287 & 0.474 & 0.40 & 60$^3$ & 4856.13 & 4430.66 & 0.958 & 1.094 \\
        &       &       & 300 &        &        & 243.48 &  282.42 & 0.430 & 0.592 &      &        & 1255.37 & 1282.49 & 0.802 & 0.862 \\
\bb     & 8$^3$ & 4$^3$ & 100 &   0.30 & 60$^3$ & 985.52 & 1213.18 & 0.287 & 0.474 & 0.30 & 60$^3$ & 4856.26 & 4430.78 & 0.958 & 1.094 \\
\bb     &       &       & 300 &        &        & 243.62 &  282.59 & 0.430 & 0.592 &      &        & 1257.75 & 1285.23 & 0.803 & 0.863 \\
        & 8$^3$ & 4$^3$ & 100 &   0.20 & 60$^3$ & 986.82 & 1214.77 & 0.287 & 0.474 & 0.20 & 60$^3$ & 4863.11 & 4437.03 & 0.957 & 1.093 \\
        &       &       & 300 &        &        & 256.59 &  300.79 & 0.434 & 0.604 &      &        & 1427.53 & 1453.02 & 0.816 & 0.879 \\
        & 8$^3$ & 4$^3$ & 100 &   0.15 & 60$^3$ & 992.62 & 1221.85 & 0.287 & 0.472 & 0.15 & 60$^3$ & 4889.02 & 4460.67 & 0.953 & 1.088 \\
        &       &       & 300 &        &        & 272.69 &  324.29 & 0.436 & 0.614 &      &        & 1622.44 & 1631.41 & 0.798 & 0.867 \\
\hline
AlAs    & 8$^3$ & 4$^3$ & 100 &   0.30 & 60$^3$ & 547.44 & 665.68 & 0.372 & 0.613 & 0.30 & 60$^3$ & 1535.47 & 1595.03 & 0.674 & 0.865 \\
        &       &       & 300 &        &        &  88.59 & 118.90 & 0.710 & 0.835 &      &        &  212.47 &  280.23 & 0.864 & 0.912 \\
\bb     & 8$^3$ & 4$^3$ & 100 &   0.20 & 60$^3$ & 548.32 & 666.75 & 0.372 & 0.613 & 0.20 & 60$^3$ & 1537.13 & 1596.76 & 0.673 & 0.864 \\
\bb     &       &       & 300 &        &        &  88.71 & 119.11 & 0.711 & 0.836 &      &        &  212.57 &  280.32 & 0.863 & 0.911 \\
        & 8$^3$ & 4$^3$ & 100 &   0.15 & 60$^3$ & 553.67 & 673.25 & 0.368 & 0.607 & 0.15 & 60$^3$ & 1544.94 & 1604.87 & 0.670 & 0.860 \\
        &       &       & 300 &        &        &  89.35 & 120.13 & 0.721 & 0.852 &      &        &  212.99 &  280.73 & 0.862 & 0.909 \\
        & 8$^3$ & 4$^3$ & 100 &   0.10 & 60$^3$ & 575.50 & 699.82 & 0.355 & 0.584 & 0.10 & 60$^3$ & 1602.74 & 1666.23 & 0.647 & 0.828 \\
        &       &       & 300 &        &        &  93.27 & 125.11 & 0.779 & 0.943 &      &        &  225.04 &  291.55 & 0.863 & 0.928 \\

\hline
AlSb    & 8$^3$ & 4$^3$ & 100 &   0.30 & 60$^3$ & 1720.10 & 1785.65 & 0.346 & 0.515 & 0.30 & 60$^3$ & 1460.25 & 1394.00 & 0.645 & 0.703 \\
        &       &       & 300 &        &        &  304.05 &  383.03 & 0.828 & 0.849 &      &        &  196.86 &  221.20 & 0.835 & 0.842 \\
\bb     & 8$^3$ & 4$^3$ & 100 &   0.20 & 60$^3$ & 1722.36 & 1787.99 & 0.345 & 0.515 & 0.20 & 60$^3$ & 1462.17 & 1395.84 & 0.644 & 0.702 \\
\bb     &       &       & 300 &        &        &  304.61 &  383.90 & 0.837 & 0.861 &      &        &  196.95 &  221.33 & 0.836 & 0.842 \\
        & 8$^3$ & 4$^3$ & 100 &   0.15 & 60$^3$ & 1735.02 & 1801.14 & 0.343 & 0.511 & 0.15 & 60$^3$ & 1471.69 & 1404.93 & 0.640 & 0.698 \\
        &       &       & 300 &        &        &  308.14 &  388.85 & 0.892 & 0.934 &      &        &  197.38 &  221.81 & 0.839 & 0.847 \\
\hline
SrO     & 8$^3$ & 4$^3$ & 100 &   0.30 & 60$^3$ & 319.91 & 331.05 & 0.795 & 0.794 & 0.30 & 60$^3$ & 166.66 & 152.33 & 0.761 & 0.793 \\
        &       &       & 300 &        &        &   5.89 &   7.64 & 0.940 & 0.931 &      &        &  18.36 &  22.15 & 0.965 & 0.941 \\
\bb     & 8$^3$ & 4$^3$ & 100 &   0.20 & 60$^3$ & 321.22 & 332.41 & 0.791 & 0.791 & 0.20 & 60$^3$ & 166.93 & 152.58 & 0.760 & 0.792 \\
\bb     &       &       & 300 &        &        &   5.93 &   7.68 & 0.935 & 0.929 &      &        &  18.36 &  22.09 & 0.964 & 0.940 \\
        & 8$^3$ & 4$^3$ & 100 &   0.15 & 60$^3$ & 327.94 & 339.35 & 0.775 & 0.774 & 0.15 & 60$^3$ & 167.90 & 153.46 & 0.756 & 0.787 \\
        &       &       & 300 &        &        &   6.59 &   8.37 & 1.028 & 0.972 &      &        &  18.41 &  22.01 & 0.965 & 0.944 \\
  \botrule
  \end{tabular}
  \caption{\label{table:energywindow}
Sensitivity of the calculated mobility and Hall factor to the size of the energy window in the BTE.
The calculations have been performed using adaptive smearing.
The numbers indicated in bold represent the smallest window such that the results are unaffected.
The same energy window has been used throughout this work.
}
\end{footnotesize}
\end{table*}

\begin{table*}[b]
  \begin{tabular}{r r r r r r r r r r r r r }
  \toprule\\
  & \multicolumn{2}{c}{Densest grids}  & \multicolumn{4}{c}{k/q extrapolation}  &   & \multicolumn{4}{c}{k/q extrapolation}  &  \\
\cline{4-7}
\cline{9-12}
  & \multicolumn{2}{c}{Drift mobility}  & \multicolumn{2}{c}{Drift mobility}      & \multicolumn{2}{c}{Hall mobility}  &  Exp. & \multicolumn{2}{c}{Drift slope}      & \multicolumn{2}{c}{Hall slope} & Exp.  \\
  Materials       & SERTA   & BTE     & SERTA   & BTE    & SERTA & BTE  &   mobility & SERTA   & BTE    & SERTA & BTE & slope \\
\hline
\\[-1.0em]
     C-e & 1613 & 1671 & 1583 & 1634 & 1664 & 1705 & \textbf{1800}~\cite{Haynes2020} & -1.81 & -1.82 & -1.79 & -1.81 & -1.4$^\dagger$~\cite{Nesladek2008}  \\
         &      &      &      &      &      &      & 1800~\cite{Madelung2002} &   &   &  &  & -1.6$^\dagger$~\cite{Gabrysch2011} \\
         &      &      &      &      &      &      & 1802$^\dagger$~\cite{Jansen2013} &   &   &  &  & \textbf{-1.88}$^\dagger$~\cite{Gabrysch2011} \\
         &      &      &      &      &      &      & 1940$^\dagger$~\cite{Gabrysch2011} \\
         &      &      &      &      &      &      & 2000~\cite{Vavilov1976} \\
         &      &      &      &      &      &      & 2750$^\dagger$~\cite{Nesladek2008} \\
     C-h & 2256 & 2290 & 2364 & 2410 & 2409 & 2467 & 1550~\cite{Vavilov1976}   & -1.97 & -1.99 & -1.88 & -1.89 & -1.72~\cite{Isberg2005} \\
         &      &      &      &      &      &      & \textbf{2100}~\cite{Madelung2002} &   &   &  &  &   -1.5~\cite{Madelung2002} \\
         &      &      &      &      &      &      & 2250$^\dagger$~\cite{Nesladek2008} &   &   &  &  & -1.5$^\dagger$~\cite{Nesladek2008} \\
         &      &      &      &      &      &      & 2300$^\dagger$~\cite{Gabrysch2011} &   &   &  &  & \textbf{-2.06}$^\dagger$~\cite{Gabrysch2011}\\
         &      &      &      &      &      &      & 2534$^\dagger$~\cite{Jansen2013} &   &   &  &  & -2.8~\cite{Madelung2002} \\
    Si-e & 1383 & 1429 & 1352 & 1383 & 1782 & 1770 & \textbf{1350}$^\dagger$~\cite{Madelung2002} & -2.19 & -2.21 & -2.06 & -2.05 & \textbf{-2.31}~\cite{Norton1973} \\
         &      &      &      &      &      &      & 1350$^\dagger$~\cite{Ludwig1956}            &   &   &  &  & -2.42~\cite{Madelung2002} \\
         &      &      &      &      &      &      & 1350$^\dagger$~\cite{Jacoboni1977}  \\
         &      &      &      &      &      &      & 1430$^\dagger$~\cite{Norton1973}  \\
         &      &      &      &      &      &      & 1450$^\dagger$~\cite{Sze2007}  \\
    Si-h &  623 &  627 &  600 &  601 &  572 &  574 & \textbf{445}$^\dagger$~\cite{Jacoboni1977}          & -2.55 & -2.53 & -3.02 & -2.93 & -2.44~\cite{Ludwig1956}  \\
         &      &      &      &      &      &      & 480$^\dagger$~\cite{Ludwig1956}            &   &   &  &  & -2.2~\cite{Madelung2002} \\
         &      &      &      &      &      &      & 480$^\dagger$~\cite{Madelung2002} &   &   &  &  & \textbf{-3.09}~\cite{Logan1960}  \\
         &      &      &      &      &      &      & 500$^\dagger$~\cite{Sze2007}  \\
  GaAs-e & 8743 &14799 & 8853 &15099 &12354 &17860 & 7200~\cite{Madelung2003}                   & -2.69 & -2.21 & -2.66 & -2.29 & -1.50~\cite{Stanley1991} \\
         &      &      &      &      &      &      & 8000~\cite{Hicks1969}                      &   &   &  &  & -2.12~\cite{Hicks1969}  \\
         &      &      &      &      &      &      & 8865~\cite{Madelung2003}                   &   &   &  &  & \textbf{-2.39}~\cite{Lin1982}    \\
         &      &      &      &      &      &      & 8500~\cite{Rode1971} \\
         &      &      &      &      &      &      & \textbf{9000}~\cite{Madelung2003}  \\
         &      &      &      &      &      &      & 9750~\cite{Madelung2003} \\
  GaAs-h &  392 &  500 &  389 &  500 &  988 & 1068 & 396~\cite{Hill1970}                        & -2.45 & -2.22 & -2.40 & -2.31 & -2.23~\cite{Hill1970} \\
         &      &      &      &      &      &      & 450~\cite{Wenzel1998}                      &   &   &  &  & \textbf{-2.39}~\cite{Kim1991} \\
         &      &      &      &      &      &      & 450~\cite{Kim1991}  \\
         &      &      &      &      &      &      & 450~\cite{Sotoodeh2000} \\
         &      &      &      &      &      &      & 460~\cite{Mears1971} \\
         &      &      &      &      &      &      & \textbf{500}~\cite{Haynes2020} \\
   SiC-e & 1885 & 2068 & 1803 & 1944 & 2204 & 2257 & 510~\cite{Madelung2003}                    & -2.90 & -2.92 & -2.85 & -2.86 & \textbf{-1.50}~\cite{Shinohara1988} \\
         &      &      &      &      &      &      & 670~\cite{Kaplan1985}  \\
         &      &      &      &      &      &      & 733~\cite{Shinohara1988}  \\
         &      &      &      &      &      &      & 800~\cite{Mnatsakanov2001}  \\
         &      &      &      &      &      &      & 890~\cite{Nelson1966}  \\
         &      &      &      &      &      &      & \textbf{1000}~\cite{Bhatnagar1993}  \\
   SiC-h &  166 &  170 &  165 &  170 &  105 &  111 & 21~\cite{Madelung2003}                     & -2.06 & -2.06 & -1.91 & -1.97 & \textbf{-1.75}~\cite{Lebedev2008}  \\
         &      &      &      &      &      &      & 40~\cite{Yamanaka1987}  \\
         &      &      &      &      &      &      & \textbf{100}~\cite{Lebedev2008}  \\
   AlP-e &  461 &  556 &  454 &  554 &  629 &  698 & 80~\cite{Madelung2003}                      & -3.15 & -2.99 & -2.89 & -2.79 & - \\
   AlP-h &   94 &  102 &   93 &  101 &   84 &   92 & -                                           & -2.96 & -2.85 & -3.43 & -3.31 & - \\
   GaP-e &  281 &  325 &  284 &  328 &  409 &  442 & 165~\cite{Craford1971}                      & -2.58 & -2.45 & -2.43 & -2.35 & -1.52~\cite{Kao1983}\\
         &      &      &      &      &      &      & \textbf{300}~\cite{Haynes2020}              &   &   &  &  & \textbf{-2.12}~\cite{Craford1971} \\
   GaP-h &  228 &  271 &  231 &  276 &  315 &  353 & 135~\cite{Madelung2002}                     & -2.96 & -2.81 & -3.00 & -2.89 & -1.95~\cite{Kao1983} \\
         &      &      &      &      &      &      & \textbf{150}~\cite{Haynes2020}              &   &   &  &  & \textbf{-2.37}~\cite{Casey1969} \\
    BN-e &  573 & 1002 &  470 &  967 &  456 &  956 & -                                           & -1.28 & -1.40 & -1.20 & -1.33 & - \\
    BN-h &  196 &  321 &  170 &  319 &  124 &  281 & 500~\cite{Madelung2002}                     & -1.31 & -1.49 & -1.18 & -1.51 &  \\
  AlAs-e &  409 &  507 &  405 &  501 &  578 &  640 & 161~\cite{Chand1984}                        & -3.14 & -2.94 & -2.84 & -2.71 & - \\
         &      &      &      &      &      &      & \textbf{294}~\cite{Madelung2003}            &   &   &  &  &  \\
  AlAs-h &  181 &  217 &  180 &  215 &  337 &  357 &  60~\cite{Sze2007}                          & -3.08 & -2.86 & -2.79 & -2.73 & -\\
  AlSb-e &  298 &  332 &  303 &  335 &  415 &  431 & 200~\cite{Madelung2002}                     & -2.38 & -2.25 & -2.36 & -2.26 & \textbf{-1.46}~\cite{Stirn1966} \\
         &      &      &      &      &      &      & \textbf{300}~\cite{Haynes2020}              &   &   &  &  & \\
  AlSb-h &  491 &  598 &  490 &  600 &  918 & 1000 & 505~\cite{Madelung2002}                     & -2.70 & -2.46 & -2.54 & -2.42 & \textbf{-1.95}~\cite{Stirn1966} \\
         &      &      &      &      &      &      & \textbf{550}~\cite{Haynes2020}              &   &   &  &  & \\
   SrO-e &   38 &   53 &   38 &   53 &   53 &   65 & -                                           & -3.90 & -3.61 & -3.48 & -3.33 & - \\
   SrO-h &  3.5 &  5.0 &  3.4 &  4.9 &  3.5 &  5.0 & -                                           & -2.33 & -2.21 & -2.31 & -2.07 & -\\
  \botrule
  \end{tabular}
  \caption{\label{table:mobility_val} Room temperature drift SERTA and BTE mobility (cm$^2$/Vs) as well as temperature exponent, as evaluated at the densest fine $\mathbf{k}$/$\mathbf{q}$-point grids from Fig.~\ref{fig:fineconvergence}, and drift SERTA and BTE Hall mobility extrapolated to infinitely dense sampling.
The experimental data in bold are the reference values used in Figs.~\ref{fig:expmob} and \ref{fig:expmobscale}.
The $^\dagger$ symbol indicates time-of-flight drift mobility experiments, while no symbols indicate Hall measurements.
}
\end{table*}

\begin{figure}[ht]
  \centering
  \includegraphics[width=0.99\linewidth]{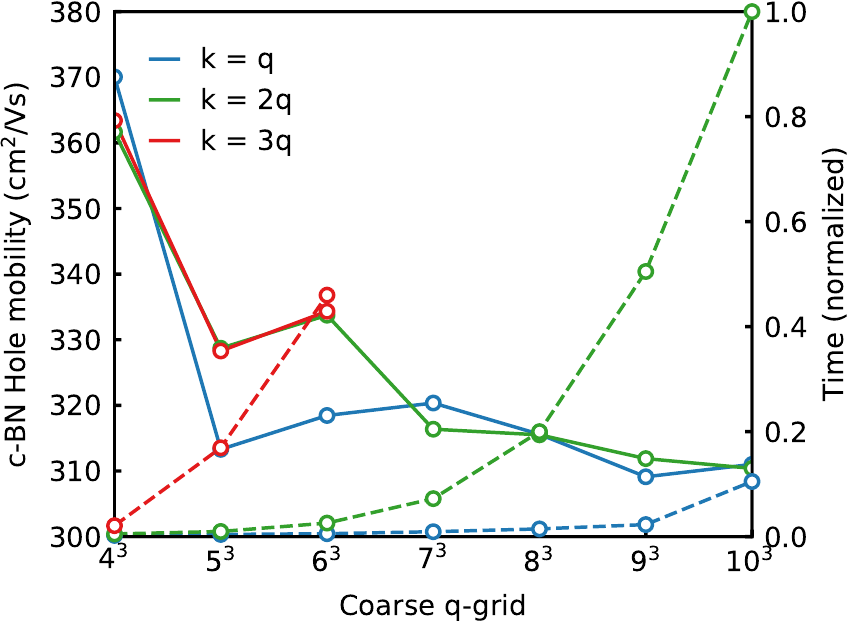}
  \caption{\label{fig:comparisoncoarse}
Calculated hole drift mobility of c-BN for various Brillouin zone grid sizes.
The horizontal axis indicates the coarse $\mathbf{q}$-point grid, and the lines correspond to different choices of the coarse $\mathbf{k}$-point grid: same as the $\mathbf{q}$-point grid (blue), twice as dense as the $\mathbf{q}$-point grid (green), and three times as dense as the \textbf{q}-point grid (red).
In all cases the fine grids were set to 60$^3$ points for both $\mathbf{k}$ and $\mathbf{q}$.
We also show with dashed lines the relative computational time needed in each case (normalized to the most time-consuming calculation).
These tests show that the results obtained by using equal coarse grids for electrons and phonons are of the same accuracy as those with denser electron grids, but much cheaper computationally.
Therefore we recommend to always use equal coarse grids.
}
\end{figure}

\begin{figure}[ht]
  \centering
  \includegraphics[width=0.95\linewidth]{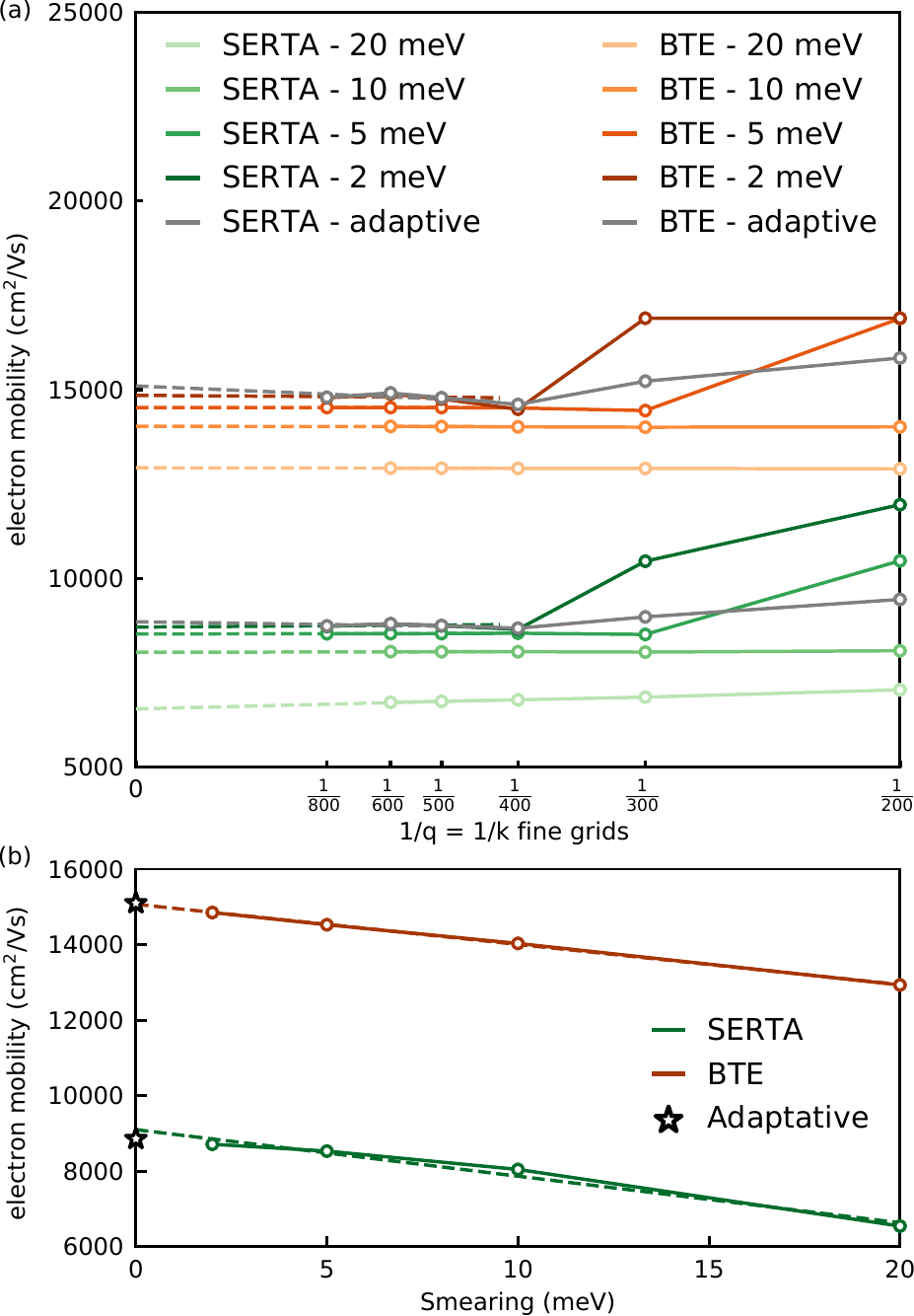}
  \caption{\label{fig:fineconvGaAs}
(a) Convergence of the calculated Hall mobility of GaAs with the fine Brillouin zone grid size.
Both SERTA and BTE results are shown using the converged coarse grids from Fig.~\ref{fig:coarseconv}.
We consider several smearing parameters as well as the adaptive smearing.
(b) Calculated electron mobility of GaAs, extrapolated to infinitely dense Brillouin zone sampling, for several smearing parameters.
The extrapolation of these results at zero smearing agrees with calculations using adaptive smearing (stars).
}
\end{figure}

\begin{figure*}[ht]
  \centering
  \includegraphics[width=0.99\linewidth]{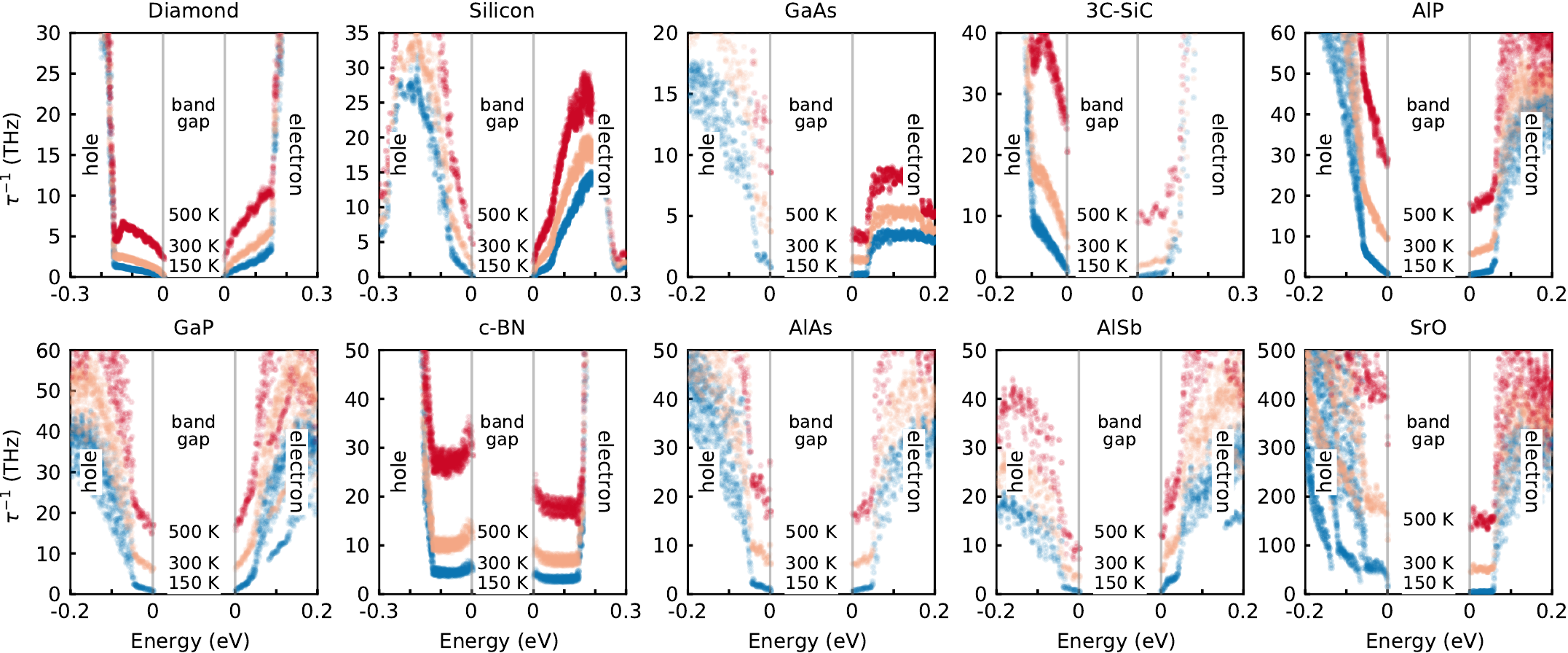}
  \caption{\label{fig:scattemp}
Calculated electron and hole scattering rates as a function of energy at 150~K, 300~K, and 500~K.
The energy range displayed here encompasses the energy window used in the calculations of carrier mobilities.
}
\end{figure*}

\end{document}